
\input epsf
\input phyzzx
\Pubnum={\vbox{\hbox{CERN-TH.7173/94}
\hbox{BONN-IR-94-01}\hbox{hep-th/9403129}}}
\pubnum={\vbox{\hbox{CERN-TH.7173/94}
\hbox{BONN-IR-94-01}\hbox{hep-th/9403129}}}
\date={February, 1994}
\pubtype={}
\def\caption#1{\vskip 0.1in\centerline{\vbox{\hsize 3.7in\noindent
     \tenpoint\baselineskip=10pt\strut #1\strut}}}
\def\a{\alpha}
\def\b{\beta}
\def\g{\gamma}
\def\d{\delta}
\def\ep{\epsilon}

\def\z{\zeta}

\def\th{\theta}

\def\k{\kappa}
\def\l{\lambda}
\def\m{\mu}

\def\x{\xi}

\def\vr{\varrho}
\def\s{\sigma}
\def\t{\tau}
\def\u{\upsilon}

\def\vp{\varphi}

\def\o{omega}

\def\G{\Gamma}
\def\D{\Delta}

\def\P{\Pi}
\def\S{\Sigma}

\def\P{\Phi}
\def\L{\Lambda}

\def\DE{{(p-x)(p-y)}}
\def\o{\over}
\def\p{\partial}
\def\ri{\rightarrow}
\def\CE{\centerline}
\def\pint{\oint _\infty {dp\over 2\pi i}}
\def\qint{\oint _\infty {dq\over 2\pi i}}
\def\DEp{{(p-x)(p-y)}}
\def\INTp{{V'(p)\over \sqrt{(p-x)(p-y)}}}
\def\h{{1\over 2}}

\def\DE{(1-tx)(1-ty)}
\def\INT{{g(t^{-1})\over \sqrt{\DE}}}

\def\inbar{\,\vrule height1.5ex width.4pt depth0pt}
\def\IB{\relax{\rm I\kern-.18em B}}
\def\IC{\relax\hbox{$\inbar\kern-.3em{\rm C}$}}
\def\ID{\relax{\rm I\kern-.18em D}}
\def\IE{\relax{\rm I\kern-.18em E}}
\def\IF{\relax{\rm I\kern-.18em F}}
\def\IG{\relax\hbox{$\inbar\kern-.3em{\rm G}$}}
\def\IH{\relax{\rm I\kern-.18em H}}
\def\II{\relax{\rm I\kern-.18em I}}
\def\IK{\relax{\rm I\kern-.18em K}}
\def\IL{\relax{\rm I\kern-.18em L}}
\def\IM{\relax{\rm I\kern-.18em M}}
\def\IN{\relax{\rm I\kern-.18em N}}
\def\IO{\relax\hbox{$\inbar\kern-.3em{\rm O}$}}
\def\IP{\relax{\rm I\kern-.18em P}}
\def\IQ{\relax\hbox{$\inbar\kern-.3em{\rm Q}$}}
\def\IR{\relax{\rm I\kern-.18em R}}
\font\cmss=cmss10 \font\cmsss=cmss10 at 7pt
\def\IZ{\relax\ifmmode\mathchoice
{\hbox{\cmss Z\kern-.4em Z}}{\hbox{\cmss Z\kern-.4em Z}}
{\lower.9pt\hbox{\cmsss Z\kern-.4em Z}}
{\lower1.2pt\hbox{\cmsss Z\kern-.4em Z}}\else{\cmss Z\kern-.4em
Z}\fi}
\def\IGa{\relax\hbox{${\rm I}\kern-.18em\Gamma$}}
\def\IPi{\relax\hbox{${\rm I}\kern-.18em\Pi$}}
\def\ITh{\relax\hbox{$\inbar\kern-.3em\Theta$}}
\def\IOm{\relax\hbox{$\inbar\kern-3.00pt\Omega$}}

\def\no{\noindent}

\def\np{{\it Nucl.~Phys. }}
\def\pl{{\it Phys.~Lett. }}

\def\mpl{{\it Mod.~Phys.~Lett. }}

\def\mko{m_k^{-}}

\def\mkI{m_k^{+}}
\def\mkpI{m_{k+1}^{+}}
\def\IT{t^{-1}}
\def\nkp{n_k^{+}}
\def\nkpp{n_{k+1}^{+}}
\def\nkm{n_k^{-}}
\def\nkpm{n_{k+1}^{-}}

\def\v{{\widehat v}}

\def\vo{{\widehat v_0}}

\def\z{z+u}
\def\DLR{{\buildrel \leftrightarrow \over \p}}

\def\DDLR{{\buildrel \leftrightarrow \over D}}
\def\DDLRt{{\buildrel \leftrightarrow \over \p_t }}

\def\px{\p _x}
\def\py{\p _y}
\def\pxy{\p ^2_{xy}}
\def\pp{\p _{+}}
\def\pms{\p _{-}}
\def\ppm{\p _{+-}^2}
\def\pt{\p _t}
\def\up{u_{+}}
\def\um{u_{-}}
\def\zp{z+\up}
\def\pint{\oint _\infty {dp\over 2\pi i}}
\def\qint{\oint _\infty {dq\over 2\pi i}}
\def\u{{\widehat u}}
\def\uo{{u_0}}

\def\h{{1\over 2}}
\def\sz{\sqrt{\zp}}

\def\tp{{\widetilde t}^{+}}
\def\tm{{\widetilde t}^{-}}

\def\to{\rightarrow}

\titlepage

\title{NON-PERTURBATIVE APPROACH TO \break 2D-SUPERGRAVITY AND
\break SUPER-VIRASORO CONSTRAINTS\foot{PhD Thesis at the University
of Bonn.}}
\vskip 0.4cm
\author{Melanie Becker}
\vskip 0.4cm
\CE{\it Theory Division, CERN}
\CE{\it CH-1211, Geneva 23, Switzerland}
\vskip 0.4cm
\CE{\it and}
\vskip 0.4cm
\CE{\it Physikalisches Institut, Universit\"at Bonn,}
\CE{\it Nussallee 12, D-53115 Bonn, Germany.}
\vskip 1.0cm

\abstract{The coupling of $N=1$ superconformal field theories of type
$(4m,2)$ to two-dimensional supergravity can be formulated
non-perturbatively in terms of a discrete super-eigenvalue model
proposed by Alvarez-Gaum\'e, Itoyama, Ma\~nes and Zadra. We derive
the superloop equations that describe,  in the double scaling limit,
the non-perturbative solution of this model. These equations are
equivalent to the double scaled super-Virasoro constraints satisfied
by the partition function. They are formulated in terms of a
$\widehat c=1$ theory, with a $\IZ_2$-twisted scalar field and a
Weyl-Majorana fermion in the Ramond sector. We have solved the
superloop equations to all orders in the genus expansion and obtained
the explicit expressions for the correlation functions of
gravitationally dressed scaling operators in the Neveu-Schwarz and
Ramond sector. In the double scaling limit, we obtain a formulation
of the model in terms of a new supersymmetric extension of the KdV
hierarchy. Integrating over the fermionic eigenvalues we obtain a
representation of the discrete supersymmetric free energy in terms of
the free energy of the one-matrix model.}

\vfill

\def\chaone{INTRODUCTION}
\def\seconeone{String Theory and 2D-Quantum Gravity}
\def\seconetwo{Superstrings}
\def\seconethree{Outline of the Thesis}
\def\chaoneone{PATH INTEGRAL APPROACH AND (SUPER-)LIOUVILLE THEORY}
\def\chaoneoneone{Bosonic String and Liouville Theory}
\def\chaoneonetwo{Non-Critical Superstring}
\def\chatwo{THE HERMITIAN ONE-MATRIX MODEL}
\def\sectwoone{Topological Expansion as a Large-$N$ Expansion
of a Matrix Model}
\def\sectwotwo{Virasoro Constraints and Loop Equations in the
Discrete
Theory}
\def\sectwothree{Solution of the Planar Loop Equation for Arbitrary
Potentials}
\def\sectwofourone{Basic Concepts on Integrable Hierarchies}
\def\sectwofourtwo{Double Scaling Limit, KdV and Continuum Virasoro
Constraints}

\def\chathree{THE SUPERSYMMETRIC EIGENVALUE MODEL FOR GENUS ZERO}
\def\secthreeone{Construction of the $N=1$ Supersymmetric Eigenvalue
Model}

\def\chafour{SUPERSYMMETRIC MATRIX~MODEL: ARBITRARY TOPOLOGIES}
\def\secfourone{Solution to the Planar Superloop Equations: General
Potential}
\def\secfourtwo{Continuum Super-Virasoro Constraints}
\def\secfourthree{Iterative Solution of the Continuum Superloop
Equations}
\def\secfourfour{Non-perturbative Solution of the Discrete
Supersymmetric Model}
\def\secfourfive{Solution to the Super-Virasoro Constraints in the
Double Scaling Limit}
\def\secfoursix{Super-Integrable Hierarchies and 2D Supergravity}
\def\chafive{CONCLUSIONS}
\endpage
\footline{\hfil}
{\vbox{\vskip 20cm\hbox{\hskip 12cm Dedicated to my mother
and}\hbox{\hskip 12cm to the memory of my father}}}
\endpage
\footline={\hss\tenrm\folio\hss}
\pageno=-3

\vbox{\vskip 2cm \hbox{\CE{\bf C{\tenpoint ONTENTS}} }}

\settabs 10 \columns
\vskip 3.5cm
\+& {\bf 1. I{\tenpoint NTRODUCTION}}&&&&&&&&&& \hfill {\bf 1} \cr
\+& 1.1. \seconeone &&&&&&&&&& \hfill 1\cr
\+& 1.2. \seconetwo &&&&&&&&&& \hfill 8\cr
\+& 1.3. \seconethree  &&&&&&&&&& \hfill 16 \cr
\vskip 1.5cm
\+& {\bf 2. P{\tenpoint ATH} I{\tenpoint NTEGRAL} A{\tenpoint
PPROACH} {\tenpoint AND} (S{\tenpoint UPER}-)L{\tenpoint IOUVILLE}
T{\tenpoint HEORY}}&&&&&&&&&& \hfill {\bf 19}\cr
\+& 2.1. \chaoneoneone &&&&&&&&&& \hfill 19\cr
\+& 2.2. \chaoneonetwo &&&&&&&&&& \hfill 27\cr
\vskip 1.5cm
\+ & {\bf 3. T{\tenpoint HE} H{\tenpoint ERMITIAN} O{\tenpoint
NE}-M{\tenpoint ATRIX} M{\tenpoint ODEL}} &&&&&&&&&& \hfill {\bf
34}\cr
\+& 3.1.  \sectwoone &&&&&&&&&& \hfill 34\cr
\+& 3.2.  \sectwotwo &&&&&&&&&& \hfill 41 \cr
\+& 3.3. \sectwothree &&&&&&&&&& \hfill 46\cr
\+& 3.4. \sectwofourone &&&&&&&&&& \hfill 54\cr
\+& 3.5. \sectwofourtwo &&&&&&&&&& \hfill 60\cr
\vskip 1.5cm
\+& {\bf 4. T{\tenpoint HE} S{\tenpoint UPERSYMMETRIC} E{\tenpoint
IGENVALUE} M{\tenpoint ODEL} {\tenpoint FOR} G{\tenpoint ENUS}
Z{\tenpoint ERO}} &&&&&&&&&& \hfill {\bf 64}\cr
\+& 4.1.  \secthreeone &&&&&&&&&& \hfill 64 \cr
\+& 4.1.1. The One-Matrix Model in Terms of Eigenvalues as Guiding
Principle &&&&&&&&&& \hfill 64\cr
\+& 4.1.2. The Generalization to the $N=1$ Supersymmetric Eigenvalue
Model &&&&&&&&&& \hfill 65\cr
\+& 4.2.   \secfourone &&&&&&&&&& \hfill 70\cr
\vskip 1.5cm
\+& {\bf 5. S{\tenpoint UPERSYMMETRIC} M{\tenpoint ATRIX} {\tenpoint
MODEL}: A{\tenpoint RBITRARY} T{\tenpoint OPOLOGIES}}&&&&&&&&&&
\hfill {\bf 78}\cr
\+& 5.1.  \secfourtwo&&&&&&&&&& \hfill 78\cr
\+& 5.2.   \secfourthree&&&&&&&&&& \hfill 85\cr
\+& 5.3.    \secfourfour &&&&&&&&&& \hfill 92\cr
\+& 5.4.   Solution to the Super-Virasoro Constraints in the Double
Scaling Limit &&&&&&&&&& \hfill 96\cr
\+& 5.5.   \secfoursix&&&&&&&&&& \hfill 100\cr
\vskip 1.5cm
\+&{\bf 5. C{\tenpoint ONCLUSIONS}} &&&&&&&&&& \hfill {\bf 109}\cr
\vskip 1.5cm
\+ {\bf Acknowledgements} &&&&&&&&&& \hfill {\bf 111}\cr
\vskip 1.5cm
\+{\bf References} &&&&&&&&&& \hfill {\bf 112}\cr

\endpage

\pageno=1
\chapter{\chaone}
\section{\seconeone}
\no String theory
\REF\gsw{M.~B.~Green, J.~H.~Schwarz and E.~Witten, ``Superstring
Theory'', Cambridge Monographs on Mathematical Physics, (1987).}
[\gsw] is our most promising candidate to be the theory that unifies
quantum field theory with general relativity. The classical motion of
a string in $D$-dimensional space-time is described by a surface
parametrized by a function $X^\m(\s,\t)$ with $\m=1,\dots,D$, which
depends on two coordinates, $\s$ and $\t$. To quantize the theory we
are interested in the calculation of the following functional
integral:

$$
{\cal Z}=\int {\cal D}X e^{-{\cal S}[X]},
\eqn\Ii
$$

\no where the integration is done over all possible surfaces
(space-time trajectories of the string). Here ${\cal S}[X]$ is the
action functional. The definition {\Ii} of the partition function is
a natural generalization of Feynman's path integral for point
particles. For a string propagating freely in flat Minkowski space,
the action is determined by the proper area swept by the string and
is given by the expression

$$
{\cal S}[X]=T\int d\s d\t\sqrt{|{\rm det}\, g_{ab}|}=T\int d \s d\t
\sqrt{(\dot{X} \cdot X')^2-\dot{X}^2 X'^2}
\eqn\IIi
$$

\no where

$$
g_{ab}=\eta_{\mu \nu} {\p X^\mu \o \p \xi^a}{\p X^\mu \o \p
\xi^b},\qquad  \dot{X}^\mu={\p X^\mu(\s,\t) \o \p \t} \qquad {\rm
and} \qquad
X'^\mu={\p X^\mu(\s,\t) \o \p \s}.
\eqn\Niii
$$

\no Here the proportionality constant $T$ is the string tension,
$(\xi^0,\xi^1)=(\t,\s)$ and the Minkowski metric is $\eta_{\mu
\nu}={\rm diag}(1,-1,\dots,-1)$.

This is known as the Nambu-Goto action. However, it is awkward to
work with this action because of the square root and of the fact that
it is highly nonlinear. A more convenient form that is classically
equivalent to the previous one can be obtained if we introduce an
auxiliary field $g_{ab}$, that is the metric tensor of the string
world-sheet

$$
{\cal S}[g,X]=-{T\o 2}\int d \s d\t \sqrt{-g}g^{ab}\eta_{\mu \nu}
\p_aX^\m \p_bX^\nu ,
\eqn\IIii
$$

\no where $g={\rm det}\, g_{ab}$. Polyakov used this action to
formulate string theory in terms of the path integral

$$
{\cal Z}=\int {\cal D}X {\cal D} g e^{-{\cal S}[g,X]}.
\eqn\IIIi
$$

\no  The action {\IIii} is conformal invariant (or local Weyl
invariant), which means that its form does not change under a
rescaling of the metric:

$$
g_{ab}\rightarrow e^{\vp (\s,\t)} g_{ab}.
\eqn\Ni
$$

\no This invariance only holds in two dimensions since in this case
$\sqrt{-g}$ is proportional to $e^\vp $ while $g^{ab}$ is
proportional to $e^{-\vp}$. The energy-momentum tensor is obtained
from the variation of the action {\IIii} with respect to the metric

$$
T_{ab}=-{2\o T\sqrt{-g}}{\d {\cal S}\o \d g^{ab}},
\eqn\IIiii
$$

\no and is given by the expression:

$$
T_{ab}=\p_aX\p_bX-\h g_{ab}g^{cd}\p_cX \p_d X=0.
\eqn\IIiv
$$

\no This energy-momentum tensor vanishes identically because of the
equation of motion of the field $g$. It is therefore also traceless.
This tracelessness reflects the fact that the action is invariant
under conformal transformations of the metric. The actions {\IIi} and
{\IIii} are both invariant under arbitrary reparametrizations
$(\s,\t)\rightarrow (f^\s (\s,\t), f^\t(\s,\t))$ of the surface,
under which

$$
g_{ab}\rightarrow \p_a f^k \p_b f^l g_{kl}.
\eqn\Niv
$$

\no The equation of motion for the field $X^\mu $ is

$$
{1\o \sqrt{-g}} \p_a\left( \sqrt{-g} g^{ab} \p_b X^\mu\right) =\D
X^\mu=0.
\eqn\Nii
$$

\no For a free closed bosonic string we impose a periodic boundary
condition

$$
X^\mu(\t,\s+\pi)=X^\mu(\t,\s).
\eqn\IIIv
$$

The simplest topology of the world-sheet is a cylinder (Fig. 1)
parametrized by $(\s,\t)$, with $0\leq \s \leq \pi$ and
$-~\infty~<\t<~\infty$.

\vskip 0.5cm
\midinsert
\epsfysize=2.0in
\centerline{\hskip 0.5cm \epsffile{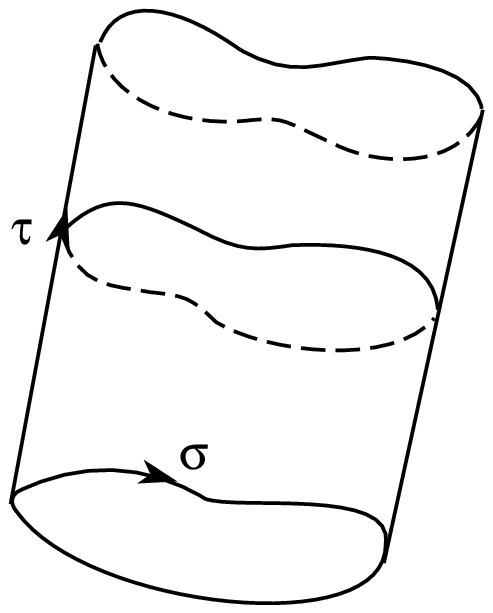}}
\caption{\hskip 1cm{\bf Fig. 1}: Propagation of a free closed
string.}
\vskip 0.5cm
\endinsert

\no We can use the reparametrization (diffeomorphism) invariance to
fix $g_{ab}=e^{\vp }\eta_{ab}$, where $\eta_{\t\t}=1$,
$\eta_{\s\s}=-1$, $\eta_{\t\s}=0$ and $e^{\vp }$ is an unknown
conformal factor. In this (conformal) gauge the action for the field
$X^\mu(z,\bar z)$ is

$$
{\cal S}=-{T\o 2} \int d^2 \s \p_a X_\mu \p^a X^\mu,
\eqn\NNv
$$

\no and the equation of motion takes the form

$$
\left( {\p^2\o \p \t^2}-{\p^2 \o \p \s^2}\right)X^\mu=0.
\eqn\Nv
$$

\no The conformal transformation $z=e^w$, where

$$
w=-2i(\t+\s)\qquad {\rm and} \qquad \bar w=-2i(\t-\s),
\eqn\IIv
$$

\no maps the cylinder to the complex punctured plane
$\IC^*=\IC-\{0\}$.
The equation of motion for $X^\mu(z,\bar z)$ is then

$$
\p_z\p_{\bar z}X^\mu (z,\bar z)=0.
\eqn\IIvi
$$

\no The most general solution compatible with the boundary conditions
{\IIIv} is given by the following expression:

$$
X^\mu(z,\bar z)=q^\mu-{i\o 4} p^\mu \log |z|^2+{i\o 2} \sum_{n\neq 0}
{a_n^\mu\o n}z^{\-n}+
{i\o 2}\sum_{n\neq 0} {\bar a_n^\mu\o n}\bar z^{\-n}.
\eqn\IIvii
$$

We now introduce two different approaches to quantize the free
bosonic string. The first one is the old covariant approach, based on
a description in terms of the $X^\m$ coordinates only, and with
restrictions to the physical Fock space in terms of the Virasoro
conditions. This quantization procedure ensures that we have no
ghosts, i.e. states with negative norm in the spectrum. It is similar
to the Gupta-Bleuler quantization of the electromagnetic field.

The energy-momentum conservation:

$$
\p_zT_{\bar z\bar z}=\p_{\bar z} T_{zz}=0
\eqn\IIvii
$$

\no implies that we can make a Laurent expansion around $z=\bar z=0$:

$$
T(z)=T_{zz}(z)=\sum_{n\in \IZ }L_n z^{-n-2},
\eqn\IIviii
$$

\no and the same for the anti-holomorphic part. Here $L_n$ and $\bar
L_n$ are the Virasoro generators that satisfy the Virasoro algebra:

$$
\{L_m,L_n\}=-i (m-n)L_{m+n},
\eqn\IIix
$$

\no where $\{\;,\; \}$ is the canonical Poisson bracket of the
classical theory. After quantization the Fourier coefficients of
$X^\mu(z,\bar z)$ become operators that satisfy the commutation
relations

$$
[a_m^\mu, a_n^\nu]=-m\d_{m+n} \eta^{\mu \nu}.
\eqn\Nvii
$$

\no The classical algebra {\IIix} becomes then the central extension
of the Virasoro algebra

$$
[L_m,L_n]=(m-n)L_{m+n}+{D\o 12} m(m^2-1) \d_{n+m},
\eqn\Nviii
$$

\no where $D$ is the anomaly term that follows from {\Nvii}.

Previously we saw that the energy-momentum tensor vanishes
classically. If we would like to keep this property also in the
quantum theory then the physical states should be annihilated by all
the modes of this field. However it turns out that this constraint is
too strong and not compatible with the anomaly $D$ of the Virasoro
algebra {\Nviii}. Therefore, the best we can demand is that physical
states are annihilated by the positive-frequency part of the
energy-momentum tensor:

$$
\eqalign{
&L_n|{\rm phys}\rangle =\bar L_n|{\rm phys}\rangle \qquad \qquad {\rm
for} \qquad \qquad n>0\cr
&\cr
& (L_0-b)|{\rm phys}\rangle =(\bar L_0-b)|{\rm phys}\rangle=0.\cr}
\eqn\IIx
$$

\no Here $b$ is a constant that comes from the ambiguity in the
normal ordering of $L_0$. It can be shown that the spectrum is
ghost-free provided that $b=1$ and $D=26$, which is the so-called
critical bosonic string
[\gsw].
Remarkably, analog Virasoro constraints will also play an important
role in the formulation of two-dimensional gravity in terms of the
one-matrix model, as we will see later.

In the modern covariant quantization of string theory we consider the
path integral {\IIIi}. A two-dimensional closed oriented manifold is
topologically completely characterized by the genus or number of
handles $h$ of the surface, so that we can write {\IIIi} as follows:

$$
{\cal Z}=\sum_{h}\l^h\int {{\cal D}g\o {\rm Vol(diff)}}\int {\cal D}
X e^{-{\cal S}[X,g]};
\eqn\Iii
$$

\no where $\l$ is a loop-counting parameter. For each topology we
have to sum over all possible embeddings (the ${\cal D}X$ integral)
and over all possible metrics in that fixed topology modulo
diffeomorphisms. This is why the volume of the diffeomorphism group
has to be divided out. In general, the action appearing in the above
expression describes a $D$-dimensional bosonic string. We will
consider the case of a two-dimensional bosonic string or more
precisely of a minimal conformal field theory coupled to gravity.

\vskip 0.5cm
\midinsert
\epsfysize=1.0in
\centerline{\hskip 0.5cm \epsffile{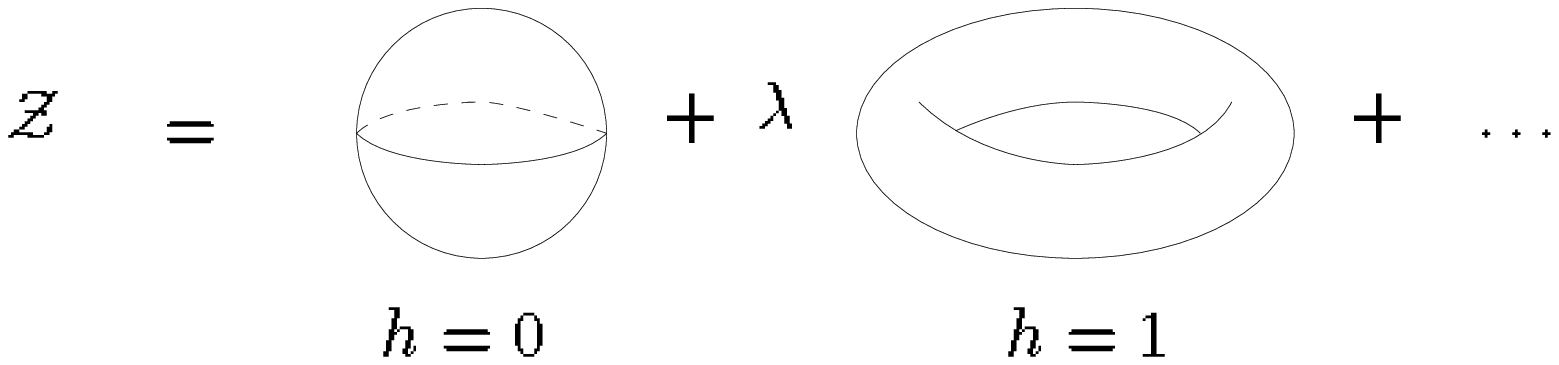}}
\vskip 0.5cm
\caption{\hskip 0.5cm {\bf Fig. 2}: Topological expansion of the
partition function.}
\vskip 0.25cm
\endinsert

To actually calculate the integral $\int {\cal D}g$, we have to
distinguish between critical and sub-critical strings. While for
critical strings the integration simplifies to a finite-dimensional
integration, in the case of sub-critical strings we are left with the
hard problem to know how to quantize Liouville theory. In two
space-time dimensions this `continuum formulation' of quantum gravity
has been first developed by Polyakov
\REF\polyakov{A.~M.~Polyakov, ``Quantum Gravity in Two Dimensions'',
{\it Mod.~Phys.~Lett.} {\bf A2} (1987) 893.}
[\polyakov]
and by Knizhnik, Polyakov and Zamolodchikov
\REF\kpz{V.~G.~Knizhnik, A.~M.~Polyakov and A.~B.~Zamolodchikov,
``Fractal Structure 2D Quantum Gravity'', {\it Mod.~Phys.~Lett.} {\bf
A3} (1988)~819.}
[\kpz] who formulated the quantization in the light cone gauge and
obtained the gravitational dimensions of gravitationally dressed
primary fields for spherical topologies. Using the conformal gauge,
David
\REF\david{F.~David, ``Conformal Field Theories Coupled to 2D Gravity
in the Conformal Gauge'', {\it Mod.~Phys.~Lett. }{\bf A3} (1988)
1651.}
[\david] and Distler and Kawai
\REF\diska{J.~Distler and H.~Kawai, ``Conformal Field Theory and 2D
Quantum Gravity or Who's Afraid of Joseph  Liouville?'', {\it
Nucl.~Phys.} {\bf B321} (1989) 509.}
[\diska]
generalized their results to Riemann surfaces of higher genus. In
general, we are interested in the calculation of string amplitudes
with external string states, so that we have to insert vertex
operators in the functional integral and we would like to calculate
the correlation functions of these operators. Several methods have
been suggested to compute correlation functions in Liouville theory
coupled to minimal conformal models.
These techniques allowed the computation of one-, two and three-point
functions \REF\gl{M.~Goulian and M.~Li, ``Correlation Functions in
Liouville Theory'', {\it Phys.~Rev.~Lett. }{\bf 66} (1991) 2051.}
\REF\dtwo{V.~S.~Dotsenko, ``Correlation
Functions of Local Operators in 2D Gravity
Coupled to Minimal Matter'', Summer School
Carg\`ese 1991, hep-th/9110030; ``Three Point
Correlation Functions of the Minimal Conformal
Field Theories Coupled to 2D Gravity'', \mpl
{\bf A6} (1991) 3601.}
[\gl,\dtwo]
 and in special cases $N$-point functions
\REF\dfk{P.~Di~Francesco and D.~Kutasov,
``World-Sheet and Space Time Physics in Two
Dimensional (Super-)String Theory'', \np {\bf
B375} (1992) 119; ``Correlation Functions in
2D String Theory'', \pl {\bf B261} (1991)
385.}
\REF\klepa{I.~R.~Klebanov, ``Ward Identities in Two-Dimensional
String Theory'', {\it Mod.~Phys.~Lett.} {\bf A7} (1992) 723; I.~R.
Klebanov and A.~Pasquinucci, ``Correlation Functions From
Two-Dimensional String Ward Identities'', {\it Nucl.~Phys.} {\bf
B393} (1993) 261.}
[\dfk,\klepa]
for spherical topologies.

More recently Mukhi and Vafa
\REF\vm{S.~Mukhi and C.~Vafa, ``Two-Dimensional Black Hole as a
Topological Coset of $c=1$ String Theory'', {\it Nucl.~Phys.} {\bf
B407} (1993) 667.}
\REF\gmm{D.~Ghoshal and S.~Mukhi, ``Topological Landau-Ginzburg Model
of Two-Dimensional String Theory'', hep-th/9312189; S.~Mukhi, ``The
Two-Dimensional String as a Topological Field Theory,
hep-th/9312190.}
[\vm,\gmm]
have shown that for the special case of the $D=1$ model coupled to
gravity it might be possible to obtain the amplitudes of tachyon
vertex operators on surfaces of arbitrary genus directly in the
continuum approach. This can be done using the equivalence to a
topological field theory formulated in terms of a Kazama-Suzuki model
or a topological Landau-Ginzburg theory
\REF\hop{A.~Hanany, Y.~Oz and R.~Plesser, ``Topological Landau
Ginzburg Formulation and Integrable Structure of 2D String Theory'',
IASSNS-HEP-94/1, hep-th/9401030.}
[\hop].

Another possibility to evaluate the sum over random surfaces in
two-dimensional quantum gravity is to discretize the problem and to
replace the integral over the metric in {\Iii} by a sum over random
triangulated surfaces of arbitrary topology. This approach reproduces
{\Iii} in the continuum limit. Surprisingly it has been found in 1989
that such a sum can be exactly evaluated using matrix model
techniques
\REF\doubles{E.~Br\'ezin and V.~A.~Kazakov,
``Exactly Solvable Field Theories of Closed Strings'',
{\it Phys.~Lett.} {\bf 236B} (1990) 144; M.~R.~Douglas and
S.~H.~Shenker,
``Strings in Less than One Dimension'',
{\it Nucl.~Phys.} {\bf B335} (1990) 635; D.~J.~Gross and
A.~A.~Migdal,
``Non-perturbative Two-Dimensional Quantum Gravity'', {\it
Phys.~Rev.~Lett.} {\bf 64} (1990) 127;
D.~J.~Gross and A.~A.~Migdal, ``A Non-perturbative Treatment of
Two-Dimensional Quantum Gravity'', {\it Nucl.~Phys.} {\bf B340}
(1990) 333.}
[\doubles].

The simplest matrix model is the one-matrix model, where the field is
an Hermitian $N\times N$ matrix. Using the planar loop equations,
Kazakov
\REF\kazakov{V.~A.~Kazakov, ``The Appearance of Matter Fields from
Quantum Fluctuations of 2D Gravity'',
{\it Mod.~Phys.~Lett.} {\bf A4} (1989) 2125.}
[\kazakov]
has carried out the analysis of the critical regimes of these models.
It has been realized that they describe the coupling of non-unitary
minimal series of conformal field theories (CFTs) of type $(2m-1,2)$
coupled to two-dimensional quantum gravity
\REF\scgm{M.~Staudacher, ``The Yang-Lee Edge Singularity on a
Dynamical Planar Random Surface'', \np {\bf B336} (1990) 349;
C.~Crnkovic, P.~Ginsparg and G.~Moore, ``The Ising Model, the
Yang-Lee Edge Singularity, and 2D Quantum Gravity'', \pl {\bf B237}
(1990) 196.}
[\scgm].

With the introduction of the double scaling limit [\doubles], it has
been possible to evaluate these models non-perturbatively, i.e. on
surfaces of arbitrary genus. It turns out that in this case there is
a close relation to integrable hierarchies
\REF\douglas{M.~R.~Douglas, ``Strings in Less than One Dimension and
the Generalized KdV Hierarchies'', {\it Phys.~Lett.} {\bf 238B}
(1990) 176.}
\REF\dbs{T.~Banks, M.~Douglas, N.~Seiberg and S.~H.~Shenker,
``Microscopic and Macroscopic Loops in Non-perturbative
Two-Dimensional Gravity'', {\it Phys.~Lett. } {\bf B238} (1990) 279.}
[\douglas,\dbs],
which is manifest through the appearance of the Korteweg-de Vries
hierarchy (KdV), which describes correlation functions of scaling
operators. The partition function is related to the tau-function of
the KdV hierarchy.
It has been shown
\REF\dvv{R.~Dijkgraaf, E.~Verlinde and
H.~Verlinde, ``Loop Equations and Virasoro Constraints in
Non-perturbative 2D Quantum Gravity'', {\it Nucl.~Phys.} {\bf B348}
(1991) 435.}
\REF\fkn{M.~Fukuma, H.~Kawai and
R.~Nakayama, ``Continuum Schwinger-Dyson Equations and Universal
Structures in Two-Dimensional Quantum Gravity'', {\it
Int.~J.~Mod.~Phys.} {\bf A6} (1991) 1385.}
[\dvv,\fkn]
that the non-perturbative loop equation that fixes the solution of
the one-matrix model in the double scaling limit can be obtained
using the KdV structure and the string equation as initial condition.
This string equation describes the form of the renormalized
cosmological constant in terms of scaling operators at the $k^{\rm
th}$ multicritical point. The loop equation is equivalent to an
infinite set of constraints satisfied by the partition function. They
form a Virasoro algebra and are therefore called Virasoro
constraints. They can be found directly in the matrix model
representation of the partition function
\REF\matsuo{Y. Matsuo, unpublished.}
\REF\mironov{A.~Mironov and A.~Morozov, ``On the Origin of Virasoro
Constraints in Matrix Models: Lagrangian Approach'', {\it
Phys.~Lett.} {\bf B252} (1990) 47.}
[\matsuo,\mironov]
or can be elegantly derived in the Witten-Kontsevich theory
\REF\witkon{E.~Witten,``Two-Dimensional
Gravity and Intersection Theory on Moduli Space", {\it Surveys in
Diff.~Geom.} {\bf 1} (1991) 243, and references therein; R.~Dijkgraaf
and E.~Witten, ``Mean Field Theory, Topological Field Theory, and
Multimatrix Models'', {\it Nucl.~Phys.}
{\bf B342} (1990) 486;
M.~Kontsevich, ``Intersection Theory
on the Moduli Space of Curves and the Matrix Airy Function", {\it
Commun.~Math.~Phys.} {\bf 147} (1992) 1;
E.~Witten, ``On~the Kontsevich Model
and Other Models of Two-Dimensional Gravity", IASSNS-HEP-91-24.}
[\witkon].
These constraints also play an important role in the generalization
of the bosonic theories to $N=1$ supersymmetric CFTs coupled to
two-dimensional supergravity, as we will later see.

\section{\seconetwo}

\no In the last section we have introduced the simplest string
theory, i.e. the bosonic string. However, it has been realized that
all phenomenological interesting string theories should incorporate
supersymmetry. Therefore we have to look at more general string
theories, the superstring theories
\REF\lt{D.~L\"ust and S.~Theisen, ``Lectures on String Theory'',
Lecture Notes in Physics, Springer (1989).}
[\gsw,\lt], which include fermionic degrees of freedom and
supersymmetry relates the bosonic space-time coordinates
$X^\m(\s,\t)$ to the fermionic partners $\Psi^\m(\s,\t)$. String
theories of this type have many advantages since undesirable features
of the bosonic string can be eliminated in the supersymmetric case
as, for example, the presence of the tachyon. The free closed
superstring (without boundaries) is described by the action\foot{The
string tension $T$ is related to the string coupling constant $\a'$
by $T=(2\pi \a')^{-1}$. In the following we choose the convention
$\a'=1/2$.}

$$
{\cal S}=-{1\o 2\pi }\int d \s d\t \left( \p_a X^\mu \p^a
X^\nu-i{\bar \Psi}^\mu \vr^a \p_a \Psi^\nu \right)\eta_{\mu \nu}.
\eqn\Iiii
$$

\no Here $\vr^a$ represent the Dirac matrices satisfying the
two-dimensional Clifford algebra:

$$
\{\vr^a,\vr^b\}=2\eta^{ab},
\eqn\Iiv
$$

\no where $\eta^{ab}$, is the flat Minkowski metric. The matrices
$\vr^a$ can be represented as follows:

$$
\vr^0=\left(\matrix{0&-i\cr i&0\cr}\right)\qquad{\rm and} \qquad
\vr^1=\left(\matrix{0&i\cr i&0\cr}\right).
\eqn\Iv
$$

\no The world-sheet spinor $\Psi^\mu$ has two components in this
basis

$$
\Psi^\mu=\left(\matrix{\Psi^\mu_-\cr \Psi^\mu_+}\right).
\eqn\Iv
$$

\no Since $\vr^a$ are purely imaginary, the Dirac operator $i\vr^a
\p_a$ is real and it makes sense to choose the components of the
world-sheet spinor $\Psi^\mu$ real. Such a two-component real spinor
is known as a Majorana spinor. The symbol $\bar \Psi$ means $\Psi^T
\vr^0$.

The action {\Iiii} is invariant under the global transformations

$$
\eqalign{\d X^\mu&={\bar \ep}\Psi^\mu\cr
\d\Psi^\mu&=-i\vr^a\p_a X^\mu \ep,\cr}
\eqn\Ivi
$$

\no where $\ep$ is a constant anticommuting Majorana spinor. Since
these transformations mix bosonic and fermionic coordinates they are
called supersymmetry transformations. If $\ep$ is not constant, the
variation of the action {\Iiii} is

$$
\d {\cal S}={2\o \pi} \int d\s d\t (\p_a \bar \ep ) J^a.
\eqn\IIvii
$$

\no Here $J^a$ is the conserved supercurrent

$$
J_a=\h\vr^b \vr_a \Psi^\mu\p_b X_\mu
\eqn\IIvi
$$

\no that satisfies $\p_a J^a=0$ because of the classical equations of
motion

$$
\p _a \p^a X^\mu=0 \qquad {\rm and} \qquad \vr^a \p_a \Psi^\mu=0.
\eqn\Nx
$$

Introducing the coordinates $(z,\bar z)$ as in {\IIv} and the
notation $\Psi_-=\Psi$ and $\Psi_+={\bar \Psi}$ the gauged fixed
action takes the form

$$
{\cal S}=-{1\o 2\pi} \int d^2 z \left( \p_{\bar z} X \p_z X-\Psi^\mu
\p_{\bar z} \Psi_\mu -{\bar \Psi}^\mu \p_z {\bar \Psi}_\mu\right) .
\eqn\Nxi
$$

\no The energy-momentum tensor and the supercurrent for the free
superstring are:

$$
\eqalign{
&T(z)=T_{zz}(z)=-\h \p_z X^\mu \p_z X_\mu -\h \p_z \Psi^\mu \Psi_\mu
\cr
&\cr
&T_F(z)=J_z(z)=-\h\Psi^\mu \p_z X_\m\cr}
\eqn\IIIvi
$$

\no and similarly for the antiholomorphic components.

Supersymmetry can be nicely formulated in the superspace language.
The world-sheet coordinates $z$, $\bar z$ are supplemented by a pair
of anticommuting variables $\th$, $\bar \th$ so that we have the
supercoordinates:

$$
{\bf z}=(z,\th),\qquad \qquad {\bf \bar z}=(\bar z, \bar \th).
\eqn\Ivii
$$

\no Taking into account the fermionic character of $\th$ and $\bar
\th$, any function $f({\bf z},{\bf \bar z})$ can be written in the
following way :

$$
f({\bf z},{\bf \bar z})=f_0(z,\bar z)+f_1(z,\bar z)\th+f_2(z,\bar
z)\bar \th
+f_4(z,\bar z)\th \bar \th.
\eqn\Iviii
$$

\no The supersymmetric derivative is defined by $D=\p / \p \th+\th\p
/ \p z$ and the integration rules are:

$$
\int d\th=0,  \qquad \qquad\int d\th \th =1\qquad {\rm and} \qquad
\int d \th d {\bar \th} \th {\bar \th}=1.
\eqn\Iix
$$

\no Introducing the superfield

$$
{\bf X^\mu}({\bf z},{\bf \bar z})=X^\mu(z,\bar z)+\th \Psi^\mu
(z,\bar z)+\bar \th \bar \Psi^\mu(z,\bar z)+\th {\bar \th} G^\mu
(z,\bar z),
\eqn\Ix
$$

\no where $G^\mu (z,\bar z)$ is an auxiliary field, we can write the
action {\Iiii} as follows:

$$
{\cal S}=-{1\o 2\pi} \int d^2 {\bf z}  \bar D {\bf X^\mu} D {\bf
X_\mu},
\eqn\Ixi
$$

\no where the infinitesimal volume element in superspace is $d^2 {\bf
z}=d^2 z d \th d{\bar \th}$. The super energy-momentum tensor ${\bf
T}({\bf z})$ has two components:

$$
{\bf T}({\bf z})=T_F(z)+\th T(z)
\eqn\IIxi
$$

\no and after quantization we obtain the following OPEs for $T(z)$
and $T_F(z)$:

$$\eqalign{
&T(z)T(w)={3D/4\o (z-w)^4}+{2T(w)\o (z-w)^2}+{\p T(w)\o
(z-w)}+\dots\cr
&\cr
&T(z)T_F(z)={3\o 2}{T_F(w)\o (z-w)^2}+{\p T_F(w)\o (z-w)} +\dots \cr
& \cr
&T_F(z)T_F(w)={D/4\o (z-w)^3}+\h {T(w)\o (z-w)} +\dots\cr}
\eqn\IIIxi
$$

\no here $D$ is the range of the index $\mu$.

An important point when dealing with fermionic strings is the
boundary condition. A closed bosonic string has periodic boundary
conditions $X^\mu(\t,\s+\pi)=X^\mu(\t,\s)$. For the superstring the
same holds for $X^\mu$, while for the world-sheet fermions $\Psi^\mu$
we can have periodic or antiperiodic boundary conditions. The first
one is called the Ramond (R) boundary condition while the second one
is the Neveu-Schwarz (NS) boundary condition:

$$
\eqalign{
\Psi^\mu(\t,\s+\pi)&=\Psi^\mu(\t,\s)\cr
\Psi^\mu(\t,\s+\pi)&=-\Psi^\mu(\t,\s)\cr}\qquad \eqalign{ &{\rm (R)}
\cr &{\rm (NS)}.\cr}
\eqn\Ixii
$$

\no Since $\Psi^\mu(z)$ is a $(1/2,0)$ field, the transformation
properties under $z\rightarrow e^{2\pi i}  z$ in the two sectors are

$$
\eqalign{
\Psi^\mu(e^{2\pi i}z)&=-\Psi^\mu(z)\cr
\Psi^\mu(e^{2\pi i}z)&=\Psi^\mu(z)\cr}\qquad \eqalign{ &{\rm (R)} \cr
&{\rm (NS).}\cr}
\eqn\IIxii
$$

\no The fields $\Psi^\mu (z)$ can be expanded in both sectors in
Fourier modes, as follows:

$$
\Psi^\mu (z)=\sum_n d_n^\mu z^{-n-\h}\qquad {\rm with} \qquad
\cases{n\in \IZ & for \quad (R) \cr n\in \IZ+\h & for \quad (NS) \cr}
\eqn\Ixiii
$$

\no The Neveu-Schwarz states therefore describe space-time bosons,
while the Ramond states describe space-time fermions. The field
$X^\mu(z,\bar z)$ has the same expansion as in the bosonic string.

If we are considering fermions on a torus, there exist four
possibilities to combine the boundary conditions around the two
non-contractible cycles

$$
{\rm (R,R)} \qquad{\rm (NS,NS)}\qquad {\rm (NS,R)} \qquad {\rm and}
\qquad {\rm (R,NS)}.
\eqn\IIxiii
$$

\no More generally, for a Riemann surface $\S_h$ of genus $h$
there are $2^{2h}$  possible choices of boundary conditions, since
there are four possibilities for each handle.  Each of these possible
assignments is called the spin structure of $\S_h$. The boundary
conditions for the holomorphic and anti-holomorphic fields $\Psi $
and ${\bar {\Psi}}$ have to be chosen in such a way that modular
invariance of the partition function is guaranteed. In the partition
functions of superconformal minimal models the holomorphic and
antiholomorphic spin structures are identified
\REF\cappelli{A.~Cappelli, ``Modular Invariant Partition Functions of
Superconformal Theories'', {\it Phys.~Lett.} {\bf 185B} (1987) 82.}
[\cappelli].

To quantize the free closed superstring in the old covariant approach
we can proceed in analogy to the bosonic theory: we impose the
(anti-)commutation relations

$$
\eqalign{
[a_m^\mu ,a_n^\nu ]&=-m \delta_{m+n}\eta^{\mu \nu}\cr
&\cr
\{d_m^\mu, d_n^\nu\}&=- \delta_{m+n}\eta^{\mu \nu}.\cr}
\eqn\IIxiii
$$

\no The Virasoro constraints are now replaced by super-Virasoro
constraints. The energy-momentum tensor has an expansion of the form:

$$
T(z)=\sum_{n\in \IZ} L_n z^{-n-2}
\eqn\Ixv
$$

\no where $L_n$ are the Virasoro generators. The supercurrent has the
same boundary condition as the fermionic fields. Therefore it has the
following Laurent expansion

$$
T_F(z)=\h \sum_r G_r z^{-r-{3\o 2}}\qquad {\rm with} \qquad
\cases{r\in \IZ & for \quad (R) \cr r\in \IZ+\h & for \quad (NS) \cr}
\eqn\Ixvi
$$

\no where $G_r$ are the generators of the superconformal
transformations. The (anti-)commutation relations follow from the OPE
{\IIIxi}:

$$\eqalign{
&[L_m,L_n]=(m-n) L_{m+n}+{\widehat c\o 8} m(m^2-1)\d_{n+m,0}\cr
&\cr
&[L_m,G_r]=\left({m\o 2}-r\right) G_{m+r}\cr
&\cr
&\{G_r,G_s\}=2L_{r+s}+{\widehat c\o 2}\left(r^2-{1\o 4} \right)
\d_{r+s,0},\cr}
\eqn\wxxvii
$$

\no where $\widehat c=D$. In the Neveu-Schwarz sector the five
generators $L_1$, $L_0$, $L_{-1}$, $G_{-1/2}$ and $G_{1/2}$ form a
closed subalgebra.

The constraint equations on physical states have the form of
super-Virasoro conditions. In order to derive them we have to
generalize the action {\Iiii} by incorporating a zweibein
$e^a_{\a}(\t,\s)$
$$
h_{\a\b}=e^a_\a e^b_\b \eta_{ab},
\eqn\wwwxxvii
$$

\no and an auxiliary field $\chi_\a(\t,\s)$ (gravitino), where the
indices satisfy $a,b=1,\dots, D$ and $\a,\b=0,1$. The complete action
is then
\REF\bvh{L.~Brink, P.~Di~Vecchia and P.~Howe, ``A Locally
Supersymmetric and Parametrization Invariant Action for the Spinning
String'', \pl{\bf 65B} (1976) 471.}
\REF\dz{S.~Deser and B.~Zumino, ``A Complete Action for the Spinning
String'', \pl {\bf 65B} (1976) 369.}
formulated in terms of these fields [\bvh,\dz]

$$
\eqalign{
{\cal S}=&-{1\o 2\pi} \int d\t d\s \sqrt{ h}\,\eta_{\mu \nu} \left(
h^{\a\b} \p_\a X^\mu \p_\b X^\nu-i e^\a_a {\bar \Psi}^\mu \vr^a \p_\a
\Psi^\nu \right)
  \cr
&\cr
& -{1\o \pi} \int d\t d\s \sqrt{ h}\, \eta_{\mu \nu} e^\a_a e^\b_b
{\bar \chi}_\a \vr^b \vr^a \Psi^\mu \p_\b X^\nu \cr
&\cr
&-{1\o 4\pi } \int d\t d\s \sqrt{ h}\, \eta_{\mu \nu} {\bar \Psi}^\mu
\Psi^\nu e^\a_a e^\b_b {\bar \chi}_\a \vr^b \vr^a \chi_\b. \cr}
\eqn\wwwxxviii
$$

\no The energy-momentum tensor and the supercurrent can be obtained
from the above action by using

$$
T_{\a\b}=-{\pi \o e} e^a_\b {\delta {\cal S} \o \delta e^a_\a }\qquad
{\rm and} \qquad J_\a=-{\pi \o 2 e} {\delta {\cal S}\o \delta
\chi^\a}.
\eqn\wwwwxxviii
$$

\no The equations of motion for the the zweibein and the gravitino
yield the constraint equations:
$$
T_{\a\b}=0 \qquad {\rm and} \qquad J_\a=0.
\eqn\wwwwxxix
$$
\no The generalization of the conformal gauge to the supersymmetric
case is the so-called superconformal gauge, where:

$$
e^a_{\a}=e^{\vp} \d_{\a}^a \qquad{\rm and} \qquad \chi_\a=\vr_a \psi.
\eqn\wwwxxix
$$

\no In the classical theory we can still use a Weyl rescaling and
super-Weyl transformation to gauge away the remaining metric and
gravitino degrees of freedom, leaving only $e^a_\a=\delta_\a^a$ and
$\chi_\a=0$. Taking into account this gauge condition, the desired
constraint equations are

$$
\eqalign{
T_{\a \b}&=\p_\a X^\mu \p_\b X_\mu -{1\o 2} \p^\g X^\mu \p_\g X_\mu
\eta_{\a\b} +{i \o 2} {\bar \Psi}^\mu \vr_\a \p_\b \Psi_\mu +{i\o 2}
{\bar \Psi}^\mu \vr_\b\p_\a\Psi_\mu=0\cr
&\cr
J_\a &={1\o 2} \vr^\b \vr_\a \Psi^\mu \p_\b X_\mu=0.\cr}
\eqn\wwwxxx
$$

These constraint equations are incorporated into the quantum theory
by requiring that the positive modes of the energy-momentum tensor
and the supercurrent annihilate the physical states

$$
\eqalign{
&G_r|{\rm phys}\rangle =0 \qquad {\rm for } \qquad r>0  \cr
&\cr
&L_n|{\rm phys}\rangle =0 \qquad {\rm for} \qquad n> 0 \cr
&\cr
&(L_0-b)|{\rm phys}\rangle =0.\cr}
\eqn\wwxxvii
$$

\no If the $G_{1/2}$ and $G_{3/2}$ constraints are satisfied, the
rest of them follow from the above algebra. The constant $b$ takes
different values depending on whether we are in the Ramond or
Neveu-Schwarz sector. It can be shown [\gsw,\lt] that $b=0$ is a
consistent choice for the Ramond sector. For the critical dimension
$D=10$, a decoupling of negative norm states takes place, if $b=1/2$
for the Neveu-Schwarz sector.

In the case of the bosonic theory we have already mentioned that the
one-matrix model provides us with a powerful non-perturbative
description. These non-perturbative effects are very important in the
context of superstring theory. They are responsible, for example, for
supersymmetry breaking.
We are therefore faced with the question whether it is possible to
find similar
(non-perturbative) powerful methods as the one-matrix model to
evaluate theories with $N=1$ supersymmetry coupled to world-sheet
supergravity.

A first approach in this direction has been taken by Alvarez-Gaum\'e
and Ma\~nes in
\REF\agm{L.~Alvarez-Gaum\'e and J.~L.~Ma\~nes, ``Supermatrix
Models'',
{\it Mod.~Phys.~Lett.} {\bf A6} (1991) 2039.}
ref.~[\agm].
Here a naive replacement of ordinary matrices by supermatrices has
been proposed to simulate the sum over supergeometries. Taking
$(N|M)$-supermatrices one is able to simulate $N$-bosonic degrees of
freedom and $M$ fermionic ones. However the result of this approach
is disappointing because these models are completely equivalent to
bosonic models based on $(N-M|0)$ matrices.

{}From the continuum point of view, Di Francesco et al.
\REF\difdk{P.~Di~Francesco, J.~Distler and
D.~Kutasov,
``Superdiscrete Series Coupled to 2D Supergravity'', {\it
Mod.~Phys.~Lett.} {\bf A5} (1990) 2135.}
[\difdk]
have analysed the connection between certain unitary superdiscrete
series of minimal models coupled to 2D-supergravity and the
Manin-Radul hierarchy
\REF\mr{Yu.~I.~Manin and A.~O.~Radul,
``A Supersymmetric Extension of
the Kadomtsev-Petviashvili Hierarchy'', {\it Commun.~Math.~Phys.}
{\bf 98} (1985) 65.}
[\mr].
This is one of the known supersymmetric extensions of the ordinary
KP-hierarchy, as we will see in detail later. However, they came to
the conclusion that odd flows are inconsistent with the string
equation, so that the relation to superintegrable hierarchies remains
obscure.

The most promising approach to the problem of finding a model that
describes $N=1$ superconformal matter coupled to 2D-supergravity has
been found by Alvarez-Gaum\'e et al.
\REF\sloops{L.~Alvarez-Gaum\'e, H.~Itoyama, J.~Ma\~nes and A.~Zadra,
``Superloop Equations and Two-Dimensional
Supergravity'' {\it Int.~J.~Mod.~Phys.} {\bf A7} (1992) 5337.}
[\sloops].
The guiding principle in ref.~[\sloops] was a set of
super-Virasoro constraints satisfied by the partition
function, which code the superloop equations. Since the
Virasoro constraints [\dvv,\fkn] played a prominent role in the
Witten-Kontsevich theory [\witkon] of the one-matrix model, it is
reasonable to
expect that a similar set of super-Virasoro constraints should
capture some important features of the supermoduli space of
super-Riemann surfaces. The model obtained in this way is an
`eigenvalue' model. It is formulated in terms of a collection
of $N$-even and $N$-odd eigenvalues $(\lambda_i,\theta_i)$.

It has been shown that in the continuum limit this model reproduces
the string susceptibility and the scaling dimensions of
$(4m,2)$-minimal superconformal models coupled to 2D-supergravity
[\sloops]. Correlation functions of gravitationally dressed scaling
operators in the Neveu-Schwarz sector agree with those computed in
the continuum super-Liouville theory in the planar approximation
\REF\sliouv{E.~Abdalla, M.~C.~B.~Abdalla,
D.~Dalmazi and K.~Harada,
``Correlation Functions in Super-Liouville Theory'', {\it
Phys.~Rev.~Lett.} {\bf 68} (1992) 1641; K.~Aoki and E.~D'Hoker,
``Correlation Functions of Minimal Models Coupled to Two-Dimensional
Quantum Supergravity'', {\it Mod.~Phys.~Lett. } {\bf A7} (1992) 333;
L.~Alvarez-Gaum\'e and P.~Zaugg,
``Some Correlation Functions of
Minimal Superconformal Models Coupled to Supergravity'', {\it
Phys.~Lett.} {\bf 215} (1992) 171.}
[\dfk,\sliouv].
A comparison between the discrete and continuum approaches can be
found in ref.
\REF\abdallabook{E.~Abdalla, M.~C.~B.~Abdalla, D.~Dalmazi and
A.~Zadra, ``2D-Gravity in Non-Critical Strings: Continuum and
Discrete Approaches'', Lecture Notes in Physics, Series M, Springer
Verlag (to appear).}
\REF\da{E.~Abdalla and D.~Dalmazi, ``Correlators in Non-critical
Superstrings Including the Spinor Emission Vertex'', {\it
Phys.~Lett.} {\bf B312} (1993) 398.} [\abdallabook,\da].
Very recently E.~Abdalla and A.~Zadra
\REF\za{E.~Abdalla and A.~Zadra, ``Non-Critical Superstrings: A
Comparison Between Continuum and Discrete Approaches'',
CERN-TH.7161/94, hep-th/9402083 (February~1994).}
[\za]
have established a precise dictionary between the operators of the
discrete super-eigenvalue model and those of the super-Liouville
theory using similar methods as those developed by Moore, Seiberg and
Staudacher
for the bosonic model
\REF\moss{G.~Moore, N.~Seiberg and M.~Staudacher, ``From Loops to
States in 2D Quantum Gravity'', {\it Nucl.~Phys.} {\bf B362} (1991)
665.}
[\moss].

An evaluation of the planar model for arbitrary superpotentials has
been carried out by Alvarez-Gaum\'e et al.
\REF\three{L.~Alvarez-Gaum\'e, K.~Becker and M.~Becker,
CERN-TH.6575/92 (July 1992).}
\REF\five{L.~Alvarez-Gaum\'e, K.~Becker, M.~Becker, R.~Empar\'an and
J.~Ma\~nes, ``Double Scaling Limit of the Super-Virasoro
Constraints'',
{\it Int.~J.~Mod.~Phys.} {\bf A8} (1993) 2297.}
\REF\salama{L.~Alvarez-Gaum\'e, K.~Becker and M.~Becker,
``Super-Virasoro Constraints and Two-Dimensional Supergravity'', talk
given at the XIX International Colloquium on Group Theoretical
Methods in Physics, Salamanca, Spain, 1992. }
[\three,\five,\salama].
The consideration of arbitrary superpotentials is important in order
to derive continuum super-Virasoro constraints. They are equivalent
to the double scaled superloop equations. The solution of these
equations up to genus two shows that the purely bosonic part of the
model (fermionic coupling constants set to zero) agrees with the
one-matrix model.

The Virasoro constraints in the double scaling limit turn out to be
described in terms of a $\widehat c=1$ theory with a $\IZ_2$-twisted
scalar field and a Weyl-Majorana fermion in the Ramond
sector\foot{These constraints have also been derived in
\REF\itoyama{H.~Itoyama, ``Integrable Superhierarchy of Discretized
2d
Supergravity'', {\it Phys.~Lett.} {\bf B299} (1993) 64.}
[\itoyama].}.
The derivation of these constraints is the first step on the way to
finding the integrable hierarchy of super-differential equations that
describes the flows in the double scaling limit.

The solution of the super-Virasoro constraints in the double scaling
limit or equivalently of the superloop equations for arbitrary genus
has been found in
\REF\two{K.~Becker and M.~Becker, ``Non-perturbative Solution of the
Super-Virasoro Constraints'', \mpl {\bf A8} (1993) 1205.}
ref.~[\two].
The supersymmetric partition function can be expressed in a simple
way through the partition function of the one-matrix model if the
fermionic degrees of freedom are integrated out. This property
appears in the discrete matrix integral even before carrying out the
continuum limit. This is important for the geometric interpretation
of the model.
The correlation functions of scaling operators in the double scaling
limit allow the study of the connection to the known superintegrable
hierarchies.

\section{\seconethree}
\no In this thesis we will solve the $N=1$ super-eigenvalue model
proposed by Alvarez-Gaum\'e et al. [\sloops] to describe the coupling
of superconformal matter of type $(4m,2)$ to 2D-supergravity. First
we will solve the model in the genus-zero approximation using an
arbitrary bosonic part of the superpotential and derive the
expression for the scaling operators in the Ramond and Neveu-Schwarz
sector of the theory. We compute the expressions for the scaling
exponents and show that we get agreement with the continuum
super-Liouville approach
\REF\pz{A.~M.~Polyakov and
A.~B.~Zamolodchikov, ``Fractal Structure of Two-Dimensional
Supergravity'',
{\it Mod.~Phys.~Lett.}
{\bf A3} (1988) 1213.}
\REF\dhk{J.~Distler, Z.~Hlousek and H.~Kawai,
``Super-Liouville
Theory as a Two-Dimensional, Superconformal Supergravity Theory'',
{\it Int.~J.~Mod.~Phys.} {\bf A5} (1990) 391.}
[\pz,\dhk].

For arbitrary potentials a careful treatment of the continuum limit
is in order. This will be done first for the case of the one-matrix
model, which we feel has not been done in the literature explicitly
enough. We are going to prove that the super-Virasoro constraints in
the continuum are described by a $\widehat c=1$ theory with a
$\IZ_2$-twisted scalar and a Weyl-Majorana fermion in the Ramond
sector.

After deriving the corresponding continuum (non-planar) loop
equations we obtain their solution up to genus two and show that the
bosonic part of the model (fermionic coupling constants set to zero)
agrees with the one-matrix model result.

We then derive the non-perturbative solution of the model in the
discrete and continuum theory. The partition function of the
supersymmetric model turns out to have a simple representation in
terms of the one-matrix model if the fermionic degrees of freedom are
integrated out. This relation is also present in the continuum
theory, as can be seen from the expressions for the scaling operators
and the string equation. We will then analyse the superintegrable
hierarchy behind this model.

The content of the subsequent chapters is the following:

{\it Chapter 2}: We introduce the (super)-Liouville approach to
2D-(super)-gravity. We compute the critical exponents and the string
susceptibility in order to compare them with the discrete approach
through the matrix model.

{\it Chapter 3}: We explain in some detail the one-matrix model. In
section 3.1 we show that the large-$N$ expansion of the matrix model
can be used to carry out the sum over topologies which we would like
to carry out in string theory. In section 3.2 we derive the discrete
Virasoro constraints and the equivalent loop equations. The one-cut
solution of the loop equations for arbitrary potentials and genus
zero is explained in section 3.3.
In section 3.4 we explain how the KdV hierarchy appears in the double
scaling limit. Related to this hierarchy are the double scaled
Virasoro constraints or equivalently the loop equation in the double
scaling limit [\dvv,\fkn].

{\it Chapter 4}: In section 4.1 we present the one-matrix model in
terms of eigenvalues from the point of view that can be easily
generalized to the supersymmetric case.
We present the model obtained in this way in ref.~[\sloops]. We
derive the form of the superloop equations, for arbitrary genus in
the discrete theory, that are equivalent to the discrete
super-Virasoro constraints. In section 4.2 we present the planar
solution of the model for arbitrary superpotentials. We derive the
expressions for the bosonic and fermionic loop operators in the
planar geometry.

{\it Chapter 5}: In section 5.1 we derive the continuum Virasoro
constraints and show that the bosonic and fermionic loop operators
become a $\IZ_2$-twisted scalar bosonic field and a Weyl-Majorana
fermion in the Ramond sector. We obtain the expressions for the
scaling operators and show that the critical exponents and the string
susceptibility agree with those of $(4m,2)$ SCFTs coupled to
supergravity. In section 5.2 we solve the continuum superloop
equations up to genus two and show that the bosonic part of the model
is represented by the one-matrix model. In section 5.3 we obtain the
non-perturbative solution of the supersymmetric model directly from
the discrete eigenvalue integral, which has a simple relation to the
one-matrix model. A similar relation to the one-matrix model appears
in the double scaling limit. We derive the non-perturbative
expressions for the correlation functions and analyse the properties
of the hierarchy behind the model in section 5.4. In section 5.5 we
review the most important properties of the known supersymmetric
extensions of KdV and compare them with the hierarchy that describes
our model.

{\it Chapter 6}: We present our conclusions.

\endpage

\chapter{\chaoneone}
\no In order to gain some basic ideas about the results we should
expect in the discrete approach to 2D gravity and its supersymmetric
generalization, we start with a short review about the path integral
approach to 2D (super)gravity. Some good articles about the Liouville
approach to 2D-quantum gravity are for example
\REF\sei{N.~Seiberg, ``Notes on Quantum
Liouville Theory and Quantum Gravity'', {\it Prog.~Theor. Phys.
Suppl.} {\bf
102} (1990) 319.}
\REF\polch{J.~Polchinski, ``Remarks on the Liouville Field Theory'',
{\it Nucl. Phys.} {\bf B357} (1991) 241.}
[\abdallabook,\sei,\polch].

\section{\chaoneoneone}
\no We begin with the bosonic theory. The partition function of a
general conformal system with central charge $c=D$ described by the
action ${\cal S_M}$ is the following integral:

$$
{\cal Z}=\int {{\cal D}g{\cal D}X\o {\rm Vol(Diff)}}\exp \left(-{\cal
S_M}-{\mu_0 \o 8\pi} \int d^2\x \sqrt{g}\right).
\eqn\ri
$$

\no Here we have divided out the volume of the diffeomorphism group
as before. For a free bosonic string moving in a flat background the
action has the form\foot{Here we have carried out an analytic
continuation to the Euclidean signature in space-time
$iX^0\rightarrow X^0$ and on the world-sheet $i\t\rightarrow \t$, so
that $g_{ab}$ gets positive definite.}\foot{In the following we
change the conventions for $\a'$: we will set $\a'=2$.}

$$
{\cal S_M}={1\o 8\pi} \int d^2 \xi \sqrt{g} g^{ab} \p_a X^\mu  \p_b
X_\mu.
\eqn\rri
$$

\no The cosmological term appearing in {\ri} is absent in the
classical theory but in the quantum theory it is necessary for
renormalizability [\polyakov]; $\mu_0$ is the bare cosmological
constant. Geometrically $A=\int d^2\x \sqrt g$ represents the area of
the surface.

The integration measures in {\ri} must be defined more precisely. The
${\cal D}X$ measure can be defined using the normalization of the
Gaussian functional integral

$$
\int {\cal D}_g\d X e^{-||\d X||_g^2}=1 \qquad {\rm with} \qquad \|\d
X\|_g^2=\int d^2\x \sqrt{g}\d X \d X.
\eqn\rii
$$

\no The ${\cal D}g$ measure can be defined similarly

$$
\int {\cal D}_g \d g e^{-\h \| \d g\|^2_g}=1 \qquad {\rm with}\qquad
\| \d g\| _g^2 =\int d^2\x \sqrt{g} (g^{ac} g^{bd}+2g^{ab} g^{cd})\d
g_{ab} \d g_{cd},
\eqn\riii
$$

\no where $\d g$ represents a metric fluctuation at some point
$g_{ab}$ in the space of metrics of a genus-$h$ surface.
These measures are invariant under the diffeomorphism group (the
group of reparametrizations) but not under conformal transformations
of the metric $g_{ab} \rightarrow e^{\varphi} g_{ab}$. It turns out
that

$$
{\cal D}_{e^\vp g}X={\cal D}_g X \exp\left({{D\o 48 \pi} {\cal
S}_L(\vp, g)}\right),
\eqn\riv
$$

\no where

$$
{\cal S}_L(\vp, g)=\int d^2 \x \sqrt{g} \left(\h g^{ab} \p_a \vp \p
_b\vp+R\vp+\mu_c e^\vp\right)
\eqn\rv
$$

\no is known as the Liouville action. Here $\mu_c $ is a (ultraviolet
divergent) non-universal term.

The space of metrics on a compact topological surface $\Sigma$ modulo
diffeomorphism and Weyl transformations is a finite dimensional
compact space ${\cal M}_h$, known as moduli space. If for each point
$\t\in {\cal M}_h$, we choose a representative metric $\widehat
g_{ab}$, then the orbits generated by the diffeomorphism and Weyl
groups acting on $\widehat g_{ab}$ generate the full space of metrics
on $\Sigma$. Thus given the slice $\widehat g(\t)$, any metric can be
represented in the form

$$
f^* g= e^\vp {\widehat g}(\t)
\eqn\rrv
$$

\no where $f^*$ represents the action of the diffeomorphism. Since
the integrand of {\ri} is diffeomorphism-invariant, the result of the
integral would be infinite, unless we gauge-fix the symmetry. This
can be done by introducing a pair of anticommuting Faddeev-Popov
ghosts $(b,c,\bar b,\bar c)$, as in ordinary gauge theories:

$$
{\cal S}_{gh}(b,c,\bar b,\bar c,g)=\int d^2\x \sqrt{g}
\left(b_{zz}\nabla _{\bar z} c^z+b_{\bar z\bar z}\nabla_z c^{\bar
z}\right).
\eqn\rvi
$$

\no The integration over the metric $g$ can be written as an
integration over the moduli $[d \t]$, an integration over the
conformal factor $\vp$ and an integration over the ghosts

$$
{\cal D}_g(gh)={\cal D}_g b {\cal D}_g c{\cal D}_g \bar b{\cal D}_g
\bar c.
\eqn\rrvi
$$

\no The partition function that we would like to calculate can then
be expressed as

$$
{\cal Z}=\int [d\t ] {\cal D}_g\vp {\cal D}_g(gh) {\cal D}_g X
\exp\left({-{\cal S_M}-{\cal S}_{gh}-{\mu_0 \o 8\pi} \int
d^2\x\sqrt{g}}\right).
\eqn\rvii
$$

\no We can now choose a metric slice $g= e^\vp \widehat g$. Under
this transformation we already know that ${\cal D}X$ transforms as
{\riv}, while a similar result can be deduced for the ghosts

$$
{\cal D}_{e^\vp {\widehat g}}(gh)={\cal D}_{\widehat g} (gh)
\exp\left({-{26\o 48 \pi} {\cal S}_L(\vp, g)}\right).
\eqn\rrvii
$$

\no The gauge fixing of the measure ${\cal D}_g \vp$ is complicated
since the norm implicitly depends on the metric. In [\david,\diska]
the assumption was made that the Jacobian of the transformation

$$
{\cal D}_{e^\vp \widehat g}\vp{\cal D}_{{e^\vp \widehat g}}(gh)
{\cal D}_{{e^\vp \widehat g}}X=J(\vp,\widehat g)
{\cal D}_{\widehat g}\vp {\cal D}_{\widehat g} (gh) {\cal
D}_{\widehat g} X
\eqn\rviii
$$

\no takes the form of a local Liouville action, so that the partition
function  becomes

$$
{\cal Z}=\int [d\t ] {\cal D}_{\widehat g} \vp {\cal D}_{\widehat g}
(gh) {\cal D}_{\widehat g} X e^{-{\cal S}}
\eqn\rix
$$

\no with

$$
{\cal S}={\cal S_M}(X,{\widehat g})+{\cal S}_{gh}(b,c,\bar b,\bar
c,{\widehat g)}
+\int d^2\x \sqrt{\widehat g} \left({\widetilde a} {\widehat
g}^{ab}\p_a \vp \p_b\vp+{\widetilde b} {\widehat R} \vp +\mu
e^{\widetilde c \vp} \right).
\eqn\rrix
$$

\no Here $\mu$ is the renormalized cosmological constant. The
constants ${\widetilde a}$, ${\widetilde b}$ and ${\widetilde c}$ are
fixed by requiring invariance under

$$
\eqalign{
{\widehat g}_{ab}&\rightarrow {\widehat g}_{ab}e^{\s(\xi)}\cr
\vp (\xi)&\rightarrow \vp (\xi)-\s(\xi).\cr}
\eqn\rrix
$$

\no This invariance comes from the fact that ${\cal Z}$ is only a
function of $e^\vp \widehat g=g$. The result is

$$
{\widetilde b}={25-D\o 48 \pi},\qquad \qquad {\widetilde a}={\tilde
b\o 2}.
\eqn\rx
$$

\no After a rescaling $\vp\rightarrow \sqrt{12/(25-D)}\, \vp$, which
gives a canonical kinetic term, the energy-momentum tensor
corresponding to this action has the form:

$$
T=-\h (\p \vp)^2+{Q \o 2} \p^2 \vp\qquad {\rm with} \qquad
Q=\sqrt{25-D\o 3}.
\eqn\rxi
$$

\no It has a central charge $c_L=1+3Q^2$. We can now determine
$\tilde c$ in {\rix}. Since $\vp$ has been rescaled, we will write
this term of the action as $e^{\g\vp}$ and determine $\g$ from the
requirement that $e^{\g\vp}$ behaves as a $(1,1)$ conformal field.
{}From the OPE with the energy-momentum tensor we obtain the
conformal weight:

$$
\D(e^{\g \vp})=\bar \D(e^{\g \vp})=-\h \g (\g-Q)=1.
\eqn\rxii
$$

\no Therefore

$$
\g={1\o \sqrt{12}}\left(\sqrt{25-D}-\sqrt{1-D}\right).
\eqn\rxiii
$$

\no Here the sign between the square roots has been chosen in such a
way that the classical limit is recovered as $D \rightarrow
{-\infty}$. From the Liouville part of the action {\rrix} it becomes
clear that $\hbar= \g^2$, so that the classical limit should coincide
with $\g \rightarrow 0$. From the above formulas it becomes clear
that the domain $D\leq 1$ is the one where the Liouville theory can
be most easily interpreted.
Summarizing, the total action is

$$
{\cal S}={\cal S_M}(X,{\widehat g})+{\cal S}_{gh}(b,c,{\bar b},{\bar
c},{\widehat g})+{1\o 8 \pi} \int d^2 \xi \sqrt{\widehat g} \left(
{\widehat g}^{ab} \p_a \vp \p_b \vp +Q {\widehat R}\vp + \mu e^{\g
\vp}\right),
\eqn\rrxiii
$$

\no with $Q$ and $\g$ given by {\rxi} and {\rxiii} respectively.

An important quantity, that we now have to consider, is the partition
function for fixed area. We can rewrite {\rvii} for fixed genus $h$
as an integral over $A$

$$
{\cal Z}_h=\int_0^\infty dA \, {\cal Z}_h (A) \, \exp\left(-{\m \o
8\pi} A\right),
\eqn\rrxiii
$$

\no where

$$
{\cal Z}_h(A)=\int {\cal D} \vp {\cal D}X e^{-{\cal S}} \d \left(\int
d^2 \x \sqrt{g}-A\right).
\eqn\rxiv
$$

\no The integration over the moduli and the ghost determinant are
included in ${\cal D}X$.
The partition function for fixed area {\rxiv} has a scaling behavior
as a function of the area that is universal. It is characterized by a
critical exponent, called the string susceptibility $\G_{str}(h)$

$$
{\cal Z}_h(A)\sim A^{\G_{str}(h)-3}.
\eqn\rxv
$$

\no Due to this behavior, it turns out that in some cases the
complete partition function {\rrxiii} is ill defined. This exponent
can be determined from {\rxiv} using simple scaling arguments.
Shifting $\vp\rightarrow \vp+\rho/\g$, where $\rho$ is a constant,
and setting $A=\exp\left( \rho\right)$ we can deduce:

$$
{\cal Z}_h(A)= A^{-Q(1-h) /\g-1}{\cal Z}_h(1).
\eqn\rrxiv
$$

\no Comparing with {\rxv} we obtain

$$
\G_{str}(h)=2-{Q\o{\g}}(1-h).
\eqn\rrxv
$$

\no For genus zero this gives

$$
\G_{str}(0)={1\o 12}\left(D-1-\sqrt{(25-D)(1-D)}\right).
\eqn\rxvi
$$

We now consider a minimal CFT coupled to 2D quantum gravity. These
models have a finite number of primary fields and are described by a
central charge

$$
D=c_{p,q}=1-{6(p-q)^2\o pq},
\eqn\rrxvi
$$

\no where $(p,q)$ are coprime integers with $q<p$. In general, the
spinless primary fields of the matter theory $\exp(ip_0X)$, with
conformal dimension $\D_0$:

$$
\D_0=\D_0^{r,s} ={1\o 4 pq} \left[ \left( rq-sp
\right)^2-\left(p-q\right )^2\right],
\eqn\rrxvii
$$

\no where $1\leq r\leq p$ and $1\leq s\leq q$, get dressed by the
Liouville field after the coupling to gravity

$$
\Phi =e^{ip_0X(\xi)} e^{\b \vp(\xi)}.
\eqn\rxvii
$$

\no The value of $\b$ is obtained by demanding that {\rxvii} is
invariant under conformal transformations, i.e. that $\Phi$ is a
$(1,1)$ field:

$$
\D_0-\h(\b-Q)\b=1.
\eqn\rxviii
$$

\no From {\rxviii} we obtain the relation

$$
\b={\sqrt{25-D}\pm \sqrt{1-D+24\D_0}\o\sqrt{12}}={Q\o 2} \pm
\sqrt{{Q^2-8 \o 4} +2\D_0}.
\eqn\rxvi
$$

\no Here the branch $\b \leq Q/2$ reproduces the semiclassical limit,
where $\b \rightarrow 0$ as $D\rightarrow -\infty$. It has been
argued in [\sei,\polch] that only these operators can exist. They are
usually called operators on the right branch. Nevertheless, operators
on the wrong branch, that violate the above Seiberg bound have an
interesting application in the context of Witten's two-dimensional
black hole
\REF\wit{E.~Witten,``On String Theory and Black Holes'', {\it
Phys.~Rev.} {\bf D44} (1991) 314.}
[\wit].

For a minimal model the value of $\G_{str}(0)$ obtained from {\rrxvi}
is

$$
\G_{str}(0)=-{p-q\o q}.
\eqn\rrxvii
$$

\no If $(p,q)=(m+1,m)$ where $m=3,4,\dots$, i.e. if we consider the
unitary models, we obtain $\G_{str}(0)=-1/m$. However, for general
$(p,q)$ models the above value of $\G_{str}(0)$ does not agree with
the expressions obtained from the matrix models, where

$$
\G_{str}(0)=-{2\o p+q-1}.
\eqn\rrxviii
$$

\no It has been suggested in
\REF\bdk{E.~Brezin, M.~R.~Douglas, V.~Kazakov and S.~H.~Shenker,``The
Ising Model Coupled To 2-D Gravity: A Nonperturbative Analysis, {\it
Phys.~Lett.} {\bf B237} (1990) 43.}
ref. [\bdk] that this discrepancy can be solved if we measure the
scaling of the partition function with respect to the operator of
lowest dimension of the CFT. For the unitary minimal models this
operator is the identity, while in the non-unitary case it has a
negative dimension

$$
\D_{min}={1-{(p-q)^2}\o {4pq}},
\eqn\rrxix
$$

\no and we obtain

$$
{Q \o \b_{min}}={2(p+q)\o p+q-1}.
\eqn\rrxx
$$

\no The expression for the string susceptibility takes then the form
{\rrxviii} (for spherical topologies) if we replace $Q / {\g}$ in
{\rrxv} by $Q/{\b_{min}}$. For the simplest non-unitary minimal
models of type $(2m-1,2)$ this exponent has the value
$\G_{str}(0)=-1/m$, so that it coincides with the expression
previously found for the unitary minimal models coupled to gravity.

Apart from the partition function, we would like to compute
amplitudes for a surface to pass a given set of points. For a surface
with pinned points $\{X_i\}$ this amplitude is given by

$$
G(X_1,\dots,X_N)=\Biggl\langle\prod_{i=1}^N \int \sqrt{g(\xi_i)}
\delta \left(X_i-X(\xi_i)\right)d^2 \xi_i \Biggr\rangle,
\eqn\rrxxi
$$

\no where the average is taken w.r.t. the functional integral {\ri}.
Passing to the momentum representation this is equivalent to

$$
G(p_1,\dots,p_N)=\Biggl\langle\prod_{i=1}^N \int \sqrt{g(\xi_i)} e^{i
p_i X(\xi_i)} d^2 \xi_i \Biggr\rangle.
\eqn\rrxxi
$$

\no These are the correlation functions of (gravitationally dressed)
scaling operators {\rxvii}. As the partition function, these
correlation functions have a universal scaling behavior as a function
of the area. The scaling behavior of the one point function for fixed
area

$$
F_{\Phi}(A)={1\o {\cal Z}_h(A)} \int {\cal D} \vp {\cal D}X e^{-{\cal
S}} \d\left(\int d^2 \x \sqrt{\widehat g} e^{\g \vp}-A\right)\int
d^2\x \sqrt{\widehat g}\, e^{ip_0 X} e^{\b \vp},
\eqn\rxivu
$$

\no  defines a critical exponent called the gravitational scaling
dimension $\D$:

$$
F_{\Phi}(A)\sim A^{1-\D}.
\eqn\rxv
$$

\no With similar scaling arguments as the ones used to determine
$\g$, it is possible to show that the one-point function has a
scaling behavior as a function of the area:

$$
F_{\Phi}(A) =A^{\b / \g} F_{\Phi}(1).
\eqn\rrxiv
$$

\no This gives $\D=1-{\b/\g}$ and therefore

$$
\D={\sqrt{1-D+24\D_0}-\sqrt{1-D}\o \sqrt{25-D}-\sqrt{1-D}}.
\eqn\rxvii
$$

Integrating over the area the one-point functions of scaling
operators,
scale as a function of the renormalized cosmological constant

$$
\int_0^\infty d A {\cal Z}_h(A) F_{\Phi}(A) \exp\left(-{ \mu \o
8\pi}\right)  \sim \m^{-\G_{str}+\D+1 },
\eqn\rxviixix
$$

\no and the partition function behaves as

$$
{\cal Z}_h \sim   \m^{2-\G_{str}}.
\eqn\rxviixixx
$$

\no These representations are useful in order to compare with the
results from matrix models.

To quantize the theory we are interested in the calculation of
$N$-point correlation functions of scaling operators. This can be
done in a flat background [\klepa,\dtwo,\gl,\dfk] or in a non-trivial
background such as the 2D black hole
\REF\wir{K.~Becker and M.~Becker, ``Correlation Functions in the {\rm
SL(2,\IR)/U(1)} Black Hole Background'', {\it Nucl.~Phys.} {\bf B},
to appear.} [\wir].
We are not going to explain these results here.

\section{\chaoneonetwo}

\no In the following we consider the $N=1$ SCFTs with central charge
$\widehat c=D$
\REF\shenk{S.~H.~Shenker, ``Introduction to Two-Dimensional Conformal
and Superconformal Field Theory'', in Proc. of Workshop on {\it
``Unified String Theories''}, Santa Barbara, CA, 1985;
M.~A.~Bershadsky, V.~G.~Knizhnik and M.~G.~Teitelman,
``Superconformal Symmetry in Two Dimensions'', {\it Phys. Lett.} {\bf
151B} (1985) 31.}
[\shenk], described by the action ${\cal S_M}$, coupled to
two-dimensional supergravity. The partition function for a fixed
topology of the base manifold is [\dhk]

$$
{\cal Z}_h=\sum_{\scriptstyle spin \atop \scriptstyle structures}
\int {\cal D}E_M^A {\cal D} {\bf X}  e^{-{\cal S_M}[{\bf X},E]}.
\eqn\ui
$$

\no Here we have used the superspace formulation introduced in
\REF\ckt{S.~Chaudhuri, H.~Kawai and S.~H.~H.~Tye, ``Path-Integral
Formulation of Closed Strings'', {\it Phys.~Rev.} {\bf D36} (1986)
1148.}
\REF\Mart{E.~Martinec, ``Superspace Geometry of Fermionic
Strings'',{\it Phys.~Rev.} {\bf D28} (1983)~2604.}
\REF\howe{P.~S.~Howe, ``Super Weyl Transformations in Two
Dimensions'', {\it J.~Phys.} {\bf A12} (1979) 393.}
[\ckt,\Mart,\howe], where $E^A_M$ means the super-zweibein. In this
formulation the action of the $N=1$ free massless superstring takes
the form

$$
{\cal S_M}={1\o 8\pi} \int d^2 {\bf z} E D{\bf X} {\bar D} {\bf X}.
\eqn\uii
$$

\no The field $E$ is the superdeterminant of the super-zweibein
$E_M^A$ and $D$ is the covariant superderivative.

In the supersymmetric action the cosmological term of the form

$$
\mu_0 \int d^2 {\bf z}\,  E,
\eqn\rrrxvii
$$

\no is not needed for renormalizability, in contrast to the bosonic
case. The only reason to include this term in the action would be
that it is compatible with all its symmetries.

The construction of the super-Liouville action is completely
equivalent to the bosonic theory and has been carried out in
[\pz,\dhk]. First we have to define a reference frame $\widehat E$:

$$
E=e^{\g \Phi_L} {\widehat  E}.
\eqn\rrrrxvii
$$

\no After gauge fixing the action in the superconformal gauge it is
assumed that Jacobian  of the transformation of the measures takes
the form of a local super-Liouville action:

$$
{\cal S}_{SL}[\Phi_L,{\widehat E}]={1\o 4\pi} \int d^2 {\bf z}
{\widehat E} \left( \h {\widehat D}_\a \Phi_L {\widehat D}^\a
\Phi_L+Q {\widehat Y}  \Phi_L \right),
\eqn\rrrxvii
$$

\no where

$$
\Phi_L({\bf z},{\bf \bar z})=\varphi(z,\bar z) +\theta \psi(z,\bar z)
+ {\bar\theta}{\bar\psi}(z,\bar z)+\theta {\bar\theta}F(z,\bar z)
\eqn\rrrrrxvii
$$

\no is the Liouville superfield, and $\psi$ and $\varphi$ are the
fields that we introduced in {\wwwxxix}; $F$ is an additional
auxiliary field and $\widehat Y$ is the supercurvature
[\ckt,\Mart,\howe].
In the component language the above action is

$$
{\cal S}_{SL}={1\o 2\pi} \int d^2 z \left( \p\vp {\bar \p} \vp+{1\o
4} Q {\widehat R} \vp-{\bar \psi}\p {\bar \psi} -\psi {\bar \p}
\psi\right),
\eqn\rrrxviii
$$

\no from which we obtain the energy-momentum tensor

$$
T_L(z)=-\h (\p \vp \p \vp -Q \p^2 \vp -\psi \p \psi).
\eqn\rrxvii
$$

\no It has a central charge

$$
c_L=1+3Q^2+\h ={3\o 2}\left(1+2Q^2\right).
\eqn\rrxx
$$

{}From the gauge-fixing procedure we get a Faddeev-Popov
superdeterminant that can be represented in terms of two ghost
superfields :

$$
B({\bf z})=\b(z)+\th b(z) \qquad {\rm and} \qquad C({\bf z})=c(z)+\th
\g (z),
\eqn\rrxviii
$$

\no where $(b,c)$ are fermionic ghost with spin $(2,-1)$ and a
central charge $c_{bc}=-26$, while $(\b,\g)$ are the bosonic
supersymmetry ghosts with a spin $(3/2,-1/2)$ and a central charge
$c_{\b\g}=11$.

The original partition function {\ui} is invariant under a
simultaneous rescaling of the frame $\widehat E$ and a shift in
$\Phi_L$

$$
\widehat E \rightarrow e^{\sigma({\bf z},{\bf \bar z})} \widehat
E,\qquad \qquad \Phi_L \rightarrow\Phi_L -{1\o \g} \s ({\bf z},{\bf
\bar z}) .
\eqn\rrxviii
$$

\no This condition can be used to determine $Q$ and $\g$. The
requirement that this symmetry is preserved in the quantum theory is
equivalent to the vanishing of the total central charge

$$
c_{tot}={\widehat c}+c_{gh}+c_L=0.
\eqn\rrxix
$$

\no The conformal anomaly of the matter and ghost system is
${\widehat c} +c_{gh}={\widehat c} -26+11$. We therefore obtain

$$
Q=\sqrt{9-{\widehat c}\o 2}.
\eqn\rrrxx
$$

\no The constant $\g$ from {\rrrrxvii} can be fixed demanding that
$\exp\left(\g\Phi_L\right)$ transforms like a density, i.e. like a
superconformal field of dimension $(1/2,1/2)$. Therefore

$$
\D(e^{\g \Phi_L})=-\h \g( \g-Q)=\h.
\eqn\rrrxxi
$$

\no This equation has the solution

$$
\g ={Q\o 2}-\h\sqrt{Q^2 -4}={1\o \sqrt{12}}\left( \sqrt{9-{\widehat
c} }- \sqrt{1-{\widehat c}}\right).
\eqn\rrrxxii
$$

\no In the above expression we have taken the branch that reproduces
the correct value of $\g$ in the semiclassical limit. We observe that
the theory is well defined provided that $\widehat c<1$. At
${\widehat c}=1$, the theory undergoes a phase transition and the
local ansatz for the Jacobian is no longer valid [\dhk].

In order to compute the critical exponents we have to rewrite the
partition function as an integral over $L$, that is the
characteristic length scale of a super random surface:

$$
{\cal Z}_h=\int_0^\infty dL {\cal Z}_h(L)
\eqn\rrrxxiii
$$

\no with

$$
{\cal Z}_h(L)=\sum_{\scriptstyle spin \atop \scriptstyle
structures}\int {\cal D} {\bf X} {\cal D}\Phi_L \delta \left( \int
d^2 {\bf z} E -L\right)e^{-{\cal S_M}-{\cal S}_{gh}-{\cal S}_{SL}}.
\eqn\rrrxxiv
$$

\no Here the dimension of $L$ is that of the length because the
Grassmann integration in $L=\int d^2 {\bf z} E$ lowers the dimension
by one.

The string susceptibility is defined as:

$$
{\cal Z}_h(L)\sim L^{\G_{str}(h)-3}.
\eqn\rrrxxv
$$

\no Like for the bosonic string, it can be computed from simple
scaling arguments:

$$
\G_{str}(h)=2-(1-h){Q\o \g} =2+{(1-h)\o 4}\left({\widehat
c}-9-\sqrt{(9-{\widehat c})(1-{\widehat c)}}\right).
\eqn\rrxxi
$$

To determine the gravitationally renormalized scaling dimensions we
have to distinguish between operators in the Neveu-Schwarz and in the
Ramond sector.
In {\IIxiii} we saw that there exist $2^{2h}$ different spin
structures for the matter theory, corresponding to the different
boundary conditions of the fermions around the non-contractible
cycles. Similarly, there exist $2^{2h}$ spin structures for the
supergravity theory. World-sheet supersymmetry requires that each
spin structure of the matter system be coupled to the same spin
structure of the supergravity system
\REF\bk{M.~Bershadsky and I.~Klebanov, ``Partition Functions and
Physical States in Two-Dimensional Quantum Gravity and
Supergravity'', {\it Nucl.~Phys.} {\bf B360} (1991) 559.}
[\bk]. The spin structure for the holomorphic and antiholomorphic
fields are identified.

The operators in the Neveu-Schwarz sector are similar to the ones of
the bosonic theory and can be written in the form:

$$
\Psi_{\rm NS}=\int d^2 {\bf z} {\widehat E}\exp \left( ik {\bf X} +\b
\Phi_L \right).
\eqn\rrxxii
$$

\no Since these operators are to be integrated they have to be
$(1/2,1/2)$ fields. The on-shell condition is then

$$
\D_0-{\b\o 2}(\b-Q)=\h.
\eqn\rrrxxii
$$

\no We use the notation $\D_0$ for the scaling dimensions of the
primary fields of the matter theory, $\exp( ik {\bf X})$. The above
equation has the solutions

$$
\b={Q\o 2} \pm \h \sqrt{Q^2-4+8\D_0}.
\eqn\rrrxxiii
$$

\no We have to keep the operators satisfying $\b\leq Q/2$.

The renormalized scaling dimension can be computed from the scaling
of the two-point function (for fixed superlength) as a function of
$L$, in analogy to the bosonic theory

$$
\D^{\rm NS}=1-{\b \o \g}={\sqrt{1-{\widehat c}}- \sqrt{1-{\widehat
c}+16\D_0}\o
\sqrt{1-{\widehat c}}-\sqrt{9-{\widehat c}}}.
\eqn\rrxxvi
$$

\no The operators in the Ramond sector are more complicated, since
they have spin [\dhk,\bk]. The renormalized scaling dimensions in
this case are given by

$$
\D^{\rm R}={\sqrt{1-{\widehat c}}- \sqrt{-{\widehat c}+16\D_0}\o
\sqrt{1-{\widehat c}}-\sqrt{9-{\widehat c}}}.
\eqn\rrxxvii
$$

The minimal $N=1$ SCFT are characterized by a pair of integers
$(p,q)$ with $q<p$ and they have a central charge

$$
{\widehat c}=1-{2(p-q)^2\o{pq}}.
\eqn\rrxxviii
$$

\no These theories have a finite number of primary fields that have a
conformal dimension

$$
\D_0=\D^{r,s}_0={{(rq-sp)^2-(q-p)^2 \o 8qp}+{1-(-)^{r-s} \o 32}}.
\eqn\rrxxviii
$$

\no with $ 1\leq r \leq p-1$ and $1\leq s \leq q-1$.

Unitary SCFTs correspond to $(p,q)=(m+2,m)$. The simplest SCFTs are
those of type $(p,q)=(4m,2)$ with $m=1,2,\dots $. They have a central
charge

$$
{\widehat c}=1-{(2m-1)^2\o m}=0, -{7\o 2}, \dots .
$$

\no These are the models that we are going to consider in the next
chapters. The case $\widehat c=0$ corresponds to pure supergravity
and it is the only unitary model of the above series.

{}From {\rrxxviii} we obtain the conformal dimension for operators in
the Neveu-Schwarz sector:

$$
\D^{2i+1,1}_{0}={i(i-2m+1)\o 4m}\qquad {\rm with} \qquad
i=0,1,\dots,2m-1;
\eqn\rrxxviii
$$

\no and in the Ramond sector

$$
\D^{2i+2,1}_{0}={{(i+1-m)^2-(m-1/2)^2 \o 4m}+{1 \o 16}}\qquad {\rm
with} \qquad i=0,1,\dots ,2m-2.
\eqn\rrxxviii
$$

The string susceptibility can be obtained from {\rrxxi}. We have to
take into account that if the theory is non-unitary, it is natural to
assume that the cosmological constant couples to the lowest dimension
operator [\bk]. Therefore, we have to replace $Q/\g$ by $Q/ \b_{min}$
in {\rrxxi}, and obtain

$$
{Q\o\b_{min}}={2(p+q)\o p+q-2}\qquad {\rm and} \qquad
\b_{min}={p+q-2\o  2\sqrt{pq}}.
\eqn\rrxxviiii
$$

\no $\b_{min}$ is the dressing of the primary field with the lowest
weight

$$
\D^{min}_0={4-(p-q)^2\o  8pq}.
\eqn\rrxxix
$$

\no For the series of type $(4m,2)$ we obtain the same value of the
string susceptibility as in the bosonic theory $\G_{str}(0)=-1/m$.

One of the advantages of the continuum formulation of 2D-supergravity
is that most of the calculations done on the sphere can be carried
out with methods similar to those of the bosonic theory. Correlation
functions of scaling operators can be computed using an analogous
Coulomb gas formalism. This has been done in [\sliouv,\abdallabook]
and the results can be taken as a guiding principle to find a more
powerful discrete formulation valid for higher genus that reproduces
the continuum results for spherical topologies. We are not going to
enter in more detail into the continuum formulation of
2D-(super)gravity. Details can be found, for example, in the review
[\abdallabook].

\vfill
\endpage

\chapter{\chatwo}

\no In chapter 1 we saw that in string theory we are interested in
the evaluation of a sum over Riemann surfaces of all possible
topologies. This can be done, using the so-called double scaling
limit, in the context of matrix models as was discovered in 1989
[\doubles]. The simplest case of a matrix model is the one-matrix
model, where
the field is an Hermitian $N\times N$ matrix. The parameter $N$ will
appear as a genus-counting parameter in the partition function and
the double lines that appear in the Feynman diagrams will be useful
to characterize the orientability of the surface, as we will later
see.

In this chapter we will derive the discrete Virasoro constraints and
the equivalent discrete loop equation that determines the solution of
the model. We will analyse carefully the case of arbitrary potentials
and prove  that in the continuum limit a set of two decoupled string
equations appears. We will furthermore present the one-matrix model
in the double scaling limit and show the appearance of the KdV
hierarchy. We introduce some basic notions of completely integrable
systems, such as the Lax-pair formalism for the KdV hierarchy and the
generalization to the KP hierarchy through the formalism of
pseudo-differential operators.

We will concentrate in this chapter on those aspects of the
one-matrix model that we consider relevant for the supersymmetric
generalization. A more complete analysis can be found in some
excellent review articles
\REF\review{V.~Kazakov, ``Bosonic Strings
and String Field Theories in One-Dimensional Target Space", lectures
given at
Carg\`ese, France, May 1990; L.~Alvarez-Gaum\'e,
{\it Helv.~Phys. Acta} {\bf 64} (1991) 359, and references therein;
P. Ginsparg, ``Matrix Models of 2d Gravity'', Los Alamos Preprint
LA-UR-9999, hep-th/9112013.}
\REF\gm{P.~Ginsparg and G.~Moore, ``Lectures on 2D Gravity and 2D
String Theory'', Lectures given at TASI Summer School, Boulder CO
(1992),
Yale Preprint YCTP-P23-92, hep-th/9304011.}
\REF\dijkgraaf{R.~Dijkgraaf, ``Intersection Theory, Integrable
Hierarchies and Topological Field Theory'', Lectures presented at the
Carg\`ese Summer School on New Symmetry Principles in Quantum Field
Theory, (1991), hep-th/9201003.}
\REF\dav{F.~David, ``Simplicial Quantum Gravity and Random
Lattices'', Lectures given at {\it Les Houches Summer School on
Gravitation and Quantizations}, Les Houches, (1992), SACLAY-T93-028.}
[\abdallabook,\review,\gm,\dijkgraaf,\dav].

\section{\sectwoone}

\no The partition function of the one-matrix model is defined as the
following matrix integral
\REF\biz{D.~Bessis, C.~Itzykson and J.~B.~Zuber, ``Quantum Field
Theory Techniques in Graphical Enumeration'', {\it Adv.~App.~Math.}
{\bf 1} (1980) 109.}
\REF\cmc{S.~Chadha, G.~Mahoux and M.~L.~Metha, ``A Method of
Integration over Matrix Variables: II'', {\it J.~Phys.} {\bf A14}
(1981) 579.}
\REF\pd{E.~Br\'ezin, C.~Itzykson, G.~Parisi and J.~B.~Zuber, ``Planar
Diagrams'', {\it Commun.~Math.~Phys.} {\bf 59} (1978) 35.}
[\biz,\cmc,\pd]:

$$
{\cal Z_B}=\int d^{N^2} \Phi \exp\left(-{N \o \L_{\cal B}} \;{\rm tr}
V(\Phi)\right).
\eqn\ai
$$

\no Here $\P$ is an Hermitian $N\times N$ matrix and $\L_{\cal B}$ is
a free parameter  related to the cosmological constant, as we will
see later. The potential is defined as:

$$
V(\Phi)={\Phi^2 \over 2} +\sum_{k \geq 3} g_k \Phi^k.
\eqn\aii
$$

\no The integration measure on Hermitian matrices is:

$$
d^{N^2}\Phi=\prod_{i=1}^{N}d\Phi_{ii}\prod_{1\leq i< j\leq
N} d({\rm Re}\,\Phi_{ij})d({\rm Im}\,  \Phi_{ij}).
\eqn\aiii
$$

\no This partition function can be evaluated perturbatively using
Feynman
rules. Expanding the exponentials, expression {\ai} is reduced to the
calculation of Gaussian integrals:

$$
{\cal Z_B}=\int d^{N^2}\Phi \exp\left(-{N \o 2\L_{\cal B}} \; {\rm
tr}  \Phi^2\right) \prod _{p\geq 3}
\left(\sum_{n_p=0}^{\infty}{g_p^{n_p}\over n_p!}\left(-{N \o \L_{\cal
B}}\right)^{n_p} ({\rm tr} \;
\Phi^p)^{n_p}\right).
\eqn\aiv
$$

\no The product runs over the different types of vertices. The
propagator is the elementary integral:

$$
\langle \Phi_{ij} \Phi_{kl}^*\rangle ={\int d^{N^2} \Phi \exp\left(
-{N \o 2\L_{\cal B} } {\rm tr} \Phi^2\right) \Phi_{ij} \Phi^*_{kl}
\o \int d^{N^2}\Phi \exp\left( -{N \o 2\L_{\cal B}} {\rm tr} \Phi^2
\right) } ={\L_{\cal B} \o N} \delta_{ik} \delta_{jl},
\eqn\aaaiv
$$

\no and can be represented by a double line (see Fig. 3).

\vskip 0.4cm
\midinsert
\epsfysize=0.45in
\centerline{\hskip 0.1cm\epsffile{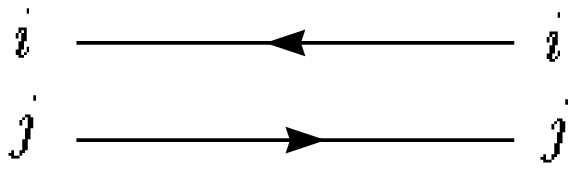}}
\vskip 0.2 cm
\caption{{\bf Fig.~3}: The elementary contraction for hermitian
matrices.}
\endinsert

\no This double line notation was first introduced by 't Hooft to
represent a gluon propagator in QCD. Since $\Phi$ is Hermitian we
have no twisted propagators and therefore the surface is orientable.
More complicated graphs can be evaluated using Wick rules.
Diagrammatically a vertex of type $\Phi^4$ takes the form shown in
Fig.~4.

\vskip 0.4cm
\midinsert
\epsfysize=1.5in
\centerline{\hskip 0.0cm \epsffile{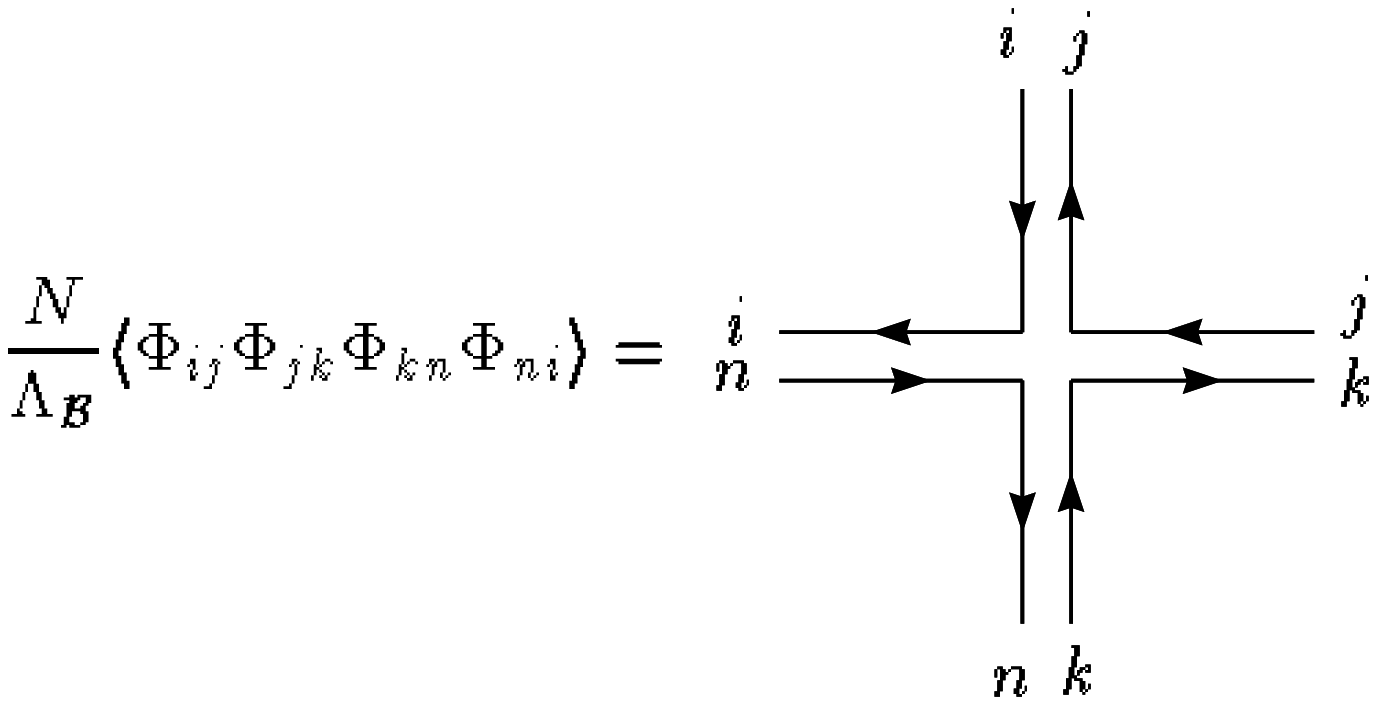}}
\caption{\hskip 1.5cm {\bf Fig. 4}: The quartic vertex representing
${\rm tr} \Phi^4$.}
\vskip 0.4cm
\endinsert

To make the above clear, we consider a particular example, where only
$g_4$ is different from zero. The partition function has the
expansion of the form
$$
\eqalign{
{\cal Z_B}(\L_{\cal B},g_4)&=\int d^{N^2}\Phi \exp\left(-{N \over
2\L_{\cal B}}\; {\rm tr}  \Phi^2 -{N \o \L_{\cal B}} g_4 {\rm tr}
\Phi^4\right)\cr
&\cr
&=1-{ N \o \L_{\cal B}} g_4 \langle {\rm tr} \Phi^4
\rangle+\left({N\o \L_{\cal B}} g_4\right)^2 \langle {\rm tr}
\Phi^4{\rm tr} \Phi^4\rangle  +\dots . }
\eqn\aaiv
$$

\no The simplest contribution, where $n_4=1$ has the following form

$$
\eqalign{
\langle {\rm tr} \Phi^4 \rangle
&=\langle \Phi_{ij} \Phi^*_{kj} \rangle\langle \Phi_{kl}
\Phi_{il}^*\rangle +\langle \Phi_{ij} \Phi_{lk}^* \rangle \langle
\Phi_{jk} \Phi_{il}^*\rangle+
\langle \Phi_{ij} \Phi_{il} ^* \rangle \langle \Phi_{jk}
\Phi_{lk}^*\rangle\cr
&\cr
&={\L_{\cal B}^2\o N^2} \left( \delta_{ik} \delta_{jj} \delta_{ki}
\delta_{ll} +\delta_{il} \delta_{jk} \delta_{ji}
\delta_{kl} +\delta_{ii} \delta_{jl} \delta_{jl} \delta_{kk}\right)
={\L_{\cal B}^2\o N^2} \left( N^3 +N+N^3\right), \cr}
\eqn\aav
$$

\no that diagrammatically can be represented as shown in Fig.~5.

\vskip 0.5cm
\midinsert
\epsfysize=2.0in
\centerline{\hskip 1.2cm \epsffile{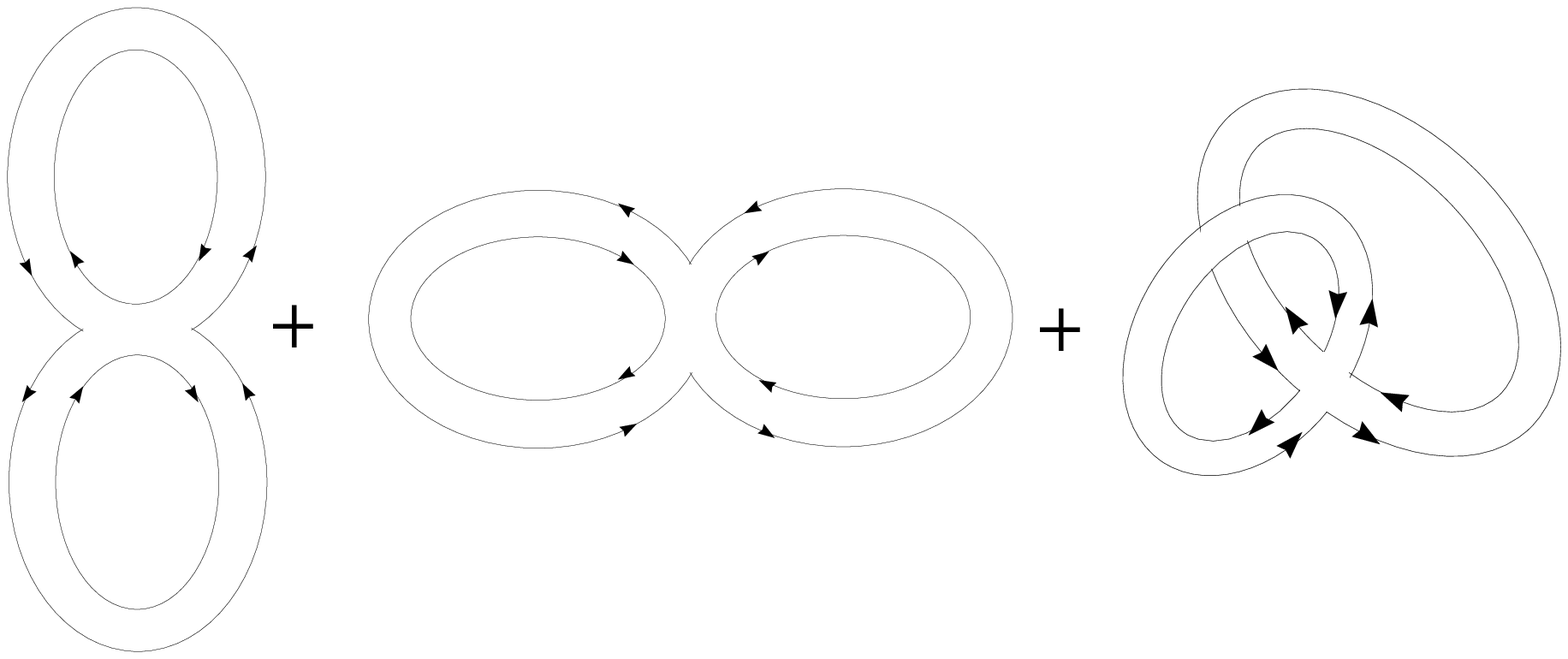}}
\caption{{\bf Fig. 5}: These are the three contributions to $\langle
{\rm tr} \Phi^4 \rangle$ of {\aav}.}
\endinsert
\vskip 0.5cm

In order compare with the Liouville theory, we are only interested in
the connected graphs of the matrix model. The generating functional
for the connected diagrams is the free energy

$$
F=\log {\cal Z_B}.
\eqn\avii
$$

\no This is the quantity that has to be compared with the partition
function of Liouville theory.

In general, a term of order $g_p^{n_p}$ in the expansion {\aiv}
defines a closed Feynman diagram with $n_p$ vertices of type $g_p$.
A graph with $(n_3,\dots,n_p)$ vertices of type $(g_3,\dots,g_p)$
will
give a contribution:

$$
(g_3^{n_3}\dots g_p^{n_p})N^{I-P+V}\L_{\cal B}^{P-V} \qquad{\rm
where} \qquad  V=\sum n_p,
\eqn\avii
$$

\no to the free energy, since each vertex contributes a factor $N
g_p/\L_{\cal B}$, each of the $P$ propagators a factor $\L_{\cal
B}/N$ and each
of the $I$ closed loops (index loops) contributes a factor $N$, due
to
the corresponding index summation. Since $p$-lines are incident
on the vertex of type $p$, the total number of propagators
$P$ of a closed graph is $2P=\sum pn_p$.
To see how a genus expansion of the free energy can be obtained we
have to introduce the dual simplex associated to each of these
diagrams.

\topinsert
\epsfysize=2.3in
\centerline{\hskip 0.5cm \epsffile{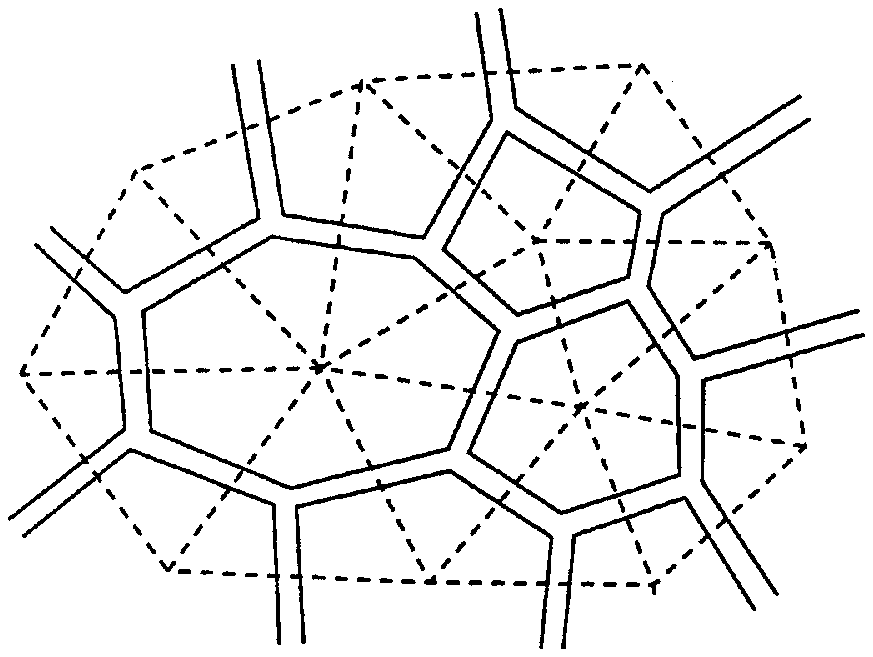}}
\vskip 0.5cm
\caption{{\bf Fig. 6}: The solid lines represent a Feynman diagram of
a $\Phi^3$ interaction, while the dual simplex is built out of
triangles and represented by broken lines.}
\vskip 0.25cm
\endinsert

Every edge of the dual simplex corresponds to a propagator of the
Feynman diagram, every $p$-gonal surface to a $p$-vertex and each
vertex to a face (Fig.~6). We can therefore associate the total
number of edges $\cal E$, faces $\cal F$ and vertices $\cal V$ of the
dual simplex to the quantities $P$, $V$ and $I$ of the Feynman
diagram.

For the dual simplex we know from the Gauss-Bonnet theorem that

$$
{\cal X}=2-2h={\cal F} -{\cal E} +{\cal V}
\eqn\aavii
$$

\no holds, where $\cal X$ is the Euler-characteristic and $h$
is the number of handles of the surface.

In this way the matrix integral {\aiv} automatically gives a random
simplicial representation of surfaces. The large $N$ expansion of
{\ai} can be interpreted a topological expansion, as we will now see.

With the previous notation {\avii} becomes

$$
\prod_{p \geq 3}g_p^{n_p} N^{2-2h} \L_{\cal B}^{ \sum_p n_p (p / 2-1)
},
\eqn\aavii
$$

\no so that we arrive at the following expansion for $F$:

$$
F=\sum_{h=0}^\infty  N^{2-2h}F_h\qquad {\rm with} \qquad
F_h=\sum_{S_h}{1\over |G(S_h)|}\L_{\cal B}^{\sum_p n_p({p/ 2}-1)}
\prod_{p} g_p^{n_p}.
\eqn\aix
$$

\no $F_h$ is the free energy for fixed genus $h$.
The sum in $F_h$ runs over all graphs of the theory, which may have a
symmetry,
so that the order of the symmetry group $G_h$ has to be divided out.
{}From
{\aix} we see that only the spherical topologies
$(h=0)$ contribute in the large-$N$ limit. However, we are interested
in summing over all topologies and it is at this point where the
notion of double scaling limit becomes relevant.

To make contact with the continuum Liouville approach, in particular
to the partition functions {\rvii}, {\rrxiii} and {\rxiv} we have to
introduce the notions of the  area and the cosmological constant in
this context. We can imagine the surface built out of equilateral
triangles with area $1/2$. It is easy to see that the total area of
the surface is given by

$$
{\cal A}=\sum _p \left({p\o 2}-1\right)n_p .
\eqn\aaaix
$$

\no The free energy for fixed genus can then be written as a function
of the free energy for fixed area

$$
F_h=\sum_{{\cal A}}\L_{\cal B}^{{\cal A}} F_h ({\cal A})\qquad {\rm
with} \qquad F_h ({\cal A})=\mathop{{\sum}'}_{S_h}{1 \o |G(S_h) | }
\prod_{p\geq 3} g_p^{n_p}
\eqn\aaax
$$

\no where the prime in the sum means that the area is held fixed.

$$
F= \sum_{h=0}^\infty  N^{2-2h}F_h.
\eqn\axxv
$$

\no For later convenience we define
$$
\L_{\cal B}=e^{-\mu_0},
\eqn\aaaxi
$$

\no where $\mu_0$ is the bare cosmological constant on the lattice.
Therefore,

$$
F_h=\sum_{\cal A}  F_h({\cal A}) e^{-\mu_0 {\cal A}}.
\eqn\aaaxii
$$

It turns out that the above sum on the lattice is ill defined
\REF\zamolo{A.~B.~Zamolodchikov, ``On the Entropy of Random
Surfaces'', {\it Phys.~Lett.} {\bf 117B} (1982) 87.}
[\zamolo]. This is because $F({\cal A})$, that is the number of
different surfaces of a given area ${\cal A}$, grows too fast as
${\cal A} \rightarrow \infty$:

$$
F_h({\cal A})\sim {\cal A}^{\G_{str} (h)-3} \exp\left({ \mu_c {\cal
A}}\right),
\eqn\aaaxiii
$$

\no where $\mu_c$ is a cutoff dependent coefficient. We will later
see that the universal critical exponent $\G_{str}(h)$ coincides with
the one introduced in the continuum theory {\rxv} for the $(2m-1,2)$
minimal models coupled to gravity.

In the vicinity of the continuum limit the sum over ${\cal A}$ can be
approximated by an integral:

$$
F_h\sim \int_0^\infty d{\cal A} e^{-\mu_0 {\cal A}} F_h({\cal A})
\sim (\mu_0-\mu_c)^{2-\G_{str}(h)}.
\eqn\aaaxiv
$$

\no This describes the behavior of the free energy for fixed genus as
a function of the cosmological constant, which is important in order
to describe the double scaling limit.

We are able to keep contributions from Riemann surfaces of arbitrary
genus in the sum {\aix}, if the limits $N\rightarrow \infty$ and $
\mu_0 \rightarrow \mu_c$ are correlated. This can be done by keeping
the `renormalized string coupling constant' $\k$ fixed as $N
\rightarrow \infty$

$$
\k^{-1}=N (\mu_0-\mu_c)^{1-\Gamma_{str}(0)/2}.
\eqn\axi
$$

\no This form of taking the continuum limit is known as the double
scaling limit
[\doubles]. The full free energy has an expansion in powers
of~$\kappa$\foot{Here $F_h$ has been rescaled.}:

$$
F=\sum_{h=0}^{\infty}\kappa^{2h-2}F_h.
\eqn\axii
$$

To define the continuum limit we have to introduce a constant $a$
with the dimension of a length, that is the lattice cut-off. This
means that the length of the basic triangulations are equal to $a$
and no longer constant.
The `renormalized cosmological constant' $t$ is defined according to:
$$
\m_0-\m_c=a^2 t,
\eqn\aaliv
$$
\no where $t$ is kept fixed by tuning $(\m_0-\m_c)$ as $a\rightarrow
0$. Taking $\m_c=0$ we obtain
$$
\L_{\cal B}=1-a^2 t.
\eqn\aalivx
$$

For the one-matrix model the value of the string susceptibility is
$\G_{str}=-1/m$, so that we effectively have to keep
$$
\k=a^{-2-{1/ m}} {1 \o N}
\eqn\alv
$$

\no fixed, as we will now see.

\section{\sectwotwo}
\no The Virasoro constraints satisfied by the partition function were
first discovered in the continuum theory [\dvv,\fkn]. These
constraints are closely related to the integrability and the KdV
structure of the one-matrix model in the scaling limit. However, it
has been realized that a similar connection to integrable hierarchies
appears in the discrete theory. Here the integrable hierarchies are
the Toda and the Volterra hierarchies
\REF\agl{L.~Alvarez-Gaum\'e , C.~Gomez and J.~Lacki, ``Integrability
in Random Matrix Models'', {\it Phys.~Lett.} {\bf B253} (1991) 56.}
\REF\martinec{E.~J.~Martinec, ``On the Origin of Integrability in
Matrix Models'', {\it Commun.~Math. Phys.} {\bf 138} (1991) 437.}
[\agl,\martinec].
The discrete Virasoro constraints can be deduced directly from the
matrix integral, as we will now see.

The infinite number of constraints satisfied by the  partition
function {\ai} follow by using the invariance of $Z$ under a change
of variables
[\matsuo,\mironov]

$$
\Phi'=\Phi+\ep\Phi^{n+1}\qquad{\rm for}\qquad n\geq -1,
\eqn\axiii
$$

\no where $\ep$ is a real parameter. Here we have taken the
analytical transformations into account, which are the ones we are
interested in, as we will see in a moment. These transformations are
classical symmetries of the potential that are equivalent to the
equation of motion of $\Phi$

$$
\sum_{k=0}^{\infty}k g_k \Phi^{k-1}=0.
\eqn\axiiii
$$

\no For the quantum theory we have to take into account the change of
the measure. Under this transformation we have $\Phi'^k=\Phi^k+\ep
k\Phi^{n+k}$ and this implies for the measure and the potential to
first order in $\ep$\foot{For $n=-1$ the measure is invariant.}:

$$
d^{N^2}\Phi ' =d^{N^2}\Phi\left(1+\ep \sum_{\a =0}^n\left({\rm tr}\;
\Phi^\a\right)({\rm tr}\;\Phi^{n-\a})\right),
\eqn\axiv
$$

$$
\sum_{k\geq 0} g_k \Phi'^k=\sum_{k\geq 0} g_k \left(\Phi^k+\ep k
\Phi^{n+k}\right).
\eqn\axv
$$

\no From the invariance of the partition function we obtain the
so-called Schwinger-Dyson equations of the one-matrix model:

$$
\biggl\langle \sum_{k=1}^\infty k g_k{\rm tr}\;
\Phi^{k-1}\biggr\rangle=0 ,
\eqn\axvi
$$
\no and
$$
\biggl\langle \sum_{\a =0}^n \left({\rm tr}\; \Phi^\a\right) \left(
{\rm tr}\; \Phi^{n-\a}\right) \biggr\rangle
={N\o \L_{\cal B}}\biggl\langle \sum_{k=1}^\infty k g_k{\rm tr}\;
\Phi^{n+k}\biggr\rangle \qquad {\rm with} \qquad n\geq 0.
\eqn\axvi
$$

\no These equations can be rewritten as the following constraint
equations, acting on the partition function of the model:

$$
L_n Z=0\qquad {\rm for} \qquad  n\geq -1,
\eqn\axviii
$$
\no where:
$$
L_{-1}=\sum_{k\geq 0}k g_k{\partial \over \partial
g_{k-1}},
\eqn\axix
$$
$$
L_0=\sum_{k\geq 0}k g_k{\partial \over \partial
g_{k}}+N^2,
\eqn\aaxviii
$$
$$
L_n={\L_{\cal B}^2\over N^2}\sum_{k=0}^{n}{\partial^2\over\partial
g_{n-k}\partial g_k}+\sum_{k\geq 0}k g_k{\partial \over \partial
g_{k+n}}.
\eqn\axx
$$

\no It is easy to check that these operators satisfy the Virasoro
algebra:

$$
[L_n,L_m]=(n-m)L_{n+m}\qquad{\rm for} \qquad n,m \geq -1
\eqn\axxi
$$

\no and are therefore called Virasoro constraints. Since $n,m\geq -1$
the term proportional to the central charge does not appear. We have
considered analytic transformations, since these are the only ones
that can be expressed through the potential and its derivatives
w.r.t. the coupling constants.

Before we derive the loop equations we have to introduce the notion
of the loop operator. The basic observables in quantum gravity are
loops [\kazakov,\dbs], since these are the only gauge-invariant
quantities. We distinguish between two kinds of operators,
microscopic and macroscopic loops.
The lattice scaling
operator has the form: $(\L_{\cal B} / N) {\rm tr} \Phi^n $. The
insertion of such an operator into the path integral creates a vertex
with $n$ spokes. It is easy to see that a hole with $n$ edges is
created in the dual surface, since the Euler characteristic has been
changed to ${\cal X}=2-2h-{\cal B}$, where ${\cal B}$ is the number
of holes and ${\cal B}=1$ for one insertion. These operators are
usually called `microscopic' loops because they have only a few
lattice spacings of length. They contain all the information about
integrals over the surface of local operators.
 The generating functional of these loops is the so-called
`macroscopic' loop operator

$$
u_0(l)={1\o Z} {\L_{\cal B} \o N} \int d^{N^2} \Phi \exp\left(-{ N\o
\L_{\cal B} }\; {\rm tr}\, V(\Phi)\right)\, {\rm tr}\, e^{l \Phi}.
\eqn\axxviii
$$

\no The macroscopic loop operators correspond to extended boundaries
on the surface and they can be expanded in terms of the lattice
scaling operators. We will omit the expectation values in our
notation:

$$
u_0(l)={\L_{\cal B}\over N}\; {\rm tr} \,e^{l\Phi}=
\sum_{n=0}^\infty {l^n\o n!} {\L_{\cal B} \o N}\; {\rm tr} \,\Phi^n=
\sum_{n=0}^{\infty}
{l^n\over n!} u_0^{(n)}.
\eqn\axxix
$$

\no The moments $u_0^{(n)}=(\L_{\cal B} / N) \,{\rm tr}\, \Phi^n $
can then be expressed through the free energy of the model\foot{Here
we have rescaled $F$ by a factor of $N^2$, i.e. we set $Z=e^{N^2 F}$;
this is useful in order to guarantee a finite zeroth-order free
energy.}:

$$
u_0^{(0)}=-\L_{\cal B} ^2{\p F \o \p g_0} =\L_{\cal B} \qquad{\rm
and} \qquad  u_0^{(n)}=-\L_{\cal B}^2{\partial F\over
\partial g_n}.
\eqn\axxx
$$

\no We have used that the dependence of $F$ on $g_0$ is trivial
$F=-g_0 / \L_{\cal B} + \dots$.

We observe, that from the $L_0$ constraint, we obtain the discrete
string equation. This means a relation between the parameter
$\L_{\cal B}$ and the operators ${\rm tr}\; \Phi^k $ :

$$
\L_{\cal B}={1\o 2} \sum_{k\geq 0} kg_k{\p_{\L_{\cal B}}}
\biggl\langle {\L_{\cal B} \o N}{\rm tr}\;\Phi^k \biggr\rangle.
\eqn\axxiv
$$

\no This equality can also be written as an integral equation, as
will be done in section 2.3. This will be important in order to
classify the different multicritical behaviors.

The $L_{-1}$ constraint takes the form:

$$
\sum_{k \geq 1}k g_k  \p_{\L_{\cal B}} \biggl\langle {\L_{\cal B} \o
N} {\rm tr}  \, \Phi^{k-1}\biggr\rangle=0.
\eqn\aaxxiv
$$

\no This relation contains a non-trivial information if the potential
of the model is arbitrary, and will be important in the next
chapters.

To find the solution of the Hermitian one-matrix model we can use the
loop equation. It has been used by Kazakov [\kazakov] in the planar
limit to analyse the critical regimes of {\ai}; they provide us with
non-perturbative information of the matrix model, as shown by David
\REF\loopdavid{F.~David, ``Loop Equations and Non-perturbative
Effects in Two-Dimensional Quantum Gravity'', {\it Mod.~Phys.~Lett.}
{\bf A5} (1990) 1019.}
[\loopdavid].

The loop equation can be determined from the Virasoro constraints.
These constraints can be expressed as non-linear differential
equations for the free energy. The $L_n$ constraint takes the form

$$
\sum_{k\geq 0}k g_k{\partial F \over \partial
g_{k+n}}+{ \L_{\cal B}^2\o N^2} \sum_{k=0}^{n}\left({\partial^2 F
\over\partial
g_{n-k}\partial g_k}+N^2 {\p F\o\p g_k}{\p F \o \p g_{n-k}
}\right)=0.
\eqn\axxvi
$$

\no In the limit $N\rightarrow \infty$ we obtain the equation:

$$
\sum_{k\geq 0}k g_k{\partial F_0 \over \partial
g_{k+n}}+
\L_{\cal B} ^2\sum_{k=0}^{n}\left({\p F_0\o\p g_k}{\p F_0 \o \p
g_{n-k} }\right)=0.
\eqn\axxvii
$$

\no This  means that {\axxvii} is equivalent to:

$$
\sum_{k\geq 1} k g_k u_0^{(n+k)}-\sum_{k=0}^n u_0^{(n-k)}
u_0^{(k)}=0.
\eqn\axxxi
$$

\topinsert
\vskip -5cm
\epsfysize=4.0in
\centerline{\hskip 0.5cm \epsffile{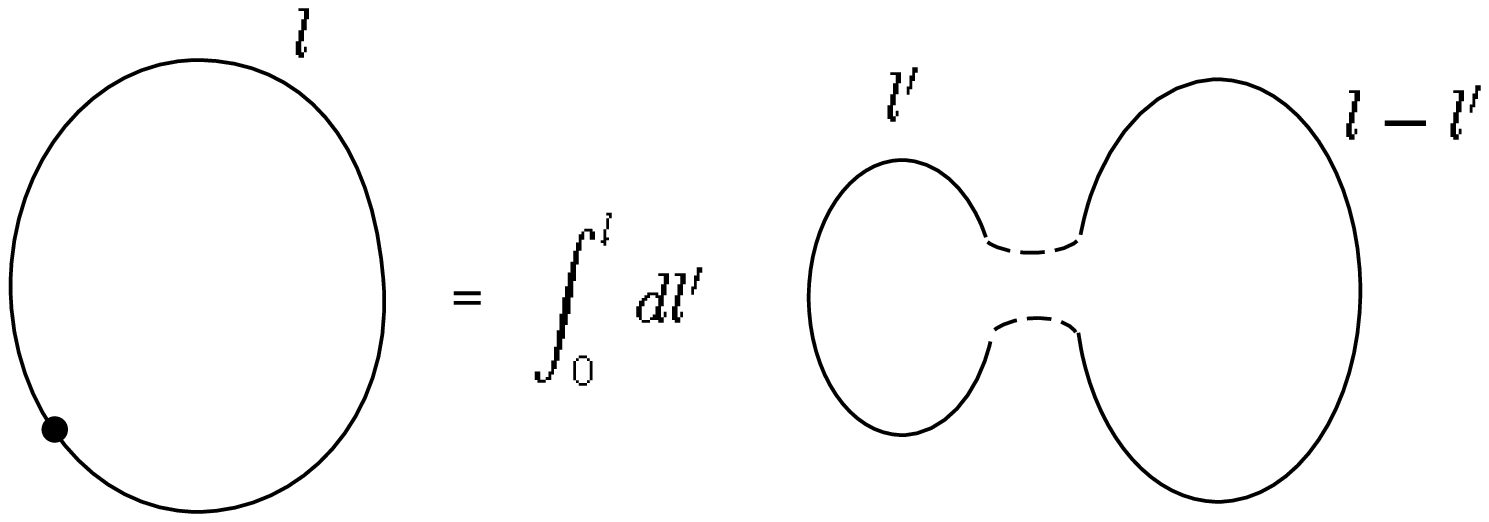}}
\caption{{\bf Fig. 7}: Geometrical representation of the loop
equation.}
\endinsert

\no Adding up we obtain, from {\axxxi}, after simple manipulations:

$$
\sum_{k\geq 1} k g_k {{\partial^{k-1}}\over{\partial l^{k-1}}}
u_0(l)=\int_0^l dl' u_0(l-l')u_0(l').
\eqn\axxxii
$$

\no This equation was first derived by Kazakov [\kazakov] using more
phenomenological arguments and is similar to the Migdal-Makeenko
equations of multicolor QCD. The loop equation has a very nice
geometrical interpretation in terms of joining and splitting of
loops, as shown in Fig. 7. It can be solved if we introduce the
Laplace transformed loop operator:

$$
u_0(p)=\int_{0}^{\infty}d l e^{-pl}u_0(l)=\sum_{k=0}^{\infty}
{u_0^{(k)}\over p^{k+1}}={\L_{\cal B} \o p}+\dots.
\eqn\axxxiii
$$

\no In terms of this operator the (non-planar) loop equation, that is
equivalent to the Virasoro constraints, is:

$$
u_0(p)^2-V'(p)u_0(p)+{1\over N^2}\chi(p,p)=Q(p).
\eqn\axxxiv
$$

\no Here $Q(p)$ is a polynomial in $p$ whose explicit form is not
necessary and we have introduced the two-loop operator:

$$
\chi(p,q)=\sum_{k,l \geq 0}{\chi_{k,l}\over p^{k+1} q^{k+1}}
\qquad{\rm with} \qquad  \chi_{k,l}=\L_{\cal B} ^4{\partial^2 F \over
\partial g_k
\partial g_l},
\eqn\axxxv
$$

\no and $V'(p)=\sum_{k\geq 1} kg_kp^{k-1}$. This equation has been
solved by Kazakov in the naive $N\rightarrow \infty$ limit. For
higher genus and even potentials the solution was found by Ambjorn
and Makeenko
\REF\amma{J.~Ambjorn and Yu.~M.~Makeenko, ``Properties of Loop
Equations for the Hermitian Matrix Model and for Two-Dimensional
Quantum Gravity'', {\it Mod.~Phys.~Lett.} {\bf A5} (1990) 1753.}
[\amma],
using the genus-expansion proposed by David [\loopdavid]. We will now
see how we can solve them on the sphere for an arbitrary potential
[\five] and how we  can identify the multicritical points, the
expressions for the scaling operators and the critical exponents.

\section{\sectwothree}

\no The solution of equation {\axxxiv} in the limit
$N\rightarrow\infty$ is

$$
u_0(p)=\h \left(V'(p)\pm \sqrt{ V'(p)^2-4Q(p)}\right).
\eqn\aaxxxvi
$$

\no The sign in {\aaxxxvi} can be determined from the requirement
that $u_0(p)$ is perturbative in the couplings and that its large $p$
expansion contains only negative powers in $p$. If we choose one
single cut dominating the scaling region, we can make the following
ansatz

$$
u_0(p)=V'(p)-M(p)\sqrt{(p-x)(p-y)}.
\eqn\axxxvi
$$

\no In the above expression $(x,y)$ are to auxiliary variables that
will be expressed through the parameter $\L_{\cal B}$. $M(p)$ is a
polynomial in $p$ that will be determined in such a way that the loop
equation is satisfied and that the loop operator has an expansion of
order $1/p$, as follows from {\axxxiii}.

The steps that we have to follow to determine the solution are:
\item{\triangleright}determine $x$ and $y$ in terms of $\L_{\cal B}$,
\item{\triangleright}fix the expression for $M(p)$,
\item{\triangleright}take the scaling limit,
\item{\triangleright}compute the expressions for the scaling
operators and the critical exponents, such as the string
susceptibility.

First we would like to derive an expression for the `cosmological
constant'\foot{Actually, this is not the cosmological constant but it
is related with it through {\aaaxi}; we will be sloppy with this
terminology.} $\L_{\cal B}$, the so-called string equation. The loop
operator $u_0(p)$ has the following expansion

$$
u_0(p)=C(x,y)+{\L_{\cal B} (x,y)\over p}+O\left({1\over p^2}\right).
\eqn\axxxvii
$$

\no From {\axxxiii} we obtain $C(x,y)=0$. Dividing this expression by
$\sqrt{(p-x)(p-y)}$ gives

$$\eqalign{
{u_0\over\sqrt\DEp}&=\INTp -M(p)\cr
&\cr
&={C(x,y)\over p}+{2\L_{\cal B} (x,y)+(x+y)C(x,y)\o 2
p^2}+O\left({1\over p^3}\right),\cr}
\eqn\axxxx
$$

\no which implies

$$
C(x,y)=-\pint \INTp=0,
\eqn\axxxxi
$$

$$
\L_{\cal B} (x,y)=-\pint\INTp \left(p-{x+y\over 2}\right).
\eqn\axxxxii
$$

\no Equations {\axxxxi} and {\axxxxii} can be used, in principle, to
rewrite $x$ and $y$ in terms of the single physical parameter
$\L_{\cal B}$.
These equations have appeared previously in the matrix model
literature (see for instance
\REF\kostov{I.~Kostov, ``Strings with Discrete Target Space'', {\it
Nucl.~Phys.} {\bf B376} (1992) 539.}
[\kostov] and references therein) in a
slightly different form.

Note that the piece proportional to $(x+y)/2$ in {\axxxxii} vanishes
when {\axxxxi} is enforced. Then one recovers the expression for
$\L_{\cal B}$, which is usually found in the literature. However, in
order
to compute partial derivatives it is essential to use the complete
form {\axxxxii}. The above equations are equivalent to the planar
discrete Virasoro constraints $L_{-1}$ and $L_0$ respectively.

 We now proceed to the computation of $M(p)$; it can be expanded in
powers of $p$

$$
M(p)=\sum _{n\geq 0}M_np^n.
\eqn\axxxxiv
$$

\no From {\axxxx} we obtain explicit expressions for the expansion
coefficients:

$$
M_n=-\qint {q^{-n-1}V'(q)\over\sqrt{(q-x)(q-y)}},
\eqn\axxxxv
$$

\no so that we obtain the integral representation

$$
M(p)=-\qint {V'(q)\over (q-p)\sqrt{(q-x)(q-y)}}.
\eqn\axxxxvi
$$

\no Therefore, we have expressed $u_0$ as a function of $\L_{\cal B}$
only, since $M(p)$ is a function of $(x,y)$ through {\axxxxvi} and
$(x,y)$ are functions of $\L_{\cal B}$, since {\axxxxi} and
{\axxxxii} hold.

In order to derive the basic observables of our theory, the scaling
operators, we would like to obtain an expression for the derivative
of $u_0(p)$ with respect to the cosmological constant $\L_{\cal B}$.
Under a variation $(dx, dy)$
compatible with the constraint {\axxxxi} $du_0$ is given by

$$
du_0(p)={1\over\sqrt{\D}}\left( -\h M(p)d\D -\D
dM(p)\right)={d\L_{\cal B}\over p}+O\left({1\over p^2}\right).
\eqn\axxxxvii
$$

\no We have introduced the notation $\D=(p-x)(p-y)$. Since the
$p$-dependence of the bracket cancels out, we obtain the expression

$$
d\L_{\cal B} =-\h M(p)d\D -\D dM(p).
\eqn\axxxxviii
$$

\no Thus, the result for $u_0$ is:

$$
{\p u_0(p)\over \p \L_{\cal B}}={1\over \sqrt{(p-x)(p-y)}}.
\eqn\axxxxix
$$

We are now in the position to analyse the theory in the scaling
limit.

 First we would like to see how this limit could be taken if the
potential is arbitrary. For this purpose we use an alternative
definition of the macroscopic loop operator {\axxix}:

$$
u_0 (l) =\lim_{n\rightarrow\infty}{N\over
\L_{\cal B} }u_0^{(n)},
\eqn\alii
$$

\no  where $l=na^{2/m}$ is the renormalized length. Here $m$ is a
positive integer related to the order of criticality as we will see
and $a$ is the cutoff. As $n$ goes to infinity the lattice spacing
$a$ will go to zero, but $l$ has to be kept fixed. This operator has
the same continuum limit as the lattice loop operator {\axxviii} up
to a multiplicative renormalization. We will therefore use the same
notation for both of them. The continuum limit has to be defined in
such a way that {\alii} makes sense, since this operator is the basic
geometric observable in 2D-quantum gravity. This will be our guiding
principle. Equations {\axxxxix} and {\alii} imply

$$
\p_{\L_{\cal B}} u_0 (l)=\lim_{n\rightarrow \infty}{1\over 2\pi i}
{N\o \L_{\cal B}}\oint_{C}{dp p^n\o \sqrt{\DEp}},
\eqn\aalii
$$

\no where the contour $C$ encloses the cut $(x,y)$. We may assume
$|y|<|x|\leq1$ without loss of generality. Taking $C$ as the unit
circle around the origin
with $p=e^{i\theta}$ we have the identity

$$
{1\over 2\pi i}\oint_C{dp\; p^n\over\sqrt\DEp} =
{1\over 2\pi}
\oint_C{d\th e^{i \th}
\exp\left({il
a^{-2/m}\th}\right)\over\sqrt{(e^{i\th}-x)(e^{i\th}-y)}}.
\eqn\aliv
$$

\no Changing to the new variable $z=a^{-2/m}\theta$, eq. {\aliv}
becomes

$$
\pt\uo (l)=-{1\o 2\pi\k}
\lim_{a\rightarrow 0}\;
a^{1/ m}\sqrt{(1-y_c)}\oint_C{dze^{ilz}
\over\sqrt{(\exp\left(ia^{2/m}z\right)-x)
(\exp\left(ia^{2/m}z\right)-y)}}.
\eqn\alvi
$$

\no Here we have taken $x_c=1$ without loss of generality. This
expression will vanish unless the integral itself is of order
$a^{-{1/ m}}$, i.e. if $x$ approaches the critical value $x_c=1$ as
$a^{2/m}$

$$
x=1-a^{2/m}\up.
\eqn\alvii
$$

\no The variable $\up$ is usually called the `specific heat'. We will
later see how it is related to the free energy of the model.
Since $x$ and $y$ are not independent variables,
$y$ will approach its critical value $y_c$ ($|y_c|<1$) at the same
time

$$
y=y_c+a^{2/n}\um.
\eqn\alviii
$$

\no Here $n$ is another positive integer that may be different from
$m$. The concrete value of $y_c$ is fixed by $x_c$ through eq.
{\axxxxi}. The previous two equations are important, since they
indicate to us how we have to take the continuum limit.

The integral in {\alvi} is dominated by the region $p\sim x_c=1$,
where
the contour can be deformed into a straight line\foot{Here we have
absorbed a factor $\sqrt{1-y_c}$ into the definition of $\kappa$.}:

$$
\kappa\pt\uo (l)=-{1\over 2\pi i}
\int_{-i\infty}^{+i\infty}dz{e^{lz}\over\sz}.
\eqn\alix
$$

\no Here we recognize the definition of the inverse Laplace transform

$$
\uo (z)\equiv\kappa{\cal L}\lbrack\uo (l)\rbrack,
\eqn\alx
$$
\no where
$$
\pt\uo (z)=-{1\over\sz}.
\eqn\alxi
$$

\no The previous result can of course also be obtained from
{\axxxxix} if we take into account the scaling law

$$
p=1+a^{2/m}z
\eqn\aalxi
$$

\no together with {\alvii} and {\alviii}.
The loop operator in the scaling limit $u_0(z)$ is connected with the
discrete one through the relation:

$$
u_0(z)=a^{-2+{1/ m}}u_0(p).
\eqn\aalxi
$$

\no There are therefore two different
ways of viewing $\uo (z)$: as the Laplace transformed of
the macroscopic loop $\uo (l)$ or as the continuum limit of the loop
operator $\uo (p)$.

{}From {\alix}, we get the expression for the macroscopic loop after
integration

$$
\pt\uo (l)=-{1\over\kappa}{e^{-l\up}\over\sqrt{\pi l}}.
\eqn\alxii
$$

\no Note that $\pt\uo (l)$ is independent of the variable $u_-$, so
that $u_+$ is the `physical' scaling variable .
This result holds as long as $|y_c|<|x_c|$ independently of the
values of $m$ and $n$. It is easy to see that for $|y_c|=|x_c|$, the
dominant end-point will be the one with the highest order of
criticality.

We would like to obtain the expressions for the scaling operators and
the form of the string equation. In analogy to the discrete partition
function {\ai} we can describe the non-perturbative partition
function of a general massive model in this limit in a formal way as
follows:

$$
Z({\widetilde t_0^+}, {\widetilde t_1^+}, \dots)=\biggl\langle
\exp\sum_{i=0}^\infty {\widetilde t_i^+}{\widetilde
\s_i^+}\biggr\rangle.
\eqn\zxxx
$$

\no Here ${\widetilde \s_i^+}$ are the scaling operators in the
continuum theory, while $ {\widetilde t_i^+}$ are the renormalized
coupling constants. Therefore we can derive the correlation functions
of the scaling operators ${\widetilde \s_i^+}$ in the background
${\widetilde t_n^+}={b_n}$, for $n\geq 0$, from $\log Z$ as follows

$$
\langle {\widetilde \s_{n_1}^+} \dots {\widetilde \s_{n_s}^+}\rangle
={\p ^s\o \p {\widetilde t_{n_1}^+}\dots \p {\widetilde
t_{n_s}^+}}\log Z({\widetilde t_0^+}, {\widetilde t_1^+},
\dots)|_{{\widetilde t_n^+}={b_n}}.
\eqn\yi
$$

\no The renormalized cosmological constant $t$ has to be identified
with $\widetilde t_0^{+}$.

We first compute the one-point functions. The macroscopic loop
operator

$$
{\widetilde u_0}(l)=\sqrt{\pi\o l}u_0(l)
\eqn\aalxii
$$

\no has an expansion in scaling operators $\langle {\widetilde
\s_k}^+ \rangle $ [\doubles,\kazakov,\dbs] dual to the renormalized
couplings ${\widetilde t_k^+}$ that can be used to compute explicit
expressions for these operators:

$$
{\widetilde u_0}(l)=\kappa \sum_{k=0}^{\infty}(-)^{k+1}{l^k\over
k!}\langle {\widetilde \sigma}_k^+\rangle  -{1\over \kappa l}.
\eqn\aaalxii
$$

\no This formula is the analog of the formula {\axxix} in the
continuum. For the scaling operators we obtain the expressions

$$
{\p \o \p t}\langle
{\widetilde \sigma^+}_k\rangle = {\p^2 F \over \p t\p {\widetilde
t^+}_k}=-{1\over
\kappa^2}{u_+^{k+1}\over (k+1)}.
\eqn\alxxvi
$$

\no From the above formula it becomes clear that the specific heat is
the two-point function of the puncture operator ${\widetilde\s}^+_0$:

$$
-{u_{+}\o {\kappa}^2}=\langle {\widetilde \s^+_0} {\widetilde  \s^+_0
}\rangle ={\p^2 F\o \p t^2}.
\eqn\yyii
$$

Next we would like to obtain the form of the string equation for
generic potentials. To do so, we have to obtain the corresponding
equations to {\axxxxi} and {\axxxxii} in the scaling limit. Comparing
the first derivatives of $\L_{\cal B}(x,y)$ and $C(x,y)$ we find that
they are not independent.
Instead,

$$\eqalign{
&\px\L_{\cal B} =\h (x-y)\px C\cr
&\cr
&\py\L_{\cal B} =\h (y-x)\py C.\cr}
\eqn\alxiii
$$

\no Similarly, we have the following identities

$$\eqalign{
&(x-y)\pxy\L_{\cal B} =-\h (\px\L_{\cal B} -\py\L_{\cal B} )\cr
&\cr
&(x-y)\pxy C =\h (\px C -\py C ).\cr}
\eqn\alxiv
$$

\no Scaling the first equation of {\alxiv} according to {\alvii} and
{\alviii} gives

$$
\ppm t\sim (a^{2/n}\pp t+a^{2/m}\pms t),
\eqn\alxv
$$

\no with an identical expression for $C$. For $a\to 0$ the r.h.s.
vanishes and we find

$$
\ppm t=0\qquad {\rm and} \qquad\ppm c=0.
\eqn\alxvi
$$

\no Here we have defined $\p_\pm\equiv \p/ \p u_\pm$ and
$c=-a^{-2}(1-y_c)C/2$. Moreover {\alxiii} implies

$$
\pp c=\pp t \qquad{\rm and } \qquad \pms c=-\pms t.
\eqn\alxvii
$$

\no One solution of {\alxvii} is given, if the renormalized
cosmological constant $t$ and $c$ are expanded in terms of scaling
operators

$$\eqalign{
&t=-\h\sum_{p> 0}\left(\tp _p\up^p +\tm_p\um^p\right), \cr
&\cr
&c=-\h\sum_{p> 0}\left(\tp _p\up^p -\tm_p\um^p\right)=0. \cr}
\eqn\alxviii
$$

\no Equations {\alxviii} are the continuum version of {\axxxxi} and
{\axxxxii} or equivalently of {\axxiv} and {\aaxxiv}. A (more
rigorous) prove of these equations can be obtained using the
continuum loop equations. This will be done in section 5.2. for the
supersymmetric case. The purely bosonic model can be obtained setting
the fermionic coupling constants to zero.

Adding and subtracting the equations in {\alxviii} we obtain the
string equation in terms of the physical variable $u_+$, which is the
one that appears in the loop operator {\alxii}

$$
t=-\sum_{p> 0}\tp _p\up^p .
\eqn\aalxviii
$$

\no Therefore, we have a parametrization of the string equation and
the loop operator in terms of the $`+'$ variables. The reason for
this is  that for $|y|<|x|$ the continuum limit is controlled by $m$.
The converse is of course true for $|y|>|x|$ and we would have a
dependence of the loop operator and the string equation on the $`-'$
variables. From now on we will omit the subindices in our formulas.

\no At the $m^{\rm th}$ multicritical point the planar string
equation has the following form

$$
t=u^m-\sum_{n>m} {\widetilde t}_nu^n.
\eqn\alxxiixi
$$

\no The cosmological constant therefore satisfies

$$
{ d t \o du}\biggr\vert_{u=0} =\dots ={d^{m-1} t \o d
u^{m-1}}\biggr\vert_{u=0}=0.
\eqn\alxxii
$$

Using the above string equation it is easy to see, that the string
susceptibility has the value $\Gamma_{str}(0)=-1/m$.
The scaling dimension of an operator $\langle {\widetilde
\s_k}\rangle  $ is given by $d_k=k/m$. These values can be compared
with those obtained from the Liouville approach in chapter~2. They
correspond to the $(2m-1,2)$ minimal CFTs coupled to 2D gravity.

\section{\sectwofourone}

\no Two-dimensional quantum gravity can be formulated
non-perturbatively if we introduce the double scaling limit
[\doubles], as mentioned in section 2.1. It turns out that in this
limit a close connection to integrable hierarchies of the KdV type
appears since, for example, correlation functions of microscopic
operators can be expressed in terms of KdV flows.

In this section we are going to explain some basic notions on
integrable hierarchies and the relation to two-dimensional quantum
gravity will be the subject of the next section. A more complete
description of the field is given in some good reviews on integrable
hierarchies, as for example
\REF\segal{
G. Segal and G. Wilson, ``Loop Groups and Equations of KdV Type'',
{\it Publ. Math. I.H.E.S.} {\bf 61} (1985) 1.}
\REF\das{A.~Das, ``Integrable Models'', World Scientific (1989),
Lecture Notes in Physics, 30.}
\REF\jimbo{E.~Date, M.~Jimbo, M.~Kashiwara and T.~Miwa,
``Transformation Groups for Soliton Equations'', in {\it Non-linear
Integrable Systems-Classical Theory and Quantum Theory}, M.~Jimbo and
T.~Miwa eds. (World Scientific, 1983).}
[\segal,\das,\jimbo,\dijkgraaf].

The following definition of an integrable system dates back to
Liouville: A Hamiltonian system with a $2N$-dimensional phase space
is integrable if and only if there exist exactly $N$ functionally
independent conserved quantities that are in involution (that is, the
Poisson brackets of these conserved quantities with one another
vanish). If the system under consideration is infinite-dimensional
this would imply the existence of an infinite number of conserved
quantities.

The KdV equation is one example of such a system. It is a non-linear
equation in one space and one time coordinate $(x,t_1)$:

$$
{\p u \o \p t_1 }=6u{\p u \o \p x}+\kappa^2 {\p^3 u \o \p x^3}.
\eqn\zi
$$

\no Here $u=u(x,t_1)$ is the dynamical variable and $\kappa$ is a
free parameter that we have introduced for later convenience. This
equation can be written as an Hamiltonian equation:

$$
{\p u \o \p t_ 1}=\{u,{\cal H}_{1}\}_2=\{u,{\cal H}_{2}\}_1.
\eqn\zii
$$

\no Here ${\cal H}_{1}$ and ${\cal H}_{2}$ are the Hamiltonians of
the system. It is a peculiarity of the KdV equation that we can find
two Hamiltonians and two Poisson brackets, so that {\zii} is
equivalent to {\zi}. The two possible definitions of the above
Poisson brackets are:

$$
\eqalign{
\{u(x),u(y)\}_1&=\p_x\delta(x-y)\cr
&\cr
\{u(x),u(y)\}_2&=\left(\kappa^2 \p_x^3+2 u \p_x+2 \p_x u
\right)\delta(x-y).\cr}
\eqn\ziii
$$

\no They satisfy the Jacobi identities and are antisymmetric. The
second structure is identical to the classical version of the
Virasoro algebra.
The corresponding two Hamiltonians are

$$
\eqalign{
&{\cal H}_1=\int_{-\infty}^\infty dx \left( {u^2\o 2}\right),\cr
&\cr
& {\cal H}_2=\int_{-\infty}^\infty dx\left( u^3-{\kappa^2 \o
2}\left({\p u\o \p x}\right)^2 \right),\cr}
\eqn\zv
$$

\no respectively.

The solution of the KdV equation is unique for a given initial
condition. It can be shown that waves without dispersion, i.e. which
maintain their shape as they move, are a class of solutions. They are
called solitons, and imply the existence of an infinite number of
conserved quantities ${\cal H}_n$. These conserved quantities
satisfy:

$$
{d{\cal H}_n\o dt_ 1 }=\{{\cal H}_n,{\cal H}_2\}_1=\{{\cal H}_n,{\cal
H}_1\}_2=  0.
\eqn\zvi
$$

\no The simplest example is of such a constant of motion is

$$
{\cal H}_0={ 1\o 2} \int_{-\infty}^\infty dx \, u .
\eqn\zvii
$$

\no Using the KdV equation {\zi}, it is easy to prove that other
constants of motion are for example ${\cal H}_1$ and ${\cal H}_2$.
One can explicitly check that these three quantities satisfy the
recursion relation

$$
\left( \kappa^2 \p_x ^3+2u \p_x+2\p_x u \right){\delta {\cal
H}_{n-1}\o \delta u}=\p_x {\d{\cal H}_n\o \d u}\qquad {\rm for}\qquad
n=0,1,2,
\eqn\zviii
$$

\no where ${\cal H}_{-1}=0$ and the functional derivative is defined
as

$$
{\d F[u(x)]\o \d u(y)}=\lim_{\ep \rightarrow 0}{1\o \ep}\left(
F[u(x)]+\ep \d(x-y)]-F[u(x)]\right).
\eqn\zix
$$

The first three conserved quantities can be obtained by taking into
account simple symmetry principles. We can identify ${\cal H}_1$ with
the momentum since it generates space translations with respect to
the first Poisson bracket

$$
\{u,{\cal H}_1\}_1={\p u\o \p x}.
\eqn\zzzzz
$$

\no Similarly ${\cal H}_2$ is the generator of time translations and
can therefore  be identified with the energy

$$
\{u,{\cal H}_2\}_1={\p u\o \p t_1 }.
\eqn\zzzzzz
$$

It can be proven inductively that we can choose all conserved
quantities ${\cal H}_n$ {\zvi}, so that {\zviii} holds for $n\in
\IN$. Furthermore this recursion relation guarantees that these
quantities are in involution:

$$
\{{\cal H}_n,{\cal H}_m\}_1=\{{\cal H}_n,{\cal H}_m\}_2=0.
\eqn\zxiii
$$

\no The relation {\zviii} is usually written as a recursion relation
for the Gel'fand-Dikii polynomials $R_k[u]$ that are closely related
to the constants of motion:

$$
\left(\kappa^2 \p_x^3+2u \p_x+2\p_x u)\right)R_{k-1}[u]=\p_x
R_k[u]\qquad {\rm with} \qquad
R_k[u]={\d {\cal H}_k\o \d u},
\eqn\zx
$$

\no with $R_0=1/2$, $R_1=u$, $R_2=3u^2+\kappa^2 \p_x^2 u$, \dots.
Starting with the expression for ${\cal H}_0 $ in eq. {\zvii} we can
use {\zx} to determine all the other charges.

Each of the conserved quantities can be considered as an Hamiltonian
that generates an evolution equation, that is a generalization of
{\zi} :

$$
{\p u\o \p t_k}=\{u,{\cal H}_{k}\}_2=\{u,{\cal H}_{k+1}\}_1,
\eqn\zxi
$$
\no or equivalently
$$
{\p u \o \p t_k}=\p_ x R_{k+1}[u].
\eqn\zxii
$$

\no The function $u$ can be thought of as depending on an infinite
number of time variables $u=u(x,t_1,\dots)$, where we define $x=t_0$.
These are known as higher flows of the KdV hierarchy.

Before we establish the connection between this formalism and
two-dimensional gravity we would like to introduce a suitable
formulation that can be used for all the $(p,q)$-minimal models
coupled to two-dimensional gravity: the Lax-pair formulation. The
formalism of pseudo-differential operators that we now explain  will
be relevant to describe the supersymmetric extensions of the KP
hierarchy in section 4.6.

The $p^{\rm th}$-reduction of the generalized KdV hierarchy is
defined in terms of a differential operator:

$$
L=\p_x ^p+\sum_{i=0}^{p-2}u_i\p_x ^i.
\eqn\zxiv
$$

\no Here $u_i$ are a priori arbitrary functions depending on the time
variables $(t_0, t_1,t_2, \dots)$\foot{It is always possible to set
the order $(p-1)$ to zero by conjugation.}. We can consider the flows

$$
{\p L \o \p t_n}=[H_n, L],
\eqn\zxviii
$$

\no generated by the infinite set of commuting Hamiltonians $H_n$.
The Hamiltonians are represented as fractional powers of the positive
part of $L^n$

$$
H_n=(L^{n/p})_+.
\eqn\zxix
$$

A fractional power of $L$ is defined as a Laurent series in the
differential operator $\p_x $. These are in general $n$-th
pseudo-differential operators, i.e. operators of the form

$$
{\cal O}=\sum_{i=-\infty}^n {\cal O}_i\p_x^i,
\eqn\zxv
$$

\no where the operator $\p_x^{-1}$ acts on a function $f$ as follows:

$$
\p_x^{-1}f=\sum_{i=0}^\infty (-)^{i}{\p^i  f\o \p x^i}\p_x^{-1-i}.
\eqn\zxxii
$$

\no The part of the sum that contains only positive powers in $\p_x $
is denoted by ${\cal O}_+$:

$$
{\cal O}_+=\sum_{i=0}^n{\cal O}_i \p_x ^i
\eqn\zxvi
$$

\no and similarly ${\cal O}_-={\cal O}-{\cal O}_+$. The residue of a
pseudo-differential operator is defined as the coefficient of $\p_x
^{-1}$ in the Laurent expansion of ${\cal O}$

$$
{\rm res}\;{\cal O}={\cal O}_{-1}.
\eqn\zxvii
$$

\no If $n$ is a multiple of $p$, the flow is trivial

$$
{\p L\o \p t_n}=0\qquad {\rm for }\qquad n=pq,
\eqn\zxx
$$

\no since $H_n=L^q$ commutes with $L$.

The simplest reduction is the 2-reduction where

$$
L=\p_x^2+u,
\eqn\zzxx
$$

\no whose first non-trivial flow is the KdV equation {\zi}. This
operator plays an important role in the non-perturbative description
of the one-matrix model as we will see later.
The flows computed at the beginning of this section can now be
obtained from {\zxviii} comparing the powers of $\p_x$ of the right-
and left hand side of this equation.

The generalized KdV hierarchies are special cases of the KP
hierarchy, which is defined in terms of an arbitrary
pseudo-differential operator

$$
Q=\p_x +\sum_{i=1}^\infty Q_i\p_x ^{-i}.
\eqn\zxxii
$$

\no The KP hierarchy is defined as the infinite set of commuting
flows on the phase space of the operator $Q$

$$
{\p Q \o \p t_n}=[Q_+^n,Q]=-[Q_-^n,Q].
\eqn\zxxiii
$$

\no The $p^{\rm th}$ KdV hierarchy satisfies the condition that
$Q^p=L$ is a pure differential operator, i.e. $Q_-^p=0$.

The absence of a term $Q_0\p_x ^0$ in $Q$ is sufficient for the
existence of a pseudo-differential operator

$$
S=1+\sum_{i=1}^\infty S_i \p_x ^{-i}\qquad {\rm with} \qquad Q=S\p_x
S^{-1}.
\eqn\zxxiv
$$

\no In terms of this operator the flow equations become

$$
{\p S\o \p t_n}=-[S\p_x ^nS^{-1}]_-S.
\eqn\zxxvi
$$

To the $p$-reduction of the generalized KdV hierarchy {\zxviii}
we can associate the so-called $\tau$-function that is related to the
differential operator $L$ as follows:

$$
{\rm res }\; L^{n/p}={\p^2 \o \p x\p t_n}\log\t.
\eqn\zxxviii
$$

\no The $\tau$-function completely determines the operator $L$ since
for $i=1,\dots,p-1$ we have

$$
{\rm res}\;L^{i/p}={i\o p}u_{p-i-1}.
\eqn\zxxix
$$

If we consider the KdV operator, the variable $u$, as a function of
which the flows are formulated, is completely determined by the
$\tau$-function:
$$
u=2\p_x ^2\log \t.
\eqn\zzzzzzz
$$

\no The $\t$-function of the KdV hierarchy satisfies the recurrence
relation

$$
\p_x ^2{\p \o \p t_{n+1}}\log \t =\left( {\k^2} \p_x^3+2 u\p_x +2\p_x
u\right) \p_x {\p \o \p t_n} \log \t.
\eqn\yiv
$$

\section{\sectwofourtwo}

\no The $\tau$-function is useful to describe the partition function
in the matrix-model approach to 2D quantum gravity as we will now
see. The important result that provides the link between the
non-perturbative approach to 2D-quantum gravity [\doubles,\dbs] and
integrable systems is that the partition function of the one-matrix
model is the square root of the tau-function of the KdV hierarchy
[\douglas]:

$$
Z(t_0, t_1, \dots)=\t^2(t_0, t_1, \dots).
\eqn\yii
$$

\no Since $u$ is related to the $\t$-function as stated in {\zzzzzzz}
it can be identified with the `specific heat' or equivalently the
two-point function of the puncture operator {\yyii}. Recall that this
means that the specific heat satisfies the KdV flow equations

$$
{\p u\o \p t_n}=D R_{n+1}[u].
\eqn\yiii
$$

\no Here $\k$ is the renormalized string coupling constant and $D=\p
/ \p t_0$.

The insertion of $\s_n$ corresponding to the differentiation with
respect to the coupling $t_n$ is then identified with the $n^{\rm
th}$ KdV flow of $u$:

$$
\langle \s_n \s_0 \s_0 \rangle ={\p u \o \p t_n}=DR_{n+1}[u].
\eqn\yv
$$

\no Therefore, the recursion relations of the $R_k[u]$'s imply
recursion relations for scaling operators. The non-perturbative
string equation can be written in terms of the Gel'fand-Dikii
polynomials

$$
t=-\sum_{k\geq 1}(2k+1)t_k R_k[u].
\eqn\yyv
$$

\no This equation is a generalization of {\aalxviii} to higher genus.
At the $m^{\rm th}$-multicritical point all $t_k$ vanish except
$t_m$. This coupling constant can be normalized in such a way that
for the first values in $m$ the string equation takes the form

$$
\eqalign{
m=1\qquad \qquad & t= u \cr
m=2\qquad \qquad & t=u^2+ {\k^2  \o 3} \p_t^2 u.  \cr }
\eqn\yyyvi
$$

\no The second equation is known as the Painlev\'e-I equation; it is
the string equation of pure gravity. For $\kappa=0$ (planar limit) we
know that the specific heat behaves in the pure gravity case as
$u(t)=-\sqrt{t}$; this is thus the initial condition for the genus
expansion of the solution. The first terms of this expansion are

$$
u(t)=-\sqrt{t}+{\k\o 24} t^{-2}+{49 \o 1152} \k^2 t^{-9/2}+\dots.
\eqn\yyyvii
$$

\no It turns out that the string equation can be written as the
following constraint equation on the square root of the partition
function, i.e. on the $\tau$-function [\dvv,\fkn]

$$
L_{-1}\t=\left(\sum_{m=1}^\infty \left(m+{1\o 2}\right) t_m{\p \o \p
t_{m-1}}+{t_0^2\o 8}
\right) \tau=0.
\eqn\yvi
$$

\no Using the relation between the partition function and the
$\t$-function, one can show that the partition function satisfies the
constraints [\dvv,\fkn]

$$
L_n\tau=0 \qquad {\rm for} \qquad n\geq -1,
\eqn\yvii
$$
\no with the $L_n$ operators defined as
$$
L_{-1}=\sum_{m=1}^\infty \left(m+{1\o 2}\right) t_m{\p \o \p
t_{m-1}}+{t_0^2\o 8\k^2 }
\eqn\yviii
$$
$$
L_0=\sum_{m=0}^\infty \left(m+{1\o 2}\right)t_m {\p \o \p t_m}+{1\o
16}
\eqn\yix
$$
$$
L_n=\sum_{m=0}^\infty \left(m+{1\o 2}\right)t_{m}{\p\o
t_{m+n}}+{\kappa^2\o 2}\sum_{m=1}^n {\p^2 \o \p t_{m-1}\p t_{n-m}}.
\eqn\yyix
$$

\no These operators satisfy the Virasoro algebra:

$$
[L_n,L_m]=(n-m)L_{n+m}\qquad {\rm for } \qquad n,m \geq -1,
\eqn\yx
$$

\no so that they are the Virasoro constraints in the double scaling
limit. They are the continuum version of the constraints {\axviii}.
Notice that for the discrete theory the constraints acted on the
partition function itself and not on the square root. The conditions
{\yvii} are due to the fact that the the recursion relation for the
$\t$-function {\yiv} implies:

$$
D^2\left( {L_{n+1} \t \o \t}\right) =
\left({\kappa^2}D^3+2uD+2Du\right)D\left({L_n \t\o \t}\right).
\eqn\yxi
$$

\no Together with the string equation {\yvi} these equations imply
the constraints {\yvii}.

The expressions {\yviii}, {\yix} and {\yyix} can be written as the
Virasoro generators of a $\IZ_2$-twisted free scalar field $x(z)$,
i.e. with $x(e^{2\pi i}z)=-x(z)$. The mode-expansion of $x(z)$ is
$$
\p x(z)=\sum_{n\in \IZ} \a_{n+{1\o 2}} z^{-n-{3\o 2}}.
\eqn\yxii
$$

\no Since the energy-momentum tensor is given by

$$
T(z)=:{1\o 2} \p x(z)^2:+{1\o 16 z^2}=\sum_{n\in \IZ} L_n z^{-n-2},
\eqn\yxiii
$$

\no we obtain the following correspondences between the modes of the
free boson and the coupling constants of the model:

$$
\a_{-n-{1\o 2}}=\left(n+{1\o 2 }\right){t_n\o \kappa} \qquad {\rm
and}\qquad
\a_{n+{1\o 2}}=\kappa {\p \o \p t_n}\qquad{\rm for}\qquad n\geq 0.
\eqn\yxv
$$

The Virasoro constraints in the double scaling limit can be shown to
be equivalent to the continuum non-perturbative loop equation of the
one-matrix model [\dvv,\fkn] in a way similar to what was already
done in the discrete theory in section 3.2. This equation has the
form

$$
{\widehat u_0}^2(z)+\k^2{\widehat \chi}(z,z)={\rm polynomial}\,(z),
\eqn\yxvi
$$

\no where we have introduced the two-loop operator
$$
{\widehat \chi}(z,z)={1\o 4z^2}+2\k^2 \sum_{n\geq 0} z^{-n-3}
\sum_{k=0}^n \langle \s_k \s_{n-k} \rangle
\eqn\yxvii
$$
\no and the loop operator takes the form
$$
{\widehat u_0}(z) =\sum_{n\geq 0} \left( n +\h\right) t_n z^{n-\h}+
\k^2\sum_{n\geq 0} z^{-n-{3\o 2}}\langle  \s_n \rangle,
\eqn\yxviii
$$
\no where ${\widehat u_0}(z)$ and ${\widehat \chi}(z,z)$ are the
continuum limit of the discrete loop and two-loop operators
respectively:

$$
{\widehat u_0}(p)=u_0(p)-{V'(p)\o 2}  =a^{2-{1\o m}} {\widehat
u_0}(z) \qquad {\rm and} \qquad
{1\o N^2} \chi(p,p) =\k^2 a^{4-{2\o m}} {\widehat \chi}(z,z).
\eqn\yxix
$$

\endpage

\chapter{\chathree}

\no In this chapter we begin to analyse how the previous formalism
could be generalized to the simplest $N=1$ SCFTs coupled to
2D-supergravity. For this generalization the Virasoro constraints
satisfied by the partition function will play a central role. We
start with the construction of the supersymmetric eigenvalue model
found by Alvarez-Gaum\'e et al. [\sloops].

\section{\secthreeone}

\no We will take as a guiding principle the one-matrix model
formulated in terms of eigenvalues and construct, in analogy, the
$N=1$ supersymmetric generalization.

\vskip 0.5cm

\no $4.1.1.$ T{\tenpoint HE} O{\tenpoint NE}-M{\tenpoint ATRIX}
M{\tenpoint ODEL} {\tenpoint IN} T{\tenpoint ERMS} {\tenpoint OF}
E{\tenpoint IGENVALUES} {\tenpoint AS} G{\tenpoint UIDING}
P{\tenpoint RINCIPLE}
\vskip 0.5cm

\no In analogy to the continuum constraints, we can write the
discrete Virasoro generators {\axviii} as the modes of the
energy-momentum tensor of a free boson $\p x$. We can introduce an
infinite set of creation and annihilation operators:

$$\eqalign{
\alpha_{-n}&=-{N\over \L_{\cal B} \sqrt{2}}n g_n\quad {\rm for }\quad
n>0 \qquad
\cr
& \cr
\alpha_n &=-{\L_{\cal B} \sqrt{2} \over N}{\partial \over
\partial g_n}\quad {\rm for} \quad n\geq 0 , \cr}
\eqn\wi
$$

\no that are the modes of $\partial x(p) =\sum_{n\in \IZ} \alpha_n
p^{-n-1}$. The modes of the energy-momentum tensor,

$$
T(p)={1\over 2}:\partial x(p) \partial x(p):=\sum_{n=0}^{\infty}
p^{-n-2} L_n ,
\eqn\wii
$$

\no are then represented by the operators {\axviii}. The
discrete-loop equation {\axxxiv} becomes:

$$
Z^{-1} T(p) Z={\rm polynomial}\,(p).
\eqn\wiii
$$

On the other hand, we can start with the free boson with
anti-periodic boundary conditions and show that we can determine the
partition function of the one matrix model. First, we have to
construct the potential of the model in terms of eigenvalues and then
determine the measure.  It is easy to see from {\wi} that precisely
the potential is related to the negative modes of the free boson
$$
\lim_{N\rightarrow \infty}{1\o N} Z^{-1} \p x^- Z \propto
V'(p)=\sum_{k>0} k g_k p^{k-1} .
\eqn\wwvi
$$

The loop operator can be determined from the positive part of the
mede expansion of $x$

$$
\lim_{N\rightarrow \infty}{1\o N} Z^{-1} \p x^+  Z \propto u_0(p) =
\sum_{n\geq 0}  { u_0^{(n)} \o p^{n+1}}.
\eqn\wwvi
$$

\no Next, we can write the partition function {\ai} in terms of
the eigenvalues of the $\Phi$-matrix $(\lambda_1,\lambda_2,\dots,
\lambda_N)$, but leaving the measure $\Delta^2(\lambda)$
undetermined:

$$
Z=\int \prod_{i} d\lambda_i \Delta^2(\lambda) \exp\left({-{N\over
\L_{\cal B}} \sum_{i} V(\lambda_i)}\right).
\eqn\wiv
$$

\no The constraints {\axviii} yield a differential equation
satisfied by $\Delta$
$$
\sum_{i}\lambda_i^{n+1}{\partial \Delta \over \partial
\lambda_i}= \Delta \sum_{i\neq j}
{\lambda_i^{n+1}\over{\lambda_i-\lambda_j}},
\eqn\wv
$$
\no whose solution up to a constant is the expected Vandermonde
determinant
$$
\Delta=\prod_{i<j} (\lambda_i-\lambda_j).
\eqn\wvi
$$
\no Therefore, if the potential of the model in terms of eigenvalues
is known, we can use the Virasoro constraints to determine the
measure.

\vskip 0.5cm

\no $4.1.2.$ T{\tenpoint HE} G{\tenpoint ENERALIZATION} {\tenpoint
TO} {\tenpoint THE} $N=1$ S{\tenpoint UPERSYMMETRIC} E{\tenpoint
IGENVALUE} M{\tenpoint ODEL}

\vskip 0.5cm

\no In the supersymmetric case we proceed by analogy with the
above arguments to obtain the expression for the partition function
in terms of eigenvalues and the corresponding loop equations. We
begin with the derivation of the discrete superloop equations. This
will be done by taking a ${\hat c}=1$ free massless superfield ${\bf
X} (p,\Pi)=x(p)+\Pi \psi(p)$ as a generalization of the free boson.
These fields have a mode expansion

$$
\partial x(p)=\sum_{n\in \IZ}\alpha_n p^{-n-1}\qquad{\rm and} \qquad
\psi(p)=\sum_{r\in \IZ+{1\o 2}} b_r p^{-r-{1\o 2}},
\eqn\wxii
$$

\no where the fermion is in the Neveu-Schwarz sector. The
energy-momentum tensor is

$$
{\bf T} (p,\Pi)\propto DX \partial_p
X=\psi\partial_px+\Pi:(\partial_p x
\partial_p x+\partial_p\psi \psi):,
\eqn\wxiii
$$

\no where $D=\p/ \p \Pi + \Pi \p_p $ is the superderivative and
$(p,\Pi)$ are the two coordinates of the superspace. The loop
operator $w(l,\theta)$ in the supersymmetric theory depends on two
variables $(l,\theta)$ that are the even and odd length respectively.
They characterize the boundary of a superdisk. We can define the
Laplace transformed loop operator:

$$
w(p,\Pi)\equiv v(p)+\Pi u(p)=\int_0^{\infty}dl \int d\theta
e^{-pl-\Pi \theta} w(l,\theta) .
\eqn\wvii
$$

\no The operator $w(p,\Pi)$ can be identified with the field ${\bf
X}^{+}(p,\Pi)$, so that it has an expansion in negative powers of
$p$:

$$
v(p)=\sum_{k\geq 0}{v^{(k)}\over p^{k+1}}\qquad{\rm and} \qquad
u(p)=\sum_{k\geq 0}{u^{(k)}\over p^{k+1}}.
\eqn\wix
$$

\no The analogy with the bosonic theory suggests that we can identify
the bosonic and the fermionic modes as bosonic and fermionic
couplings respectively

$$
\eqalign{
\alpha_p &=-{\Lambda_S \over N}{\partial \over \partial
g_p},\qquad\qquad \hskip 0.8cm\alpha_{-p}=-{N\over \Lambda_S}p
g_p,\qquad\qquad
\hskip 0.0cm {\rm for}\qquad  p=0,1,2,\dots \cr &\cr & \cr
b_{p+{1\o 2}}&=-{\Lambda_S \over N}{\partial \over\partial
\xi_{p+{1\o 2}}}
,\qquad\qquad b_{-p-{1\o 2}}=-{N\over \Lambda_S }\xi_{p+{1\o
2}}\qquad \qquad {\rm for } \qquad p=0,1,2,\dots\cr}
\eqn\wx
$$

\no If we write $F=N^{-2}\log\, Z$ then the moments $u^{(k)}$ and
$v^{(k)}$ can be identified with derivatives of the free energy

$$
u^{(0)}=\L_S \qquad \qquad u^{(n)}=-\L_S^2{\p F\over \p g_n}\qquad
\qquad
v^{(n)}=-\L_S^2{\p F \over \p \xi_{n+{1\o 2}}}.
\eqn\wxxiv
$$

\no As a basic postulate we take the (non-planar) discrete superloop
equations
to be described by the above energy-momentum tensor [\sloops]

$$
Z^{-1}{\bf T}(p,\Pi)Z={\rm polynomial}\,(p).
\eqn\wxiv
$$

\no In terms of $u(p)$ and $v(p)$ this is equivalent to a system of
two equations:

$$
(u(p)-V'(p))^2+(v(p)-\xi(p))'(v(p)-\xi(p))+{1\o N^2}\left(\chi^{{{\rm
BB}}}(p,p) +\chi^{{\rm FF}}(p,p)\right)=Q_0,
\eqn\wxv
$$

$$
u(p)v(p)-V'(p)v(p)-\xi(p)u(p)+{1\o N^2}\chi^{{\rm BF}}(p,p)=
Q_1.
\eqn\wxvi
$$

\no Here we have introduced the potentials:

$$
V(p)=\sum_{k\geq 0}g_k p^{k} \qquad{\rm and} \qquad
\xi(p)=\sum_{k\geq 0}
\xi_{k+1/2} p^k
\eqn\wxvii
$$

\no and the two-loop operators are defined as:

$$\eqalign{
& \chi^{{\rm BB}}(p,q)=\sum_{k,l\geq 0}{\L_S^4\over p^{k+1}
q^{l+1}}{\p^2 F\over
\p g_k\partial g_l},\cr
& \cr
&\chi^{{\rm BF}}(p,q)=\sum_{k,l\geq 0}{\L_S^4 \over p^{k+1}
q^{l+1}}{\p^2 F \over \p \xi_{k+{1\o 2}} \p g_l},\cr
&\cr
& \chi^{{\rm FF}}(p,p)=\sum_{n\geq 1}\sum_{r={1\o 2}}^{n-{1\o 2}}
{\L_S^4\over p^{n+2}}\left({n\over2}-r\right)
{\p^2 F\over\p \xi_r \p \xi_{n-r}}.\cr}
\eqn\wxviii
$$

\no The quantities $Q_0$ and $Q_1$ are polynomials in $p$. Although
their explicit form can be computed they will not be needed. In terms
of the original loop operator
$w(l,\theta)$ eqs. {\wxv} and {\wxvi} take a form
similar to {\axxxii}:

$$
{\cal P}{\cal K} w(l,\theta)+2{\cal K} {\cal
P}w(l,\theta)=(w\circ {\cal P}w)(l,\theta),
\eqn\wxix
$$

\no with

$$
{\cal K}\equiv \sum_{k\geq 1}\left(k g_k {\partial \over \partial
\theta}-\xi_{k-{1\o 2}}\right){\partial^{k-1}\over\partial
l^{k-1}}\qquad {\rm and} \qquad {\cal P}=\theta+l{\p \o \p \theta}.
\eqn\wxxi
$$

\no The convolution between two superfunctions $f_1(z,\theta)$ and
$f_2(z,\theta)$ is defined according to:

$$
(f\circ g)(z,\theta)\equiv\int d\theta'\int_0^z f(z',\theta')
g(z-z',\theta-\theta') dz'.
\eqn\wxxii
$$

Now we would like to see what the discrete super-Virasoro constraints
and the partition function on which they act look like. The analogue
of the potential of the one-matrix model is identified with the
negative mode expansion of the superfield $DV(p,\Pi)\sim D {\bf X}^-$

$$
V(p,\Pi)=\sum_{k\geq 0} (g_k p^k+\x_{k+\h} \Pi p^k).
\eqn\wwxxii
$$

\no The previous formula suggests the introduction of a
superpotential that in terms of the even eigenvalues $\l_i$ and the
odd eigenvalues $\theta_i$ takes the form

$$
V(\lambda,\theta)=\sum_{k\geq0}\sum_{i=1}^{N}(g_k\lambda_i^k
+\xi_{k+{1\o 2}}\theta_i\lambda_i^{k}).
\eqn\wxxiii
$$

\no Writing the partition function as:

$$
Z=\int \prod_{i=1}^N d\lambda_i d\theta_i \Delta(\lambda,\theta)
\exp\left(-{N\over \Lambda_S} V(\lambda,\theta)\right),
\eqn\wxxv
$$

\no we can determine the explicit form of the measure
$\Delta(\lambda,\theta)$ by imposing the super-Virasoro
constraints. The explicit representation of the
super-Virasoro operators using {\wx} as differential operators is

$$\eqalign{
G_{n-{1\o 2}}&=\sum_{k=0}^\infty \xi_{k+{1\o 2}}{\partial \over
\partial g_{k+n}}+\sum_{k=0}^\infty kg_k{\partial\over\partial
\xi_{k+n-{1\o 2}}}+{\L_S^2\over
N^2}\sum_{k=0}^{n-1}{\partial^2\over \partial
\xi_{k+{1\o 2}}\partial g_{n-1-k}} \qquad {\rm for}\qquad n\geq 0,\cr
&\cr&\cr
L_n& ={\L_S^2\over 2N^2}\sum_{k=0}^n{\partial^2\over\partial
g_k \partial g_{n-k}}+\sum_{k=1}^\infty
kg_k{\partial\over\partial g_{n+k}}
+\sum_{r={1\o 2}}^\infty\left({n\over2}+r\right)\xi_r{\partial\o
\partial\xi_{r+n}}\cr
&\cr
& +{\L_S^2 \o 2N^2}\sum_{r={1\o 2}}^{n-{1\o 2}}\left( {n\o
2}-r\right) {\p^2 \o \p \x_r\p \x_{n-r}}\hskip 6.6cm {\rm for} \qquad
n\geq -1.\cr}
\eqn\wxxvi
$$

\no These operators satisfy a closed subalgebra of the $N=1$
superconformal algebra in the Neveu-Schwarz sector. The partition
function of the supersymmetric model is annihilated by $L_n$ for
$n\geq-1$ and by $G_{n-{1\o 2}}$ for $n\geq 0$. Since the last
equation in {\wxxvii} holds, it suffices to implement the $G_{n-\h}$
constraint.
This leads to a set of equations of the form:

$$
\sum_{i}\lambda_i^n\left(-{\partial\over\partial\theta_i}
+\theta_i{\partial\over\partial\lambda_i}\right)\Delta=\Delta
\sum_{i\neq j} \theta_i{\lambda_i^n-\lambda_j^n\over
\lambda_i-\lambda_j},
\eqn\wxxviii
$$

\no whose unique solution, up to a multiplicative constant, is:

$$
\Delta(\lambda,\theta)=\prod_{i<j}(\lambda_i-\lambda_j
-\theta_i\theta_j).
\eqn\wxxix
$$

\no Hence the model we would like to solve is:

$$
Z=\int \prod_{i=1}^Nd\lambda_i d\theta_i \prod_{i<j}(\lambda_i
-\lambda_j-\theta_i\theta_j)\exp\left({-{N\over\Lambda_S}
\sum_{k=0}^\infty\sum_{i=0}^\infty\left(g_k\l_i^k+\x_{k+{1\o
2}}\th_i\l_i^k\right)}\right).
\eqn\wxxx
$$

\no Finally, we would like to remark that the loop operator can be
explicitly written in terms of the eigenvalues as:

$$
w(l,\theta)\equiv{\Lambda\over N}\sum_{i} e^{l\lambda_i+\theta
\theta_i}.
\eqn\wxxxi
$$

\no In the next section we begin the analysis of this model in the
large-$N$ limit.

\section{\secfourone}

\no We are going to calculate the expressions for the bosonic and
fermionic loop operators that follow from the previous model in the
spherical topology. The planar loop equations follow from {\wxv} and
{\wxvi} for $N \rightarrow \infty$:

$$
\left(u(p)-V'(p)\right)^2+\left(v(p)-\xi(p)
\right)'\left(v(p)-\xi(p)\right)=Q_0(p),
\eqn\wxxxii
$$

$$
\left(v(p)-\xi(p)\right)\left(u(p)-V'(p)\right)=Q_1(p).
\eqn\wxxxiii
$$

\no As in the bosonic case we look for the one-cut solution. We will
consider a general bosonic part of the superpotential so that the
one-cut ansatz takes the form:

$$
u(p)=u_0(p)+u_2(p)=V'(p)-M(p)\sqrt{\Delta}-{A(p)\over\D^{3/2}},
\eqn\wxxxiv
$$

$$
v(p)=\xi(p)-{N(p)\over\sqrt{\Delta}},
\eqn\wxxxv
$$

\no with $\D=(p-x)(p-y)$. Here $u_0(p)$ coincides the solution of the
purely bosonic model {\axxxxix} and determines the planar solution in
order zero of the fermionic couplings. The subindex in {\wxxxiv}
indicates the order in fermionic couplings. As we will now see, the
superloop equations can be solved taking for the loop operator an
expansion that is at most quadratic in fermions. It has been argued
in [\sloops] that the solution to the superloop equations is unique,
so that this property is inherent in the model. It will also appear
non-perturbatively [\two], as we will see later, and means that
correlation functions involving more than two fermionic scaling
operators will vanish.

To obtain the complete solution of the model we have to determine the
functions $M(p)$, $A(p)$ and $N(p)$ in {\wxxxiv} and {\wxxxv}; $M(p)$
has already been determined in {\axxxxvi}. Here it will be written in
a simpler form, in terms of recurrence relations. We introduce the
notation

$$
x=S+\sqrt{R} \qquad {\rm and} \qquad y=S-\sqrt{R}.
\eqn\wwxxxv
$$

\no To write down the explicit form of $M$, $N$ and $A$, we note that
any analytic function
$f(p)$ can be written in the form:

$$
f(p)=f^-(\D)+\D'f^+(\D)\qquad{\rm with} \qquad
\qquad \D'={d\D\over dp}=2(p-S),
\eqn\wxxxvi
$$

\no where we have split the function $f(p)$ into two terms of
opposite parity
with respect to the change $(p-S)\rightarrow (S-p)$. More concretely

$$\eqalign{
&M(p)=M^-(\D)+\D'M^+(\D),\cr
&N(p)=\D N^-(\D)+\D'N^+(\D),\cr
&A(p)=A^-(\D)+\D'A^+(\D).\cr}
\eqn\wxxxvii
$$

\no We can expand $M$, $N$ and $A$ in powers of $\D$

$$
M^{\pm}(\D)=\sum_{k\geq 0} m_{k}^{\pm}\D^k, \qquad
N^{\pm}=\sum_{k\geq 0}n_{k}^{\pm}\D^k, \qquad
A^{\pm}(\D)=\sum_{k\geq 0}a_{k}^{\pm}\D^k ,
\eqn\wxxxviii
$$

\no and we have to calculate the expansion coefficients. To determine
$A$ we substitute the above expressions in {\wxxxiv} and {\wxxxv} and
require that the left-hand side of {\wxxxii} and {\wxxxiii} be
polynomials in $p$. After some computations we obtain that $A^-(\D)$
and $A^+(\D)$ are completely determined by $a_0^{-}$ and $a_0^{+}$.
The results are

$$
A^{-}=a_0^{-}=-{2R\over
({m^{-}_0})^2-4R({m^{+}_0})^2}({m_0^{-}}n_0^{-}
n_0^{+}-4Rm_0^{+}n_1^{+}n_0^{+}),
\eqn\wxxxix
$$

$$
A^{+}=a_0^{+}=-{2R\over({m^{-}_0})^2-
4R({m^{+}_0})^2}({m_0^{-}}n_1^{+}
n_0^{+}-m_0^{+}n_0^{-}n_0^{+}).
\eqn\wxxxx
$$

\no Therefore, once we have determined the coefficients $m_k^{\pm}$
and
$n_k^{\pm}$ we will have the complete solution to the planar model.
We split $M(p)$ from {\axxxxvi} in $M^{\pm}(\D)$, and find:

$$
M^-(\D)=\oint_0{dt\over 2\pi i}\INT{t(1-tS)\over\DE-t^2\D}
\eqn\wxxxxi
$$

$$
M^+(\D)={1\over2}\oint_0{dt \over 2\pi i}\INT{t^2\over\DE-t^2\D}.
\eqn\wxxxxii
$$

\no Here we have introduced $g(p)=pV'(p)$ and transformed the
integrals so that the contour goes around zero. Expanding in powers
of $\D$ we obtain for $m_k^{\pm}$ explicit formulae that satisfy the
following recursion relations:

$$
{\p m_k^{\pm} \over\p R}=\left(k+{3\over2}\right)m_k^{\pm}, \qquad
\qquad \qquad
{\p \mkI\over \p S}=\left(k+{3\over2}\right)m_{k+1}^{-},
\eqn\recuone
$$
\vskip 0.5cm
$$
{\p \mko \over \p S}=4(k+1)\mkI+4R\left(k+{3\over2}\right)\mkpI.
\eqn\recutwo
$$

\no Therefore, we only have to determine the initial values
$m_0^{\pm}$, and the values of the rest of the $m_k^{\pm}$'s follow
from the previous relations. This can be done by differentiating
{\axxxxii} with respect to $x$ and $y$ and comparing to {\axxxxvi}.
This gives:

$$
\px\L_S={1\over 4}(x-y)M(x)\qquad{\rm and} \qquad \py\L_S =
{1\over 4}(y-x)M(y).
\eqn\wxxxxv
$$

\no On the other hand by {\wxxxviii}:

$$
M(x)=m_0^- +(x-y)m_0^+, \qquad{\rm and}  \qquad
M(y)=m_0^- +(y-x)m_0^+.
\eqn\wxxxxvi
$$

\no This gives $m_0^\pm$ in terms of the single function $\Lambda_S
(x,y)$

$$
m_0^-={2\over (x-y)}(\px -\py )\L_S\qquad {\rm and} \qquad
m_0^+={2\over (x-y)^2}(\px +\py )\L_S.
\eqn\wxxxxvii
$$

\no Therefore, $M(p)$ has been determined as a function of $\L_S$.

We continue now with the calculation of $N_{\pm}(\D)$ that goes along
the same lines. We have the following representations as contour
integrals:

$$
N^{-}(\D)=\oint_0{dt \o 2\pi i}{\xi(\IT)\over
\sqrt{\DE}}{1-St\over\DE-t^2\D},
\eqn\wxxxxviii
$$

$$
N^{+}(\D)={1\over 2}\oint_0{dt \o 2\pi
i}\xi(\IT)\IT{\sqrt{\DE}\over\DE-t^2\D}.
\eqn\wxxxxix
$$

\no Expanding in powers of $\D$ we obtain that the $n^{\pm}_k$'s
satisfy the
recursion relations:

$$\eqalign{
&{\p \nkm \over\p R}=\left(k+{3\over2}\right)\nkpm, \qquad\qquad {\p
\nkp\over \p S}=\left(k+{1\over2}\right)\nkm,\cr
& \cr
& {\p \nkp\over\p R}= \left(k+{1\over2}\right)\nkpp , \qquad
\qquad {\p \nkm \over \p S}=4(k+1)\nkpp+4R\left(k+{3\over2}\right)
n_{k+2}^{+}.\cr}
\eqn\wxxxxi
$$

\no Here we see that all the $n^{\pm}_k$'s can be written in terms of
$n_{0}^{+}$ (the fermionic scaling variable), which can be defined
through a contour integral in the same way as the cosmological
constant in {\axxxxii}:

$$
n_0^+={1\over 2}\oint_0{dt \o 2\pi i}{\xi(\IT)\over
t\sqrt{(t-x)(t-y)}}.
\eqn\wxxxxiv
$$

\no The function $n_0^{+}$ has a property that will later be relevant
in the continuum limit

$$
(x-y)\p_{xy}n_0^+={1\over 2}(\p_x n_0^+-\p_yn_0^+).
\eqn\xixixi
$$

\no The complete solution of the planar model has now been computed.

We can use these results to obtain explicit expressions for the loop
operators in the discrete theory. The above recursion relations and
some amount of algebra determine
the total derivative of the fermionic loop operator $v(p)$ with
respect to $R$:

$$
{dv(p)\over dR}={1\over \sqrt{\D}}\left({\p n_0^+\over \p
S}+4 R{dS\over dR}{\p n_0^+\over \p R}-{ 2R \over \D}{dS\over
dR}n_0^+
\right)-{n_0^+\o 2}{\D'\over\D^{3/2}}.
\eqn\wxxxxiii
$$

\no The order zero in fermionic couplings of the bosonic loop
operator $u_0(p)$ has already been determined and is given by
{\axxxxix}.
The expression for $u_2(p)$ follows from considerations similar to
those for the fermionic loop operator:

$$
u_2(p)={4 R\over (m_0^-)^2-4R(m_0^+)^2}\left(
m_0^-{\p n_0^+\over \p S}-4Rm_0^+{\p n_0^+\over \p R}
+
\D'\left(m_0^-{\p n_0^+\over \p R}-m_0^+{\p n_0^+\over \p
S}\right)\right){n_0^+\over \D^{3/2}}.
\eqn\wxxxxv
$$

\no In order to take the continuum limit we want to transform the
derivatives with respect to $R$ and $S$ appearing in $u_2(p)$ into
derivatives with
respect to $\L$. Using implicit derivatives we obtain

$$
{\p x\over\p \L}={R^{-1/2}\over m_0^{-}+2\sqrt{R}m_0^{+} }
\qquad\qquad
{\p y\over\p \L}=-{R^{-1/2}\over m_0^{-}-2\sqrt{R}m_0^{+}}
\eqn\wxxxxvi
$$

\no and since

$$
{\p \o \p R}={1\o x-y} \left( {\p \o \p x}-{\p \o \p y}\right)
\qquad {\rm and} \qquad
{\p \o \p R}={1\o x-y} \left( {\p \o \p x}-{\p \o \p y}\right)
\eqn\wwxxxxvi
$$

\no we obtain the result

$$
u_2(p)={4R^{3/2}\over\D^{3/2}}\left({\p x\over\p\L}{\p
n_0^+\over\p x}-{\p y\over\p\L}{\p n_0^+\over\p y}
+{\D'\over 2 \sqrt{R}}{\p n_0^+\over\p \L}\right)n_0^+.
\eqn\wwxxxxvii
$$

\no Since the continuum limit of $u_0(p)$ has been computed in
{\alxi}, we now determine the continuum limits of $u_2(p)$ and
$v(p)$. All we have to do for that is to determine the behavior of
$n_o^+$ in the scaling limit. We denote by $\tau$ the scaling
function corresponding to $n_0^+$.
To determine the scaling of $n_0^{+}$ we notice that the scaling of
$u_2(p)$ should be

$$
u_2(z)=a^{-2+{1\o m}}u_2(p)
\eqn\wwwxxxxviii
$$

\no as determined from the scaling of $u_0(p)$ in {\aalxi}. Therefore
$n_0^{+}$ should behave as:

$$
n_0^{+}={a^{2+{1\over m}}\over (1-y_c)}\tau(u)
\eqn\wwwwxxxxviii
$$

\no where the sign is fixed by the continuum limit of the
fermionic-loop
operator. We notice that eq. {\xixixi} gives, in the continuum limit,
an equation for $\tau$:

$$
{\p^2 \tau \over \p u_+ \p u_-}=\p_{+-}\tau=0.
\eqn\wwxxxxviii
$$

\no A solution to this equation is given by

$$\tau={\widetilde
\tau}_0+\sum_{k>0}({\widetilde \tau}_k^+u_+^k+{\widetilde
\tau}_k^-u_-^k).
\eqn\wwxxxxviii
$$

\no Here ${\widetilde \tau}_k^\pm =0$ are the renormalized fermionic
couplings.
For simplicity can we set consistently ${\widetilde \tau}_k^-=0$. We
will omit the $+$ and $-$ signs introduced for the specific heat in
section 3.3. The complete bosonic-loop operator is then:

$$
{ \p u (z)\o \p t}=-{1\over \sqrt{z+u}}+{\p\o \p t}
\left({\t\p_t\t\over
(\z)^{3\o 2}}\right).
\eqn\wxxxxx
$$

\no In a similar way we get the continuum limit of the fermionic
loop operator. Using {\wxxxxiii} and transforming the
derivatives, we get in the scaling limit:

$$
{\p v(z)\o \p t}=a^{1/ m}\left({1\over
\sqrt{\z}}{\DLR \o \p t} \t \right)\qquad {\rm where}\qquad  \qquad
v(z)=a^{-2+{1/m}}
\sqrt{1-y_c} v(p).
\eqn\wli
$$

\no The factor $a^{1/m}$ can be absorbed into the bare fermionic
superlength $\th^B$

$$
w(l,\th^B)=u(l)+\th^B v(l),
\eqn\wwli
$$

\no because we can define renormalized superlength

$$
\th=\th^B a^{1/m},
\eqn\wwlii
$$

\no that behaves dimensionally as $\rm (length)^{1/2}$.

Now we would like to obtain the expressions for the macroscopic loop
operators from which we can obtain the bosonic and fermionic scaling
operators. They are defined in a similar way as in {\alii}:

$$
u(l)=u_0(l)+u_2(l)=\lim _{n\rightarrow \infty}\biggl\langle
\sum_{i=1}^{N} \lambda_i^{n}\biggr\rangle =\lim_{n\rightarrow \infty}
{N\o \L_S} u^{(n)},
\eqn\wwli
$$

$$
v(l)=\lim _{n\rightarrow \infty}\biggl\langle
\sum_{i=1}^{N}\theta_i\lambda_i^n\biggr\rangle =
\lim_{n\rightarrow \infty} {N\o \L_S} v^{(n)}.
\eqn\wwlii
$$

\no We can calculate the asymptotic behaviors using some standard
formulas from the theory of Legendre and Gegenbauer polynomials
\REF\ederly{A.~Erdelyi, ``Higher
Transcendental Functions'', New York, McGraw-Hill,
1954.}
\REF\grad{I.~S.~Gradshteyn and I.~M.~Ryzhik, ``Table of Integrals,
Series and Products'', New York,
Academic Press, 1980.}
[\ederly,\grad].
To give an example, we start with the derivation of the $u_0$
operator, which has already been done in section 3.3 through a
contour integral. We can expand in terms of Legendre polynomials

$$
{\p u_0(p)\over \p\L_S}=\sum_{n\geq 0}{1\over p^{n+1}}{\p u_0^{(n)}
\over \p\L_S}=\sum_{n\geq0}(xy)^{n/2}
P_n\left({x+y\over2\sqrt{xy}}\right){1\over p^{n+1}},
\eqn\wwliii
$$

\no so that the expansion coefficients become

$$
{\p u_0^{(n)}\over \p
\L_S}=(xy)^{n/2}P_n\left({x+y\over2\sqrt{xy}}\right).
\eqn\cxlvii
$$

\no Using the relation to the hypergeometric function $F$

$$
P_n(z)=F(-n,n+1,1,(1-z)/2),
\eqn\wwliv
$$

\no and the asymptotic formula for $F$ as $n\rightarrow\infty$, we
obtain:

$$
{\p u_0^{(n)}\over \p\L_S}\simeq {1\over\sqrt{\pi n}}{1\over
\sqrt{x-y}}(x^{n+1/2}+iy^{n+1/2})\;\;.
\eqn\cxlviii
$$

\no If we choose $y_c<x_c$, $|y_c|<1$, we see that the critical
value for $x$ is $x_c=1$ and that the second term of the sum vanishes
in the limit $n\rightarrow\infty$. Defining the renormalized length
as $l=na^{2/m}$, we obtain the expression {\alxii}. The computation
of $u_2(p)$ and the fermionic loop operator $v(p)$ can be done with
the same method. We use several formulas for Gegenbauer polynomials:

$$
{1\over \D^{3/2}}=\sum_{n=0}^\infty
(xy)^{n\over 2}C_n^{3/2}(z) p^{-n-3}\qquad z={x+y \over 2 \sqrt{xy}}
$$

$$
C_n^{\nu}(z)={\Gamma(n+2\nu)\over \Gamma(n+1)\Gamma(2\nu)}
F\left(n+2\nu,-n,\nu+{1\over2},{1\over 2}-{z\over 2}\right),
\eqn\clxxi
$$

\no and the asymptotic behavior of the Legendre polynomials. The
final answers are

$$
{\p\o \p t} \langle {\widetilde u}(l)\rangle =-{e^{-lu}\over \kappa
l }+ {2\over
\kappa}{\p\o \p t}\left(e^{-lu} \tau {\p\tau\o \p t}  \right)
\eqn\wlii
$$

$$
{\p \o\p t}\langle {\widetilde v}(l)\rangle={2\over
\kappa}a^{{1\o m}}e^{-lu} \left( {\p\t\o \p u}+l\t\right) {\p u\o \p
t} ,
\eqn\wliii
$$

\no where we have defined

$$
{\widetilde u}(l)=\sqrt{\pi \o l}u(l) \qquad {\rm and} \qquad
{\widetilde v}(l)=2\sqrt{\pi l}v(l).
\eqn\wliv
$$

\no We notice that these operators have an expansion at most
quadratic in the fermionic couplings. This means that correlation
functions involving more than two fermions vanish. In the next
section we are going to derive the expressions for the scaling
operators that follow from these loop operators.

\endpage

\chapter{\chafour}

\no In this chapter we will analyse the super-eigenvalue model in the
double scaling limit and the connection to superintegrable
hierarchies. One of the basic ingredients to obtain the
non-perturbative solution of the model are the super-Virasoro
constraints in the double scaling limit. In order to calculate the
non-universal parts of the loop operators we will compute explicit
expressions for the scaling operators on the sphere. We will find
agreement with the scaling dimensions of $(4m,2)$ superconformal
field theories coupled to 2D-supergravity.

\section{\secfourtwo}

\no We will now see that, in the continuum limit, the bosonic and
fermionic loop operators become a $\IZ_2$-twisted scalar
bosonic field and a Weyl-Majorana fermion in the Ramond sector.

We redefine the purely bosonic part of the loop operator [\dvv]
according to:

$$
{\widehat u}_0(p)=u_0(p)-V'(p)=-M(p)\sqrt{(p-x)(p-y)}.
\eqn\di
$$

\no Using the relations {\recuone}, {\recutwo} and {\wxxxxvii} we
obtain in the scaling limit

$$
{\widehat u}_0(z)=a^{-2+{1\over m}}{\widehat u}_0(p)=
\sqrt{1-y_c}\sum_{k\geq 0}{\Gamma({1\over 2})\over \Gamma(k+
{3\over 2})} (-)^{k+1}{\p^{k+1} t\over \p u^{k+1}}(\z)^{k+{1\over
2}}.
\eqn\diii
$$

\no This expression coincides with the Laplace transform of the loop
operator.
We will show that ${\widehat u}_0(z)$ can be identified with
a free bosonic scalar field $x(z)$ in two dimensions with
antiperiodic boundary conditions

$$
\p x(z)=\sum_{n\in \IZ} \alpha_{n+{1\o 2}}z^{-n-{3\o 2}} \qquad{\rm
with} \qquad
x(e^{2\pi i}z)=-x(z).
\eqn\dii
$$

\no The Laplace transform of the bosonic loop
can be decomposed into two pieces

$$
\u_0 (z)\equiv t(z)+u_0(z).
\eqn\dii
$$

\no Here $t(z)$ denotes the non-universal part of the loop operator
that is not dependent on the topology. The operator $u=u_0+u_2$ can
be expanded in terms of scaling operators (see [\dvv] and references
therein) in the following way:

$$
u(z)=\k^2 \sum_{k\geq 0}z^{-k-{3\o 2}}\langle \sigma_k\rangle
\qquad{\rm where} \qquad
\langle \sigma_k\rangle ={\p F\over \p t_k}.
\eqn\div
$$

\no Therefore, if we compute the Laurent expansion of the operator
{\wxxxxx}\foot{Now and in the
following we redefine $u(z)$, $\t$ and $t$ in
order to absorb $\sqrt{1-y_c}$ factors, and take effectively
$y_c=0$.}

$$
{\p u(z)\over \p t}=-{1\over \sqrt{z}}-
\sum_{k\geq 0}{(-)^{k+1}\o k!}{\Gamma(k+{3\over 2})\over
\Gamma({1\over 2})}
\left( {u^{k+1}\over k+1}+2\p_t(u^k \t \p_t \t)
\right)z^{-k-{3\o 2}}
\eqn\dv
$$

\no  for $z\rightarrow\infty$, we obtain the following expression for
the scaling operators $\langle \sigma_k\rangle $

$$
{\p \langle {\sigma}_k\rangle \o \p t}={1\over \k^2 }{(-)^{k} \Gamma
(k+{3\over 2})\over
k!\Gamma({1\over 2})}\left({u^{k+1}\over k+1}+2\p_t (u^k\t \p_t \t)
\right).
\eqn\dvi
$$

\no The expansion of the macroscopic loop {\wlii}

$$
\langle {\widetilde u}(l)\rangle =\kappa \sum_{k=0}
^{\infty}(-)^{k+1}{l^k\over k!}\langle {\widetilde
\sigma}_k\rangle  -{1\over \kappa l},\qquad{\rm where}\qquad \langle
{\widetilde \sigma}_k\rangle = {\p F \over \p {\widetilde t}_k},
\eqn\dviii
$$

\no  determines the form of the scaling operators that are dual to
the original couplings ${\widetilde t}_k$ that differ from the
couplings $t_k$ by a multiplicative constant:

$$
\p_t\langle {{\widetilde \sigma}}_k\rangle =-{1\over \k^2 }
\left({u^{k+1}\over k+1}+2\p_t (u^k\t\p_t \t) \right).
\eqn\dix
$$

\no The relation between the coupling constants $t_k$ and
${\widetilde t}_k$ is
therefore

$$
{\widetilde t}_k=(-)^{k+1}{\Gamma(k+{3\over 2}) \over
k!\Gamma({1\over 2})} t_k.
\eqn\dx
$$

\no The non-universal part of ${\widehat u}_0(z)$ can be calculated
using the planar string equation {\alxxiixi}. After differentiation
w.r.t. $u$ we obtain

$$
{\p^{k+1} t\over \p u^{k+1}}=-{\widetilde t}_{k+1}(k+1)!+O(u).
\eqn\dv
$$

\no Now taking into account the relation
between ${\widetilde t}_k$ and $t_k$  we obtain the same result for
the non-universal part of ${\widehat u_0}$ as in the purely bosonic
model:

$$
t(z)=\sum_{k\geq 0}\left(k+{1\over 2}\right)t_k z^{k-{1\o 2}}.
\eqn\ddv
$$

\no The bosonic-loop operator is therefore a bosonic scalar field
with antiperiodic boundary conditions:

$$
u(z)=\sum_{k\geq 0}\left(k+{1\over 2}\right)t_kz^{k-{1\o 2}}+\k^2
\sum_{k\geq 0}z^{-k-{3\o 2}} \langle \sigma_k\rangle.
\eqn\dvii
$$

\no The relations between the coupling constants of the model and
the modes of the bosonic field are

$$
\alpha_{n+{1\o 2}}=\kappa {\p \over \p t_n} \qquad {\rm and} \qquad
\alpha_{-n-{1\o 2}}=\left(n+{1\over 2}\right){t_n\over \kappa}
\qquad {\rm for} \qquad n\geq 0.
\eqn\dxx
$$

We can handle in a similar way the fermionic-loop operator ${\widehat
v(z)}$ that will be identified with a Weyl-Majorana fermion in the
Ramond sector

$$
\psi(z)=\sum_{n\in \IZ}\psi_n z^{-n-{1\o 2}}.
\eqn\dx
$$

\no First we redefine

$$
{\widehat v(p)}=v(p)-\xi(p)=-{N(p) \over \sqrt{\D}}.
\eqn\dviii
$$

\no With the recursion relations {\wxxxxi} we get, in the continuum
limit, an equation analogous to {\diii}:

$$
{\widehat v(z)}=a^{{1\o m}}\left(-{\t \over
\sqrt{z}}+2\sum_{k\geq 0}(-)^k{\Gamma({1\over 2})\over
\Gamma(k+{3\over 2})} \p_u^{k+1}\t (\z)^{k+{1\o 2}}\right).
\eqn\dix
$$

\no The scaling operators are calculated from the expansion of the
loop operator and its Laplace transform

$$
v(z)=\kappa^2 a^{{1\o m}} \sum_{n = 0}^\infty z^{-n-{1\o 2}}\langle
\nu_n\rangle\qquad {\rm and }\qquad
{\widetilde v}(l)=\kappa a^{{1\o m}}\sum_{k=0}^\infty
(-)^{k+1}{l^k\over k!}\langle {\widetilde \nu_k}\rangle,
\eqn\dix
$$

\no and are given by

$$
{\p\o \p t}\langle \nu_k\rangle ={(-)^k\over \k^2 }
{\Gamma(k+{1\over 2})\over k! \Gamma({1\over 2})}u^k{\DLR\o \p t}
\t,\qquad {\rm and}\qquad
{\p\o \p t}\langle \widetilde \nu_k\rangle =-{2\over \k^2}u^k {\DLR\o
\p t}\t .
\eqn\dxi
$$

\no Note that the two-sided derivative in {\dxi} guarantees
Fermi statistics for the fermionic two-point functions

$$
{\p\o \p t}\langle {\widetilde \nu}_k {\widetilde \nu}_n
\rangle=-{2\over
\k^2 }u^k{\DLR \o \p t} u^n.
$$

\no The relation between ${\widetilde \tau_k}$ and $\tau_k$ is
therefore

$$
{\widetilde \tau_k}={(-)^{k+1}\over 2 k!}{\Gamma(k+{1\over 2})
\over \Gamma({1\over 2})} \tau_k.
\eqn\dxii
$$

\no Using the planar equation $\t=\sum_{k\geq 0}{\widetilde
\tau_k}u^k$, we fix the non-universal part of the fermionic loop:

$$
\eta(z)={\tau_0\over 2}z^{-{1\o 2}}+\sum_{k\geq 0}\tau_{k+1}z^{k+{1\o
2}}.
\eqn\dxiii
$$

\no Finally we obtain for the whole fermionic-loop operator a
Weyl-Majorana fermion in the Ramond sector:

$$
{\widehat v(z)}={\tau_0\over 2}z^{-{1\o 2}}+\sum_{k\geq 0}
\tau_{k+1}z^{k+{1\o 2}}+\k^2 \sum_{k\geq 0}
z^{-k-{1\over 2}}\langle \nu_k\rangle .
\eqn\dxv
$$

\no In terms of the mode expansion {\dx}, we obtain the following
representation of the coupling constants:

$$
\psi_0={\tau_0\over 2\kappa }+\kappa {\p \over \p \tau_0},\qquad
\qquad
\psi_n=\kappa {\p \over \p \tau_n} \qquad{\rm and} \qquad
\psi_{-n}={\tau_n \over \kappa} \qquad{\rm for} \qquad n>0.
\eqn\dxxi
$$

\no The identification of the zero mode guarantees $\psi_0^2=1/2$.
The super-Virasoro constraints in the continuum are therefore
described by the super-energy-momentum tensor of a ${\widehat
c}=1$ superconformal field theory,

$$
T_F(z)={1\over 2}\p x(z)\psi(z),
\eqn\dxxiii
$$
$$
T(z)={1\over 2}:\p  x (z)\p  x (z):+{1\over 2}:\p
\psi(z) \psi(z):+{1\over 8z^2}.
\eqn\dxxiv
$$

\no With the mode expansion

$$
T_F(z)=\h \sum_{n \in \IZ}z^{-n-2}G_{n+{1\o 2}}
\qquad { \rm and}  \qquad T(z)=\sum_{n\in \IZ}z^{-n-2}L_n,
\eqn\dxxv
$$

\no we obtain in terms of the coupling constants:

$$
G_{n+{1\o 2}}={t_0\tau_0\over 4 \k^2 }\delta_{n,-1}+\sum_{k\geq
0}\left(k+{1\over 2}\right)t_k{\p \over \p \tau_{n+k+1}}+
{\tau_0\over 2}{\p  \over \p t_n}+\sum_{k\geq
0}\tau_{k+1}{\p \over \p t_{n+k+1}}+\k^2  \sum _{k=0}^{n}{\p^2\over
\p t_k \p \tau_{n-k}}.
\eqn\dxxvi
$$

\no The modes of $T(z)$ are fixed by the
anticommutation relation {\wxxvii}:

$$
\eqalign{
L_n=&(t_0^2-2\tau_0\tau_1){\delta_{n,-1}\over 8 \k^2  }
+{\delta_{n,0}\o 8}
+\sum_{k\geq 0}
\left(k+{1\over 2}\right)t_k{\p \over \p t_{n+k}}+\sum_{k\geq
0}\left({n\over 2}+k+1\right)\tau_{k+1}{\p  \over \p
\tau_{n+k+1}}+\cr
&\cr
&{n\over 4}\tau_0{\p\over\p\tau_n}
+{ \k^2 \over 2}\sum
_{k=1}^{n}{\p^2\over \p t_{k-1} \p t_{n-k}}
-{ \k^2 \over
2}\sum_0^n\left(k+\h\right){\p^2\over\p\tau_k \tau_{n-k}}.\cr}
\eqn\dddxxvi
$$

\no After the double scaling limit we obtain that the superloop
equations in the continuum are equivalent to:

$$
Z^{-1}{\bf T}(z,\th )Z={\rm polynomial}\,(z,\th),
\eqn\dxxix
$$

\no with ${ \bf T} (z,\theta ) =T_F(z)+\theta T(z)$ determined by
{\dxxiii} and
{\dxxiv}.

The continuum limit of the superloop equations are described by a
$\IZ_2$-twisted massless scalar field and a Weyl-Majorana fermion in
the Ramond sector. Notice that we only need to derive {\dxxiii} since
{\dxxiv}
follows from simple commutation relations. This makes the
computation simpler than in the bosonic case, because $T_{F}(z)$
has no normal-ordering ambiguities.

Since we have explicit expressions for the scaling operators and the
string equation for genus zero we can determine the critical
exponents. For this purpose we have to consider the zero coupling
regime, where ${\widetilde t_n}=0$ and ${\widetilde \tau_k}=0$. The
expressions for the scaling operators follow from the free energy

$$
\k^2 D^ 2 F_S  =u-2\t D^2   \t,
\eqn\ddxxix
$$

\no where $D=\p /\p t_0$. From the scaling of the free energy we
obtain the string susceptibility $\G_{str}$ as we saw in
{\rxviixixx}:

$$
\p_t^2 F_S\sim t^{-\G_{str}}.
\eqn\dxxx
$$

\no We can read off the gravitational scaling dimension $\D$ of some
scaling operator ${\cal O}$ from the relation {\rxv}

$$
\p_t^2\biggl\langle \prod_i{\cal O}_i \biggr\rangle \sim t^{-\G_{str}
+\sum_i(\D _i-1)}.
\eqn\dxxxi
$$

\no Using the form of the string equation for the $m$-multicritical
point {\alxxiixi} we find

$$
\p_t^2 F_S\sim u=t^{1/ m}.
\eqn\dxxxii
$$

\no We thus have the same family of susceptibilities $\G_{str}=-{1/
m}$, $m=1,2,3,\dots$, as the purely bosonic theory. For the bosonic
operators we find

$$
\p_t^2\langle \s_n\rangle\sim {\p u\o \p t_n}=u^n\p_t u
\sim t^{{n / m}+{1/ m}-1}=t^{-\G_{str} +{n/ m}-1}.
\eqn\dxxxiii
$$

\no We obtain the same spectrum of dimensions as in the Hermitian
one-matrix model:

$$
\D _{\s_n}={n\o m}.
\eqn\dxxxiv
$$

\no The scaling dimension of the fermionic operators can be
determined from

$$
\p_t^2\langle\nu_k\nu_n\rangle\sim t^{{k/ m}+{n/ m}-2}=t^{-\G_{str}
+(\D _{\nu_n}-1)+(\D _{\nu_m}-1)}.
\eqn\dxxxv
$$

\no We obtain the result

$$
\D _{\nu_n}={n\o m}-{1\o 2m}.
\eqn\dxxxvi
$$

\no These are the gravitational scaling dimensions of
operators in the Neveu-Schwarz and Ramond sector of $(4m,2)$ $N=1$
superconformal minimal models coupled to two-dimensional supergravity
[\pz,\dhk].

\section{\secfourthree}

\no In the following we solve the continuum loop equations
determined by {\dxxix}, expanding the equations in the genus in a
similar way as proposed by David for the bosonic theory [\loopdavid].
In the continuum the loop equations
get more transparent than in the discrete theory. It is interesting
to consider also the planar case and see how the results for the
loop operators and the string equation found in the previous sections
appear in a simple way, as a solution of the continuum equations. For
higher genera, i.e. for the
torus, the double-torus, $\dots$, we obtain a systematic expansion
determining all correlators beyond the planar limit.

In the double scaling limit, the two superloop equations, that
are equivalent to the continuum super-Virasoro constraints are:

$$
\u(z)\v(z)+\k^2\chi^{{\rm BF}}(z)={\rm
polynomial}\,(z) ,
\eqn\ei
$$
\no and
$$
\u(z)^2-\v(z) \p \v(z)
+\k^2\left(\chi^{{\rm BB}}(z)+\chi^{{\rm FF}}(z)+
{1\over 4z^2}\right)={\rm polynomial}\, (z),
\eqn\eii
$$

\no where the two-loop operators are defined by:

$$\eqalign{
\chi^{\rm BF}(z)&= \sum_{k,l\geq 0}
z^{-k-l-2}{\p^2 F_S\over \p t_k\p \tau_l}\cr
&\cr
\chi^{\rm BB}(z) &=
\sum_{k,l\geq 0}z^{-k-l-3}{\p^2 F_S\over \p t_k \p t_l}, \cr
& \cr
\chi^{\rm FF}(z)&=
-\sum_{k,l\geq 0}\left(k+{1\o 2}\right)z^{-k-l-2}{\p^2 F_S \o \p
\t_k\p \t_l}.
\cr}
\eqn\eiii
$$

\no The loop operators and the free energy have the following genus
expansion

$$
\eqalign{
u(z)&=u_0(z)+\k u_1(z)+\dots=\sum_{k\geq 0}
\left(u_0^{(2k)}(z)+\k u_1^{(2k)}(z)+\dots \right),\cr
&\cr
v(z)&=v_0(z)+\k v_1(z)+\dots=\sum_{k\geq 0}\left( v_0^{(2k+1)}(z)+\k
v_1^{(2k+1)}(z)+\dots\right) ,\cr
&\cr
\chi^{a}(z)&=\chi^{a}_0(z)+\k\chi^{a}_1(z)+\dots \qquad{\rm with}
\qquad
a={\rm BF,BB,FF}, \cr
&\cr
F_S & =F_0+\k F_1+\dots .\cr}
\eqn\eiv
$$

\no The sub-indices in our notation indicate the genus, while for the
order in fermionic couplings we introduce upper indices. We now
analyse the solution as a function of the genus.

\vskip 0.5cm

\no {\it Planar solution}

\vskip 0.5cm

\no The leading terms in the genus expansion of eqs. {\ei}
and {\eii} are the planar loop equations

$$
\u_0(z)\vo(z)={\rm polynomial}\,(z),
\eqn\ev
$$
$$
\u_0(z)^2-\vo(z)\p_z \vo(z)={\rm polynomial}\, (z).
\eqn\evi
$$

\no We follow closely the steps of the discrete case with a one-cut
ansatz for ${\widehat u}_0(z)$ and a similar expression for the
fermionic-loop operator

$$
\u_0(z)=M(z)\sqrt{\z}+{A(z)\over(\z)^{3/2}}\qquad {\rm and} \qquad
\vo(z)={N(z)\over \sqrt{\z}}.
\eqn\eviii
$$

\no Expanding in powers of $(\z)$

$$
N(z)=\sum_{k\geq 0}n_k(\z)^k ,\qquad\qquad
M(z)=\sum_{k\geq 0} m_k (\z)^k,
\eqn\eix
$$

\no we see that $A(z)$ is determined by demanding the r.h.s. of {\ev}
to be polynomial in $z$

$$
\u_0(z)=M(z)\sqrt{\z}-{1\over 2m_0}{n_1 n_0\over {(\z)}^{3/2}};
\eqn\ex
$$

\no $M(z)$ is determined from $u^{(0)}_0\sim O(z^{-3/2})$, which
gives

$$
{\p u_0^{(0)}\over \p u}={1\over \sqrt{\z}}\left({1\over
2}M(z)+(\z)\p_uM(z)\right)-{1\over 2\sqrt{z}}{\p t_0\over \p u}.
\eqn\exv
$$

\no The previous equation determines $M(z)$ from the vanishing of the
$z^{-1/2}$ contribution

$$
{1\over 2}M(z)+(\z)\p_u M(z)={1\over 2}{\p t_0\over \p u}.
\eqn\exvi
$$

\no Inserting {\eix} in this equation we obtain a relation between
the coefficients $m_k$ and the renormalized cosmological constant
$t_0$:

$$
m_0={\p t_0 \over \p u} \qquad {\rm and}\qquad  m_k={(-)^k\over
2}{\Gamma({1\over 2}) \over \Gamma(k+{3\over
2})}\;{\p^{k+1}t_0\over \p u^{k+1}}
\qquad{\rm for} \qquad k\geq 1 ,
\eqn\exvii
$$

\no and this fixes $M(z)$ completely.
 The modes of $N(z)$ are determined by demanding that $v_0(z)$ has
for $z\rightarrow\infty$ the following expansion:

$$
v_0(z)={N(z)\over \sqrt{\z}}-\eta(z)=O(z^{-1/2}).
\eqn\exii
$$

\no This implies the expansion $\p v_0 / \p u \sim O(z^{-1/2})$.
Therefore we obtain the recursion relation

$$
{\p n_k\over \p u}=-\left(k+{1\over 2}\right)n_{k+1} \qquad{\rm
for}\qquad  k\geq 1,
\eqn\exiii
$$

\no that gives the value of all the $n_k$'s it terms of $n_0$.
Fermi statistics and compatibility between the bosonic and
fermionic loops give the following relation between $n_1$ and $n_0$:

$$
n_1=-4\p_u n_0.
\eqn\eexiii
$$

\no Therefore as a function of $M(z)$ and $n_0$ the bosonic and
fermionic loop operators have the following form:

$$
\u_0(z)=M(z)\sqrt{\z}-{2\over m_0}{n_0 \p_u n_0 \over(\z)^{3/2}}
\qquad {\rm and}\qquad  {\p v_0(z)\o \p t}=-{1\over\sqrt{\z}}{\DLR\o
\p t} n_0.
\eqn\exix
$$

\no With the ansatz  $n_0=\sum_{n\geq 0}\beta_n u^n\;\;$ and {\exii}
we obtain
a relation to the fermionic coupling constants $\tau_n$

$$
\beta_n={\widetilde \tau_n}={(-)^n\Gamma(n+{1\over 2}) \over 2 n!
\Gamma({1\over
2})}\tau_n.
\eqn\eeexiii
$$

\no Thus $n_0$ is determined by $\tau(u)$ in the spherical topology
as follows:

$$
n_0=-\tau(u)\qquad {\rm with} \qquad \t(u) =\sum {\widetilde
\tau_n}u^n.
\eqn\exiv
$$

\no The final result for the loop operators that follows  from
{\exix} is then

$$
\u_0(z)=M(z)\sqrt{\z}+{\tau \p_t \tau \over (\z)^{3/2}}\qquad {\rm
and} \qquad
\p_t v_0(z)={1\over \sqrt{\z}} \DLR \t .
$$

\no This coincides with the results of sections 4.2 and 5.1. We
obtain the planar string equation from the purely bosonic part of
$\u_0(z)$:

$$
{u_0^{(0)}\over \sqrt{\z}}=M(z)-{t(z)\over \sqrt{\z}}=O(z^{-2}).
\eqn\exx
$$

\no The string equation {\aalxviii} follows from the vanishing of the
terms
proportional to $z^{-1}$:

$$
\sum_{k\geq 0}{(-)^k\over
k!}{\Gamma(k+{3\over2})\over  \Gamma({1\over 2})} u^{k} t_k =0.
\eqn\exxi
$$

\vskip 0.5cm

\no {\it Genus-one solution}

\vskip 0.5cm

\no The two superloop equations obtained from {\ei} and {\eii}
for genus one are\foot{In the following we will omit the $z$
dependences.}:

$$
2\u_0 u_1-v_1\p\vo-\vo\p v_1+
\chi_0^{\rm BB}+\chi_0^{{\rm FF}}+{1\over 4z^2}=
{\rm polynomial}\,(z),
\eqn\exxii
$$

$$
v_1\u_0+\vo u_1+\chi_0^{\rm BF}={\rm polynomial}\,(z);
\eqn\exxiii
$$

\no ${\widehat u}_0$ and ${\widehat v}_0$ are already determined
since they are the solution on the sphere.
The two-loop correlators are  also  determined by the results for
genus zero and are obtained from {\eiii} for $F_S=F_0$

$$
\eqalign{
\chi_0^{\rm BF}&=-{1\over2(\z)^{{3/ 2}}}\left(
{1\over\sqrt{\z}}{\DLR\o \p t} \tau
\right),\cr
&\cr
\chi_0^{\rm BB}&= {1\over 8}\left( {1\over
(\z)^2}-{1\over z^2} \right) -
{1\over 2}{\p\o \p t}\left({\tau\p_t\tau\over  (\z)^3}\right),\cr
&\cr
\chi_0^{\rm FF}&={1\o 8}\left( {1\o (\z)^2}-{1\o z^2}\right).\cr}
\eqn\exxiv
$$

\no We obtain for the fermionic-loop operator

$$
v_1=\sum_{k=0}^3 (\z)^{-k-1/2}V_{k+{1/ 2}},
\eqn\eexxiv
$$

\no with

$$\eqalign{
&V_{{1/ 2}}=-{1\over 3}D \left({ D ^2 \tau  \over
D  u}\right) \qquad\qquad \;\;\;\;\;\; \;\;\;\;\;\; V_{{5/2}}=
-{D u\DDLR  \tau \over 2} \cr &V_{{3/ 2}}={2\over 3}D \left(
{V_{5/ 2}\over D u}\right) -\tau D\left({D ^2u\over 2 D u}\right)
\qquad V_{{7/ 2}}= -{5\o 8} (D u)^2\tau .
\cr}
$$

\no The bosonic-loop operator can be determined in the same way. With
the expansion

$$
u_1(z) = \sum_{k=1}^4 (\z)^{-k-1/2}U_{k+{1/ 2}},
\eqn\exxvii
$$

\no we obtain for the coefficients

$$\eqalign{
&U_{3/2}={D ^2u\over 12D u}-2\tau\DDLR  V_{1/2}
\qquad\qquad\;\;\;\;\;\;\;\;\;\; U_{7/2}=5\tau D (D u D \tau)
\cr  &U_{5/2}=-{D u\over 8}+6V_{3/2}D \tau-2\tau \DDLR  D ^2 \tau
\qquad  U_{9/2}=7V_{7/2}\DDLRt \tau. \cr}
\eqn\exxxiii
$$

\no The value of $V_{1/2}$ cannot be
determined by the requirement that the l.h.s. of
{\exxiii} is a polynomial in $z$.
It follows from the consistency between $v_1$ and $u_1$.
This condition is given by the requirement of commuting
derivatives:

$$
{\p^2 F_1\over \p t\p \tau_k}={\p^2 F_1 \over \p \tau_k \p
t}
\eqn\exxxiiibis
$$

\no where $F_1$ denotes the genus-one contribution to the free
energy.

For the purely bosonic part of $u_1$ we easily see that this is the
same
result as for the one-matrix model. The equation to be solved in this
case is

$$
{\widehat u}_0(z)u_1(z)+\chi_0(z)+{1\over 8z^2}={\rm
polynomial}\, (z),
\eqn\exxxiv
$$
$$
\chi_0(z)={1\over 8}\left( {1\over (\z)^2}+{1\over z^2}\right).
\eqn\exxxv
$$

\no Equation {\exxxiv} coincides with the order-zero equation in
fermionic couplings coming from {\exxii} just because the bosonic
two-point correlators {\exxiv} and
{\exxxv} coincide in the bosonic part.
The reader may have noticed that the solution presented for
$v_1$ is only linear in the fermionic couplings, as was the case
for genus zero. This comes from the fact that the third order in
fermionic couplings

$$
v_1^{(3)} {\widehat u}_0^{(0)}+{\widehat v}_0u_1^{(2)}+v_1^{(1)}
{\tau \p_t \tau
\over \D^{3/2}} ={\rm polynomial}\, (z)
\eqn\exxxvi
$$

\no is solved by $v_1^{(3)}=0$. This means that also
on the torus we have a maximal coupling of two fermions.

\vskip 0.5cm
\no {\it Genus-two solution}
\vskip 0.5cm

\no Now we consider the situation in genus two to see if the heat
capacity changes with respect to the one-matrix model. The two
superloop equations to be solved are

$$
2{\widehat u_0}u_2+u_1^2-v_2\p {\widehat v_0} -{\widehat v_0}\p v_2
-v_1\p v_1 +\chi_1^{\rm BB}+\chi_1^{\rm FF}={\rm polynomial}\,(z),
\eqn\exxxvii
$$

$$
{\widehat v_0}u_2+{\widehat u_0}v_2+u_1 v_1+\chi_1^{\rm BF}=
{\rm polynomial}\,(z).
\eqn\exxxviii
$$

\no To calculate the heat capacity we are only interested in
the purely bosonic piece of the two-loop correlators that
are given by (take {\eiii} for $F_S =F_1$):

$$
\chi_1^{\rm FF}={\chi_1^{\rm BB}}^{(0)}=
{13\over 16 m_0^2}{1\over \D^5}-{3\over 4}{m_1\over
m_0^3}{1\over \D^4}+\left({3\over 8}{m_1^2\over
m_0^4}-{5\over 16}{m_2\over m_0^3}\right){1\over \D^3} .
\eqn\exli
$$

\no We see that again these two-loop correlators coincide with
the one-matrix model values. This immediately implies that, in genus
two we have the same value in the purely
bosonic part of the free energy as in the one-matrix model.
The complete solution to order zero in fermionic couplings is

$$\eqalign{
&u_2^{(0)}={a \over \D^{3/2}}+{b\over \D^{5/2}}+{c\over
\D^{7/2}}+{d\over \D^{9/2}}+{e\over \D^{11/2}}\cr
&a=-{63 m_1^4\over 32 m_0^7}+{75
m_1^2 m_2\over 16 m_0^6}-{145 m_2^2 \over 128 m_0^5}-{77 m_1m_3\over
32
m_0^5}+{105 m_4 \over 128 m_0^4}\cr
&b={63 m_1^3 \over 32 m_0^6}-{87 m_1 m_2 \over 32 m_0^5}+{105
m_3 \over 128 m_0^4}\cr
&c=-{63 m_1^2 \over 32 m_0^5}+{145 m_2 \over 128 m_0^4}\qquad d={203
m_1 \over 128 m_0^4}\qquad e=-{105 \over 128 m_0^3}.\cr}
\eqn\exlii
$$

\no To compare this result with the one-matrix model we continue the
calculation for the case of pure gravity
($m_k=0,\;\;k\geq 2$). Our results reproduce up to genus two the
values expected for the Painlev\'e-I equation. Since

$$
m_0={1\over Du} \qquad{\rm and} \qquad  m_1={2\over 3}m_0^3D^2 u,
\eqn\exliii
$$

\no we get for the heat capacity the expansion

$$
\langle \sigma_0 \sigma_0 \rangle=
-{u\over 4}+{\k^2 \over 12}D\left({D^2 u \over
Du}\right)-\kappa^4 {63\over 162}D\left((D^2u)^4\over
(Du)^5\right)+\dots .
\eqn\exliv
$$

\no For pure gravity we take $u=\sqrt{t}$, so that we obtain

$$
\langle \sigma_0 \sigma_0 \rangle=-{1\over 4}
\left(\sqrt{t}-{1\over 24}{\k^2 \over t^2}-{49 \over 1152}{\kappa^4
\over t^{9/2}}+\dots \right).
\eqn\exlv
$$

\no This agrees with the first three terms of the solution to the
Painlev\'e-I equation appearing in pure gravity {\yyyvii}.

\section{\secfourfour}

\no In the last section we have solved the superloop equations up to
genus two. However the procedure is rather cumbersome and it would be
desirable to find a more efficient method to obtain the solution for
arbitrary genus. In this section we will analyse the discrete theory
and find a powerful relation to the one-matrix model that will allow
us to obtain the non-perturbative solution for the discrete theory
[\two].

We analyse the solution of the discrete superloop equations, or
equivalently the discrete super-Virasoro constraints, through
the representation {\wxxx} of the partition function. We first
consider the
case without fermionic couplings $\xi_{k+{1\over 2}}=0$. The
partition
function can be written as a function of the Pfaffian of the
antisymmetric $2N\times 2N$ matrix $M_{ij}$:

$$\eqalign{
&M_{ij}=\l_{ij}^{-1}={1\o \l_i-\l_j}\qquad {\rm for} \qquad i\neq j ,
\qquad M_{ii}=0 \cr
& \cr
&Pf^{2N}(M_{ij})=\sqrt{\det M_{ij}}\cr}
\eqn\to
$$

\no and the usual Vandermonde $\D(\l_{ij})=\prod_{i<j}\l_{ij}$:

{$$
Z_S(g_k,\xi_{k+{1\over
2}}=0,2N)=\int\prod_{n=1}^{2N}d\l_n \D(\l_{ij}) Pf^{2N}(M_{ij}){\rm
exp} \left(-{2N\over \L_S}\sum_k\sum_{i=1}^{2N}
g_k\lambda_i^k\right).
\eqn\ti
$$

\no Here the product of the Vandermonde and the Pfaffian is totally
symmetric
under exchange of two eigenvalues. We would like to show that the
relation of the supersymmetric partition function at zero fermionic
couplings {\ti} to the bosonic
partition function of the one-matrix model {\ai} is given by:

$$
Z_S(g_k,\xi_{k+{1\over 2}}=0,2N)={1\over 2^N} {2N \choose N} {\cal
Z_B^{\rm 2}}(g_k,N).
\eqn\tii
$$

\no To prove this equality we show that the following equivalence for
the measures of both models holds

$$
 Pf^{2N}(M_{ij})\prod_{i<j}^{2N}(\l_i-\l_j) ={1\over 2^{N}}
\sum_{I,J}
S^{I,J}\Delta^2(I)\Delta^2(J).
\eqn\tiii
$$

\no Here we have introduced the notation
$I=(\l_{i_1},\dots,\l_{i_N})$,
$i_1<\dots<i_N$ (same for $J$), and
$S^{I,J}$ means symmetric permutations between the elements of $I$
and $J$.
The equality {\tii} is then a consequence of {\tiii} since the
partition
function {\ti} with $2N$ eigenvalues will factorize into $ 2N \choose
N$
products of two independent one-matrix models with $N$ eigenvalues.

Obviously {\tii} holds for the case of two eigenvalues, so that to
illustrate
the situation we start with the first non-trivial case of four
eigenvalues $N=2$. We will consider the left- and r.h.s. of {\tiii}
as polynomials $P(\l)$ and $Q(\l)$, respectively, in $\l=\l_4$:

$$
P(\l)=(\l-\l_2)(\l-\l_3)\l_{12}\l_{13}-
(\l-\l_1)(\l-\l_3)\l_{12}\l_{23}+
(\l-\l_1)(\l-\l_2)\l_{13}\l_{23},
\eqn\tiv
$$

$$
Q(\l)={1\over 2}\left((\l-\l_3)^2\l_{12}^2+(\l-\l_2)^2\l_{13}^2+
(\l-\l_1)^2\l_{23}^2\right).
\eqn\tv
$$

\no Because both polynomials agree for $\l=\l_i$, $i=1,2,3$, and
since they are of degree two in $\l$, it follows that $Q(\l)\equiv
P(\l)$ for all $\l$.

We show that {\tiii} is correct by induction over $N$, i.e. assuming
that {\tiii} holds for $2N$ eigenvalues we prove the case of $2N+2$
eigenvalues. For $N>2$ we set $\lambda=\lambda_{2N+2}$; $Q(\l)$ and
$P(\l)$ are then polynomials of degree $2N-2$ in $\lambda$. To show
that both polynomials agree, we prove
that they agree in $2N+1$ points. It is enough to show that they
coincide for $\lambda=\lambda_{2N+1}$ since the symmetry under the
interchange of $\l_{2N+1}$ and $\l_i$, for $i=1,\dots,2N$  guarantees
$P(\l_i)=Q(\l_i)$ for $i=1,\dots,2N$. Using the induction hypothesis
we see that $Q(\l_{2N+1})$ agrees with $P(\l_{2N+1})$:

$$
Q(\l_{2N+1})=
{1\over 2^{N}}
\sum_{I,J}
S^{I,J}\Delta^2(I)\Delta^2(J)
\prod_{\scriptstyle i_1< \dots <i_{N}\atop \scriptstyle j_1<\dots
<j_{N}}
(\l_i-\l_{2N+1})^2(\l_j-\l_{2N+1})^2
$$

$$=
Pf^{2N}(M_{ij})\prod_{i<j}^{2N}(\l_{i}-\l_j)
\prod_{i_1<\dots<i_N}(\l_i-\l_{2N+1})^2
\prod_{j_1<\dots<j_N}(\l_j-\l_{2N+1})^2=P(\l_{2N+1}) .
\eqn\tvi
$$

\no This completes the proof that $Q(\l)\equiv P(\l)$ in the case of
$2N+2$ eigenvalues. We conclude that the bosonic part of
the supersymmetric $2N\times 2N$ model is proportional to the square
of the
corresponding $N\times N$ bosonic one-matrix model {\tii}. With this
result we can already find the form of the discrete string equation
for
arbitrary genus.

Writing $Z=e^{N^2 F_S}$, eq. {\tii} implies that the fermionic
independent piece of the supersymmetric free energy $F_S^{(0)}$ is
related
to the free energy of the one-matrix model ${\cal F_B}$ (up to an
irrelevant
additive constant):

$$
F_S^{(0)}={{\cal F_B}\over 2}.
\eqn\tvii
$$

\no Since the dependence of the free energy on $g_0$ is trivial, ${\p
 F/ \p
g_0}=-1/\L$, {\tvii} implies a rescaling of the cosmological
constant $\L_S=2\L_{\cal B}$. This relation can also be obtained, if
we
evaluate the bosonic piece of the $L_0$ constraint {\wxxv}:

$$
\L_S=2\sum_{k=1}^\infty kg_k\p_{\L_{\cal B}}\left(-\L_{\cal B}^2{\p
{\cal
F_B}\over \p g_k}\right).
\eqn\tviii
$$

\no Up to a factor of two, the cosmological constant satisfies the
same string equation as in the one-matrix model {\axxiv}.

Including the fermionic couplings, the following equality for the
second order in fermionic couplings holds:

$$
{\p^2 Z_S(2N) \over \p \xi_{k+{1\over 2}}\p
\xi_{n+{1\over 2}}}={1\over 2^{N-1}}{ 2N \choose N}
\left(
{\p^2 {\cal Z_B}(N)\over \p g_{k+1}\p g_n}{\cal Z_B}(N)-{\p {\cal
Z_B}(N) \over \p g_{k+1}}{\p {\cal Z_B}(N)\over \p g_n}\right)
-(k\leftrightarrow n).
\eqn\tix
$$

\no This formula is purely bosonic in the sense that after derivation
we set the fermionic couplings to zero. The simplest example is $N=1$

$$\eqalign{
{\p^2 Z_S (2)\over \p \xi_{k+{1\over 2}}\p \xi_{n+{1\over 2}}}&=
\int d\l_1 d\l_2(\l_1-\l_2)(\l_1^n\l_2^k-\l_1^k
\l_2^n)e^{\sum g_k (\l_1^k+\l_2^k)}\cr
&\cr
&=\int d\l_1 d\l_2(\l_1^{k+1}\l_2^n+\l_2^{k+1}\l_1^j -\l_1^{n+1}
\l_2^k-\l_2^{n+1}\l_1^k)e^{\sum g_k (\l_1^k+\l_2^k)}\cr
&\cr
&=2\left(
{\p^2 {\cal Z_B}(1)\over \p g_{k+1}\p g_n}{\cal Z_B}(1)-{\p {\cal
Z_B}(1) \over \p g_{k+1}}{\p {\cal Z_B}(1)\over \p g_n}\right)
-(k\leftrightarrow n).\cr}
\eqn\ttix
$$

\no The general proof can be done with similar methods as the
previous ones. We have to take into account the identity:

$$
\sum_{\alpha,\beta=1}^{2N} \l_\alpha^k \l_\beta^n \int
\prod_{i=1}^{2N}
d\theta_i \D (\l ,\theta) \theta_\alpha \theta_\beta=
$$
$$
{1\over 2^{N-1}}
\sum_{I,J}
S^{I,J}\Delta^2(I)\Delta^2(J)
\sum_{\alpha,\beta=1}^N (\l_{i_{\alpha}}^{k+1}\l_{i_{\beta}}^n
-\l_{i_{\alpha}}^{k+1}\l_{j_{\beta}}^n)-(k\leftrightarrow n) .
\eqn\tx
$$

\no This relation can be proved by comparing a general coefficient
${\l_i}^n{\l_j}^k$ from the left and r.h.s. of the above equation. We
have to use the property {\tiii}. Equations {\tvii} and {\tix} imply
that the free energy, until the second order in fermionic couplings,
is given by:

$$
F_S={{\cal F_B}\over 2}-{1\over 2}\sum_{k,n}\xi_{k+{1\over
2}}\xi_{n+{1\over 2}}{\p^2 {\cal F_B}\over \p g_{k+1} \p g_n}.
\eqn\txi
$$

Since the solution to the continuum super-Virasoro constraints will
be
analysed in a moment, it will become clear that {\txi} is the
complete
expansion of the free energy and no dependence on more than two
fermionic
couplings appears. This characteristic has been confirmed by McArthur
\REF\mcaone{I.~N.~McArthur, ``The Partition Function of the
Supersymmetric Eigenvalue Model'', {\it Mod.~Phys.~Lett.} {\bf A8}
(1993) 3355.}
[\mcaone]
using methods similar to the previous ones. Equation {\txi} gives a
representation of the model {\wxxx} in terms of the one-matrix model
since the fermionic variables $\theta_i$ have been integrated
out.

\section{\secfourfive}

\no In section~4.1 we saw that the super-Virasoro constraints are
described by the super-energy-momentum tensor of a ${\widehat c}=1$
superconformal field theory. The modes of the supercurrent
$T_F={1\over2}\sum G_{n+{1\over 2}}z^{-n-2}$ are described in terms
of the coupling constants of the model for $n\geq -1$:

$$
G_{n+{1\over 2}}={t_0\tau_0\over 4\k^2 }\delta_{n,-1}+\sum_{k\geq
0}\left(k+{1\over 2}\right)t_k{\p \over \p \tau_{n+k+1}}+
{\tau_0\over 2}{\p
\over \p t_n}+\sum_{k\geq 0}\tau_{k+1}{\p \over \p t_{n+k+1}}
+\k^2  \sum _{k=0}^{n}{\p^2\over \p t_k \p \tau_{n-k}}.
\eqn\yi
$$

\no The constraints imposed on the partition function
$Z_S(t_k,\tau_k)=\exp(F_S)$ are:

$$
G_{n+{1\over 2}} Z_S(t_k,\tau_k)=0 \qquad {\rm for} \qquad n\geq -1.
\eqn\yii
$$

\no We now present the solution to these constraints in the double
scaling limit. The constraints determine that in this limit the model
{\wxxx} is described by the free energy

$$
\k^2 DF_S=D^{-1}u-2\tau D \tau,
\eqn\yiii
$$

\no where $D=\p / \p t_0$ and $t$ is the renormalized cosmological
constant satisfying the string equation

$$
t=-\sum_{k\geq 1}(2k+1) t_k R_k[u].
\eqn\yiv
$$

\no The non-perturbative form of the fermionic scaling variable
$\t=\p F_s/ \p \t_0-\t_0/2$ is

$$
\tau=-\sum_{k\geq 0}\tau_k R_k[u].
\eqn
\yyvi
$$

\no As usual, the bosonic piece of the two-point function of the
puncture operator $u$  satisfies the KdV-flow equation

$$
{\p u \over \p t_n}=D R_{n+1}[u],
\eqn\yv
$$

\no where $R_k[u]$ are the  Gel'fand-Dikii polynomials.

To show that eqs. {\yiii}, {\yiv} and {\yyvi} are the solution to the
super-Virasoro conditions {\yii}, it is enough to show that the
$G_{-{1/ 2}}$ and the $G_{3/ 2}$ constraints are satisfied, since the
algebra {\wxxvii} guarantees for all the remaining constraints. That
this is indeed the case can be shown by simple calculations, where we
have used well-known relations of the one-matrix model, like the
recurrence relations of the Gel'fand-Dikii polynomials.

This form of the free energy, the string equation {\yiv} and the form
of $\t$ {\yyvi} coincide with the solution presented in sections 4.2
and 5.2 for the first few orders of the genus expansion. The free
energy {\yiii} can be written as a function of the free
energy of the one-matrix model; this is very similar to the relation
{\txi}
of the discrete model and has an expansion at most quadratic in the
fermionic couplings:

$$
F_S={\cal F_B}-\sum_{{\scriptstyle n\geq 0 \atop \scriptstyle k \geq
1}}
\tau_n \tau_k
{\p^2{\cal F_B}\over \p t_n\p t_{k-1}}.
\eqn\yvii
$$

The form of the free energy and the string equation obtained imply
that the bosonic piece of the model is equivalent to an ordinary
one-matrix model in the double scaling limit, where the constraints
act on the square root of the partition function $L_n
\sqrt{{\cal Z_B}}=0$ for $n\geq -1$ [\dvv].

The fermionic dependence of the
free energy determines that correlation functions involving more than
two
fermionic operators vanish. These
properties were known to hold for the first orders of the genus
expansion, as shown in sections 4.2 and 5.2, and are in agreement
with the previous analysis of the discrete model.

The correlation functions in the double scaling limit follow from
{\yiii}, {\yiv} and {\yyvi}. As a function of the
two-point function of the puncture operator $U=u-2\tau D^2 \tau$,
they are given by:

$$
{\p  U \over \p \tau_k} =2D \left(R_k[u] \DDLR \tau \right),\qquad
\qquad
{\p U \over \p t_k} =DR_{k+1}[u]-2{\p \over \p t_k}\left(\tau
D^2\tau\right).
\eqn\yviii
$$

The scaling dimensions of the scaling operators {\yviii}
and the string susceptibility obtained from {\yiv} are in agreement
with
the results of $(4m,2)$-minimal superconformal models coupled to
2D-supergravity [\dhk,\pz]. From {\yviii} we obtain for the first
non-trivial bosonic flow a set of two
equations:

$$
{\p u \over \p t_1}=D(3 u^2+\k^2 D^2 u),\qquad\qquad
{\p \tau \over \p t_1}=6uD\tau +\k^2  D^3 \tau,
\eqn\yix
$$

\no where the first one is the ordinary KdV equation. These equations
are invariant under the global supersymmetric transformations

$$
\delta u =\epsilon D \tau \qquad {\rm and} \qquad \delta
\tau=\epsilon u,
\eqn\yyviii
$$

\no where $\epsilon$ is a constant anticommuting parameter. This
property is known to hold for the supersymmetric
extensions of KdV of Manin-Radul [\mr] or Mulase-Rabin
\REF\rabin{J.~M.~Rabin, ``The Geometry of the KP Flows'', {\it
Commun.~Math.~Phys.} {\bf 137} (1991) 533;
M.~Mulase, `` A New Super KP System and a Characterization
of the Jacobians of Arbitrary Algebraic Curves'',
{\it J.~Diff.~Geom.} {\bf 34} (1991) 651.}
[\rabin].
We will introduce the supersymmetric extensions of the general KP
hierarchy in the next section and restrict ourselves for the moment
to describing some properties of the correlation functions.
We can write the form of our generalized KdV equation in terms of
$U$ the variable that is, perhaps, more convenient:

$$
{\p U\over \p t_1}=D(3U^2+\k^2 D^2U-12DU\tau D\tau +6\k^2 D\tau
D^3\tau ),
\eqn\yx
$$

$$
{\p \tau \over \p t_1}=6UD\tau -12\tau D\tau D^2\tau +\k^2  D^3\tau.
\eqn\yxi
$$

\no Equations {\yx} and {\yxi} can independently be derived from the
first orders of the genus expansion, with the method proposed in ref.
[\five].
The bosonic flows satisfy the following  recursion relations that can
be obtained
from the recursion relations of the $R_k[u]$'s:

$$
\pmatrix{ {\p u/ \p t_{k+1}} \cr {\p \tau / \p t_{k+1}}\cr}=
\pmatrix{ \k^2 D^2+2u+2DuD^{-1} & 0 \cr
2D\tau D^{-1} +2D^{-1} D\tau & \k^2 D^2+2u+2D^{-1}u D \cr }
\pmatrix{ {\p u / \p t_k} \cr {\p \tau / \p t_k} \cr }_.
\eqn\yxii
$$

\no The fermionic flows satisfy the relations:

$$
{\p u \over \p \tau_k}=0 \qquad {\rm and} \qquad
{\p \tau \over \p \tau_{k+1}}=(\k^2 D^2+2u+2D^{-1}uD){\p \tau \over
\p \tau_k}.
\eqn\yxiii
$$

The aim of the next section is to introduce the supersymmetric
extensions of the KP hierarchy that are known in the literature, and
to see whether it is possible to find agreement with our correlation
functions.

\section{\secfoursix}

\no In the last section we obtained the form of the correlation
functions of scaling operators in the double scaling limit. To gain
insight into the geometric interpretation of the solution obtained it
would be interesting to see whether we are able to identify the
hierarchy of non-linear differential equations obtained from our
model with one of the supersymmetric extensions of the KdV (or more
general KP) hierarchies known in the literature.

The program of supersymmetrization was initiated by Manin and Radul
[\mr]. They introduced a supersymmetric generalization of the KP
system in the Lax formalism.
The Manin-Radul hierarchy [\mr] is a set of flow equations deforming
the pseudo-differential operator of the form:

$$
Q=\delta+Q_0+Q_1\delta^{-1}+Q_2 \delta^{-2}+\dots
\eqn\qi
$$

\no where

$$
\delta={ \p \over \p \x}+\x{\p \o \p x},
\eqn\qi
$$

\no and the coefficients $u_i(x,\x,T)$ are formal power series in the
even variables $x,T_{2n}$ and the odd variables $\x, T_{2n-1}$, with
$n=1,2,\dots$. We will use the notation $T=(T_1,T_2,...)$.  The
expression for $\d^{-1}$ can be obtained using $\d^{-1}=\d
\p_x^{-1}$, where we have taken into account that $\d^{-2}=\p
_x^{-1}$.

The condition $\delta Q_0+2Q_1=0$ is necessary and sufficient for the
existence of a pseudodifferential operator:

$$
S=1+S_1\delta ^{-1}+S_2\delta ^{-2}+\dots\qquad {\rm with} \qquad
S^{-1}QS=\d.
\eqn\qii
$$

\no In terms of the operator $Q$, the Manin-Radul hierarchy reads:

$$
{\p Q\o \p T_{2n}}=[Q_+^{2n},Q]=-[Q_-^{2n},Q],
\eqn\qiii
$$
$$
{\p Q\o \p T_{2n-1}}=-[Q_-^{2n-1},Q]+\sum_{k=1}^\infty T_{2k-1}{\p
Q\o \p T_{2n+2k-2}},
\eqn\qiv
$$

\no where $Q^n=Q^n_++Q^n_-$ is the decomposition into non-negative
and negative powers of $\delta $. We can also describe the hierarchy
in terms of the operator $S$:

$$
{\p S\o \p T_{2n}}=-[S \delta ^{2n} S^{-1}]_-S
\eqn\qv
$$

$$
{\p S\o \p T_{2n-1}}=-\left[S\left(\delta ^{2n-1}+\sum_{k=1}^\infty
T_{2k-1}\delta ^{2n+2k-2}\right)S^{-1}\right]_-S.
\eqn\qvi
$$

\no The term involving the infinite sum was necessary to make the
Manin-Radul system completely integrable, but this term also made it
quite difficult to understand its geometric meaning.

To understand whether there is a relation between our correlation
functions and this hierarchy, it is important to know the
$\tau$-function. This super-$\t$-function was computed by Ueno and
Yamada
\REF\ue{K.~Ueno and H.~Yamada, "Supersymmetric Extension of the
Kadomtsev-Petviashvili Hierarchy and the Universal Grassman
Manifold", {\it Adv.~Stud.~Pur.~Math.} {\bf 16} (1988) 373.}
[\ue] for the Manin-Radul hierarchy
in terms of superdeterminants. The partition function of the matrix
model can then be identified with the $\tau$-function of the
hierarchy, so that we have a unique way of identifying the variables
from the matrix model with those of the superhierarchy. Di-Francesco
et al. [\difdk] have analysed the role that this
super-$\tau$-function plays in the description of $N=1$ unitary SCFTs
coupled to 2D-supergravity. Using the Lax-pair formalism they
identified the string equation and the relevant primary fields.
However, they came to the conclusion that odd flows are inconsistent
with the string equation, so that it does not seem so clear that this
hierarchy plays a role in 2D-supergravity.

A different supersymmetric generalization of KP, that precisely
differs from the previous one in the odd flows, is the Mulase-Rabin
hierarchy [\rabin]. It is formulated in terms of the operator $S$
rather then in terms of a Lax pair. It is a more natural
supersymmetric generalization of ordinary KP, from the geometric
point if view. The flows are given by:

$$
{\p S\o \p T_{2n}}=-[S \delta ^{2n} S^{-1}]_-S
\eqn\qvii
$$

$$
{\p S\o \p T_{2n-1}}=-[S\p_\x\p^{n-1}S^{-1}]_-S=-[S(\delta ^{2n-1}-\x
\delta ^{2n})S^{-1}]_-S.
\eqn\qviii
$$

\no If we compare these equations with the Manin-Radul hierarchy in
{\qv} and {\qvi}, we realize that the even flows are the same while
the odd flows are completely different. Both hierarchies share the
characteristic that if the fermionic variables are set to zero we
obtain the standard KP hierarchy. This is a common feature of all the
supersymmetric extensions of KdV (KP).

Since the above formulation is rather abstract, we would like first
to compute  explicitly the simplest even and odd flows of the most
general two reduction.
The sKdV equations, i.e. the supersymmetric generalization of KdV,
are obtained by considering the flows obtained from the operator
[\mr]:

$$
L=\delta ^4+v_1\delta +v_0,
\eqn\qix
$$

\no where $v_1$ is fermionic and $v_0$ bosonic and $L=Q^4$. This is
the most general two reduction of both supersymmetric hierarchies,
since the definition of $S$ in terms of $Q$ {\qii} determines
$v_3=v_2=0$. To calculate the flows we have to take into account the
generalized Leibniz rule:

$$\eqalign{
\d(ab)&=\d(a)b+(-)^{\widetilde a}a\d(b), \cr
&\cr
\d^{-2}f&=\sum_{j=0}^\infty (-)^j{\p ^j f\o \p x^j}\p_x^{-1-j}.\cr}
\eqn\qqix
$$

\no Here $\widetilde a=0$ if $a$ is a bosonic function and
$\widetilde a=1$ if $a$ is fermionic. From the definition of $S$ in
terms of $L$,

$$
LS=S{\d^4},
\eqn\www
$$

\no we can obtain a representation of $S$ in terms of $v_1$ and
$v_0$:

$$\eqalign{
&\d ^2S_1=-{v_1\o 2},\cr
&\cr
& \d ^2S_2={v_1 S_1\o 2}-{v_0\o 2},\cr
&\cr
&\d ^2S_3={\d ^2v_1\o 4}-{1\o 2}(v_1\d S_1 +v_1 S_2+v_0 S_1).\cr}
\eqn\qxiv
$$

\no For arbitrary $k$ we have the recursion relation

$$
\d^2 S_{k+1}={(-)^{k+1}\o 2}v_1 S_{k}-{1\o 2}\left(\d^4 +v_1 \d
+v_0\right)S_{k-1}\qquad {\rm for}\qquad k=2,3,4,\dots
\eqn\qqqix
$$

\no These relations are valid for both the Manin-Radul and the
Mulase-Rabin hierarchies, since we only need, to obtain them, the
definition of $S$ in terms of $L$ and not the special form of the
flows.

The first bosonic flow for these hierarchies is:

$$
{\p v_0\o \p T_2}={\p v_0\o \p x}\qquad {\rm and}\qquad
{\p v_1\o \p T_2}={\p v_1\o \p x}.
\eqn\qqix
$$

\no These are the equations for a chiral wave moving to the left. A
similar equation appears also in the ordinary KdV hierarchy. The flow
with respect to $T_4$ is trivial, taking into account the special
form of the operator that we are considering {\qix}. The next even
flow leads us to the generalization of the KdV equation to the
supersymmetric case

$$
{\p v_0 \o \p T_6}={\p \o \p x}\left({1\o 4} {\p^2 v_0 \o \p
x^2}+{3\o 4}v_0^2+{3\o4} v_1\delta v_0\right),
\eqn\qx
$$

$$
{\p v_1 \o \p T_6}={\p \o \p x}\left({1\o 4} {\p^2 v_1 \o \p
x^2}+{3\o 4}v_1\delta v_1+{3\o2} v_1v_0\right).
\eqn\qxi
$$

\no Since the Manin-Radul and the Mulase-Rabin hierarchies agree in
the bosonic flows, these equations are valid for both systems.
Setting $v_1=0$ we get the ordinary KdV equation. Setting $v_0=0$ and
$v_1=w_1+\x w_0$, where $\x$ is a purely fermionic parameter, {\qx}
takes the form:

$$
{\p w_0 \o \p T_6}={\p \o \p x}\left({1\o 4} {\p^2 w_0 \o \p
x^2}+{3\o 4}w_0^2-{3\o4} w_1{\p w_1\o \p x}\right),
\eqn\qxii
$$

$$
{\p w_1 \o \p T_6}={\p \o \p x}\left({1\o 4} {\p^2 w_1 \o \p
x^2}+{3\o 4}w_0w_1\right).
\eqn\qxiii
$$

\no Setting $w_1=0$ we again get the KdV equation. This means that
there are two possible operators that will lead us to the KdV
hierarchy:

$$
L_1=\d^4+\x w_0 \d \qquad {\rm and} \qquad L_2=\d^4+v_0.
\eqn\qqxiii
$$

\no In general, it is not consistent to consider the odd flows of
these operators. The coefficients $v_1$ and $w_1$ are uniquely
determined by the $\tau$-function, so that we have no freedom to set
them to zero and still keep odd flows consistently in the theory.

In the following we will write down the expressions for the simplest
fermionic flows of the operator \qix . First we consider the
Mulase-Rabin case. The form of the first fermionic flow is then given
by

$$
{\p S_1\o \p T_1}={\p S_1\o \p \x}\qquad {\rm and}\qquad {\p S_2\o \p
T_1}
=-S_3+S_1S_2+{\p S_2 \o \p\x}.
\eqn\qxv
$$

\no The second fermionic flow satisfies

$$
{\p S_1\o \p T_3}=\d\left(\d ^2S_1+\d S_2+S_3 -S_1 S_2\right)-(\d
S_1)^2.
\eqn\qxvi
$$

\no The fermionic flows are local in terms of the variables $S_i$,
but not in terms of $v_1$ and $v_0$. For $L_1$ we can write the
second fermionic flow in the following way:

$$
{\p \x w_0\o \p T_3}={1\o 4}\left[\left(\d^{-1}\x w_0\right) \d\x
w_0+\d^3\x w_0\right].
\eqn\qxv
$$

\no This equation can be expanded in the number of fermionic
couplings. The purely bosonic part of the resulting equation leads us
to $\d^2w_0-w_0\d^{-2}w_0=0$, which clearly is inconsistent. Thus we
cannot allow the operator $L_1$ as a reduction of Mulase-Rabin and
take the odd flows into account at the same time. The operator
$L=\d^4+v_1\d$ was considered in
\REF\mcarthur{I.~N.~McArthur, ``Two Reduction of the Super-KP
Hierarchy'', {\it Commun.~Math.~Phys.} {\bf 159} (1994) 121.}
ref. [\mcarthur], with a similar conclusion.

The first fermionic flows of the Manin-Radul hierarchy are:

$$
{\p S_1\o \p T_1}=\delta S_1-\sum_{k=1}^\infty T_{2k-1}{\p S_1\o \p
T_{2k}}
\qquad {\rm or}\qquad {\p v_1\o \p T_1}=\delta v_1-\sum_{k=1}^\infty
T_{2k-1}{\p v_1\o \p T_{2k}}
\eqn\qxvi
$$

$$
{\p S_1\o \p T_3}=\d \left(\d^2 S_1+\d S_2 +S_3-S_1S_2-S_1 \d
S_1\right)-\sum_{k=1}^\infty T_{2k-1}{\p S_1\o \p T_{2k+2}}.
\eqn\qxvi
$$

\no The expressions for $\p S_1/ \p T_{2k}$ are easy to write down
because the bosonic flows for $L_1$ have the same expressions as for
ordinary KdV.
More general flows can be obtained (in principle) using recursion
relations
\REF\figueroa{
J.~M.~Figueroa-O'Farrill, J.~Mas and E.~Ramos, ``Integrability and
Bi-Hamiltonian Structure of the Even Order SKDV Hierarchies'', {\it
Rev.~Math.~Phys.} {\bf 3} (1991) 479.}
[\figueroa,\mcarthur].

As we said previously, to see if one of these hierarchies describes
our correlation functions in the double scaling limit we have to find
the identification between the variables of the supersymmetric model
with those of the hierarchy. The easiest way will of course be the
identification of the $\tau$-function with the partition function.
The first even and odd flows of the Manin-Radul hierarchy in terms of
the $\tau$-function take the form [\ue]

$$
\eqalign{
\d(\log \t)& =\d_1(\log \t) =S_1, \cr
&\cr
\d_2(\log \t)& = -\d S_1,\cr
&\cr
\d_3(\log \t)& = -\left( \h (3\d +\d_1)S_2 +\p_x S_1-S_1 \d S_1
\right),\cr
&\cr
\d_4(\log \t)&=2 \d S_3 +\d \p_x S_1-(\d S_1)^2 -2\d( S_1 S_2).\cr}
\eqn\qqxvi
$$

\no Here we have used the notation

$$
\eqalign{
\d_{2l}&={\p \o \p T _{2l}}, \cr
&\cr
\d_{2l-1}&={\p  \o \p T_{2l-1}}+\sum_{k=1}^\infty T_{2k-1}{\p \o \p
T_{2l+2k-2}}.\cr}
\eqn\qqxvii
$$

\no It is easy to show that the following identities hold:

$$
\eqalign{
\d_1(S_1)& =-\d (S_1)\cr
&\cr
\d_1(S_{2j})&=-\d(S_{2j}) -2S_1 S_{2j}+2 S_{2j-1}\cr
&\cr
\d_1(S_{2j-1})&= -\d(S_{2j-1})-2S_1 S_{2j-1}.\cr}
\eqn\qqviii
$$

\no Eqs. {\qqxvi} give a relation between the coefficients $v_0$ and
$v_1$ of the operator $L$ {\qix} and the $\tau$-function. The first
equation implies the condition $v_1=-2\d^3(\log \tau)$. Requiring
compatibility among both KdV equations fixes $v_0=2\d^4(\log \tau)$.
This means that the operator we are going to consider in the
comparison with the Manin-Radul hierarchy is

$$
L=\d^4-2\d^3 (\log \t)  \d + 2\d^4 (\log \t).
\eqn\qqix
$$

\no If we now postulate that the partition function is the
$\tau$-function of the KdV hierarchy, this would mean that (up to
multiplicative factors coming from the identification of the coupling
constants of the matrix model) the two-point function of the puncture
operator

$$
U=u-2\tau D^2 \t
\eqn\qqx
$$

\no satisfies the KdV-flow equations.

To see if we can find agreement with the matrix model, we can check,
for example, if the KdV equations {\qx} and {\qxi} are satisfied. We
already know how to interpret $v_0$ and $v_1$ in terms of correlation
functions. If we now check the KdV equations we arrive at the
conclusion that the fermionic dependence in these equations is
inconsistent!.

We have no other choice for the identification of the variables
between the two theories. We conclude that the Manin-Radul hierarchy
does not reproduce the flows of the super-eigenvalue model.

 This inconsistency will appear also in a comparison with the
Mulase-Rabin hierarchy, since we are considering an even flow.
However in this case we have the difficulty that there does not exist
a formulation in terms of a $\tau$-function yet
\REF\Mulase{M.~Mulase and J.~Rabin, private communication.}
[\Mulase],
so that in principle the connection between $v_0$ and $v_1$ and the
matrix-model variables could be different.

The above identification is the simplest generalization we could
think of, since it is a formulation in terms of the two-point
function of the puncture operator.

Next, we would like to remark that there exists another extension of
the KdV hierarchy that has been proposed by Kupershmidt
\REF\ck{M.~Chaichian and P.~P.~Kulish, ``Superconformal Algebras and
their Relation to Integrable System'', {\it Phys.~Lett.} {\bf 183B}
(1987) 169; P.~Mathieu, `` Superconformal Algebra and Supersymmetric
Korteweg-de Vries Equation'', {\it Phys.~Lett.} {\bf 203B}
(1988) 287; P.~Mathieu, ``Supersymmetric Extension of the Korteweg-de
Vries Equation'', {\it J.~Math.~Phys.} {\bf 29} (1988) 2499;
P.~Mathieu, ``Hamiltonian Structure of Graded and Super Evolution
Equations'', {\it Lett.~Math.~Phys.} {\bf 16} (1988) 199; A.~Bilal
and J.~L.~Gervais, ``Superconformal Algebra and Super KDV Equation:
Two Infinite Families of Polynomial Function with Vanishing Poisson
Brackets'', {\it Phys.~Lett.} {\bf 211B} (1988) 95.}
\REF\kuper{B.~A.~Kupershmidt, ``Super Korteweg-de Vries Equations
Associated to Super-Extensions of the Virasoro Algebra'', {\it
Phys.~Lett.} {\bf 109A} (1985) 417; B.~A.~Kupershmidt, ``Bosons and
Fermions Interacting Integrably with the Korteweg-de Vries Field'',
{\it J.~Phys.} {\bf A17} (1984) L869; B.~A.~Kupershmidt, ``A Super
Korteweg-De Vries Equation: an Integrable System'', {\it Phys.~Lett.}
{\bf 102A} (1984) 213.}
[\ck,\kuper].
In this hierarchy the KdV equation takes the following form:

$$
u_t={\p \o \p x}\left( 3u^2-u_{xx}+3\vp\vp_{x}\right)
\eqn\qxiv
$$

$$
\vp_t=3u_x\vp+6u\vp_x-4\vp_{xxx},
\eqn\qxv
$$

\no where $u$ is a commuting and $\vp$ is the anticommuting variable.
We notice that while the first equation coincides with the first one
of Manin-Radul (after a trivial rescaling), the second one is
different.
In analogy to the bosonic KdV hierarchy, this one has a formulation
in terms of two Hamiltonian structures and a non-trivial Lax
representation. However this system of differential equations is less
interesting for the comparison with the correlation functions of the
super-eigenvalue model, since the odd flows are not present in this
formulation.

We therefore take a more radical point of view and claim that the
hierarchy of super-differential equations presented is a new
supersymmetric extension of the KdV hierarchy. The properties of this
hierarchy have been studied by Figueroa-O'Farrill and Stanciu
\REF\ffsone{J.~M.~Figueroa-O'Farril and S.~Stanciu, ``On a new
Supersymmetric KdV Hierarchy in 2D Quantum Supergravity'', {\it
Phys.~Lett.} {\bf B316} (1993) 282; J.~M.~Figueroa-O'Farril and
S.~Stanciu,
``New Supersymmetrizations of the Generalized KdV hierarchies'',
{\it Mod.~Phys.~Lett.} {\bf A8} (1993) 2125.}
\REF\mcatwo{I.~N.~McArthur, ``Odd Poisson Brackets and the Fermionic
Hierarchy of Becker and Becker'', {\it J.~Phys.~A:~Math.~Gen.} {\bf
26} (1993) 6379.}
[\ffsone] and by McArthur [\mcatwo]. In
\REF\mcarthurthree{I.~N.~McArthur, ``A Discrete Integrable Hierarchy
Related to the Supersymmetric Eigenvalue Model'', Western-Australia
preprint, February 1994.}
ref. [\mcarthurthree] a discrete analogue to this integrable
hierarchy, that is related to the Toda lattice, is found.

It would be interesting to see whether it is possible to find a
connection between this super-hierarchy and topological supergravity
in the same sense as it holds for the one-matrix model [\dvv].
Furthermore it would be interesting to find the connection to the
Landau-Ginzburg approach presented by Di Francesco et al. [\difdk]
and to see whether the new hierarchy explains the discrepancies
found.

\endpage

\chapter{\chafive}

\no The (discrete) Virasoro constraints, which are closely related to
the integrability of the one-matrix model, can be generalized to the
$N=1$ supersymmetric case in terms of the modes of the
energy-momentum tensor of a $\widehat{c}=1$ free massless superfield.
Imposing the super-Virasoro constraints on the partition function,
Alvarez-Gaum\'e et al. [\sloops] have constructed a discrete
super-eigenvalue model formulated in terms of $N$ even and $N$ odd
eigenvalues $(\lambda_i,\theta_i)$ and bosonic and fermionic coupling
constants $(g_k,\xi_{k+1/2})$ respectively. This model can be solved
by evaluating the superloop equations, which nicely generalize the
loop equations found by Kazakov for the one-matrix model [\kazakov].

As a starting point, we have solved this model for generic potentials
and genus zero by evaluating the planar superloop equations. The
solution led to a set of multicritical points described by a planar
string equation that has the same form as in the one-matrix model.
These multicritical points are labeled by an integer $m=1,2, \dots$,
where the string susceptibility takes the value $\G_{str}=-1/m$. We
have obtained explicit expressions for the macroscopic superloop
operators and the microscopic scaling operators, $\langle \s_n
\rangle $ and $\langle \nu_n\rangle $, corresponding to the operators
in the Neveu-Schwarz sector and  Ramond sector of the SCFT coupled to
2D-supergravity. The values of the gravitational scaling dimensions
and the string susceptibility correspond to those of $(4m,2)$-minimal
SCFTs coupled to 2D-supergravity [\pz,\dhk].
The correlation functions of operators in the Neveu-Schwarz sector
agree with those computed in the super-Liouville approach for
spherical topologies [\dfk,\sliouv,\abdallabook].

Considering arbitrary potentials was important in order to derive the
Virasoro constraints in the double scaling limit. These constraints
are formulated in terms of a $\widehat{c}=1$ theory with a
$\IZ_2$-twisted scalar field and a Weyl-Majorana fermion in the
Ramond sector.
The superloop equations in the continuum limit are equivalent to the
double scaled super-Virasoro constraints.

We have solved these superloop equations to all orders in the genus
expansion. Up to genus two we have used the expansion of these
equations in the string coupling constant to obtain the form of the
microscopic scaling operators. For arbitrary genera we have been able
to express the free energy of the super-eigenvalue model in terms of
Gel'fand-Dikii polynomials. Using their recurrence relation it is
possible to show that the super-Virasoro constraints are satisfied.
It turned out, that the bosonic part of the model (fermionic
couplings set to zero) reproduces the solution of the one-matrix
model. We have also been able to write the super-eigenvalue model as
a matrix model in terms of purely bosonic hermitian matrices. The
non-perturbative expression for the free energy of the supersymmetric
theory had a simple connection to the free energy of the discrete
one-matrix model. This relation is important in order to find the
geometric interpretation of this model and to obtain the Feynman
rules for the discrete theory.

The non-perturbative solution of the continuum theory showed that the
integrable structure that appears behind this model is a new
supersymmetric generalization of the KdV hierarchy.

\endpage

\no {\bf Acknowledgements}

\no I would like to thank my advisor W.~Nahm for giving me the
opportunity to carry out the research work of this thesis at the
Theory Division of CERN. The collaboration with my advisor
L.~Alvarez-Gaum\'e and his guidance has been very important for me
and I would like to give my warmest thanks to him. I thank very
specially my sister K.~Becker for an intensive and fruitful
collaboration during the years. It is a pleasure to acknowledge
discussions on different aspects of string theory and 2D-quantum
gravity with L.~Alvarez-Gaum\'e,  J.~F.~Barb\'on, M.~Bershadsky,
M.~Douglas, R.~Empar\'an, K.~Kiritsis, D.~Kutasov, W.~Lerche,
J.~L.~Ma\~nes, G.~Moore, M.~Mulase, W.~Nahm, J.~Rabin, C.~Vafa,
E.~Verlinde and A.~Zadra.
I thank W.~Nahm for a careful reading of the manuscript.
I am grateful to E.~Abdalla and M.~C.~B.~Abdalla for providing me
with a copy of their book [\abdallabook] prior to publication.
Finally, I thank the Theory Division of CERN for hospitality and the
Graduiertenf\"orderung des Landes NRW for financial support.
\endpage
\refout
\end